\documentstyle[aps]{revtex}

\def\setRevDate $#1 #2 #3${#2}
\def\TeXdrawId{\setRevDate $Date: 1995/12/19 16:40:42 $ TeXdraw V2R0}

\chardef\catamp=\the\catcode`\@
\catcode`\@=11

\long                              
\def\centertexdraw #1{\hbox to \hsize{\hss
                                      \btexdraw #1\etexdraw
                                      \hss}}

\def\btexdraw {\x@pix=0             \y@pix=0
               \x@segoffpix=\x@pix  \y@segoffpix=\y@pix
               \t@exdrawdef
               \setbox\t@xdbox=\vbox\bgroup\offinterlineskip
                   \global\d@bs=0           
                   \global\t@extonlytrue    
                   \global\p@osinitfalse
                   \s@avemove \x@pix \y@pix 
                   \m@pendingfalse
                   \global\p@osinitfalse    
                   \p@athfalse
                   \the\everytexdraw}

\def\etexdraw {\ift@extonly \else
                 \t@drclose      
               \fi
               \egroup           
               \ifdim \wd\t@xdbox>0pt
                 \t@xderror {TeXdraw box non-zero size,
                             possible extraneous text}%
               \fi
               \vbox {\offinterlineskip
                      \pixtobp \xminpix \l@lxbp  \pixtobp \yminpix \l@lybp
                      \pixtobp \xmaxpix \u@rxbp  \pixtobp \ymaxpix \u@rybp
                      \hbox{\t@xdinclude 
                        [{\l@lxbp},{\l@lybp}][{\u@rxbp},{\u@rybp}]{\p@sfile}}%
                      \pixtodim \xminpix \t@xpos  \pixtodim \yminpix \t@ypos
                      \kern \t@ypos
                      \hbox {\kern -\t@xpos
                             \box\t@xdbox       
                             \kern \t@xpos}%
                      \kern -\t@ypos\relax}}

\def\drawdim #1 {\def\d@dim{#1\relax}}

\def\setunitscale #1 {\edef\u@nitsc{#1}%
                      \realmult \u@nitsc \s@egsc \d@sc}
\def\relunitscale #1 {\realmult {#1}\u@nitsc \u@nitsc
                      \realmult \u@nitsc \s@egsc \d@sc}
\def\setsegscale #1 {\edef\s@egsc {#1}%
                     \realmult \u@nitsc \s@egsc \d@sc}
\def\relsegscale #1 {\realmult {#1}\s@egsc \s@egsc
                     \realmult \u@nitsc \s@egsc \d@sc}

\def\bsegment {\ifp@ath
                 \f@lushbs
                 \f@lushmove
               \fi
               \begingroup
               \x@segoffpix=\x@pix
               \y@segoffpix=\y@pix
               \setsegscale 1
               \global\advance \d@bs by 1\relax}
\def\esegment {\endgroup
               \ifnum \d@bs=0
                 \writetx {es}%
               \else
                 \global\advance \d@bs by -1
               \fi}

\def\savecurrpos (#1 #2){\getsympos (#1 #2)\a@rgx\a@rgy
                         \s@etcsn \a@rgx {\the\x@pix}%
                         \s@etcsn \a@rgy {\the\y@pix}}
\def\savepos (#1 #2)(#3 #4){\getpos (#1 #2)\a@rgx\a@rgy
                            \coordtopix \a@rgx \t@pixa
                            \advance \t@pixa by \x@segoffpix
                            \coordtopix \a@rgy \t@pixb
                            \advance \t@pixb by \y@segoffpix
                            \getsympos (#3 #4)\a@rgx\a@rgy
                            \s@etcsn \a@rgx {\the\t@pixa}%
                            \s@etcsn \a@rgy {\the\t@pixb}}

\def\linewd #1 {\coordtopix {#1}\t@pixa
                \f@lushbs
                \writetx {\the\t@pixa\space sl}}
\def\setgray #1 {\f@lushbs
                 \writetx {#1 sg}}
\def\lpatt (#1){\listtopix (#1)\p@ixlist
                \f@lushbs
                \writetx {[\p@ixlist] sd}}

\def\lvec (#1 #2){\getpos (#1 #2)\a@rgx\a@rgy
                  \s@etpospix \a@rgx \a@rgy
                  \writeps {\the\x@pix\space \the\y@pix\space lv}}
\def\rlvec (#1 #2){\getpos (#1 #2)\a@rgx\a@rgy
                   \r@elpospix \a@rgx \a@rgy
                   \writeps {\the\x@pix\space \the\y@pix\space lv}}
\def\move (#1 #2){\getpos (#1 #2)\a@rgx\a@rgy
                  \s@etpospix \a@rgx \a@rgy
                  \s@avemove \x@pix \y@pix}
\def\rmove (#1 #2){\getpos (#1 #2)\a@rgx\a@rgy
                   \r@elpospix \a@rgx \a@rgy
                   \s@avemove \x@pix \y@pix}

\def\lcir r:#1 {\coordtopix {#1}\t@pixa
                \writeps {\the\t@pixa\space cr}%
                \r@elupd \t@pixa \t@pixa
                \r@elupd {-\t@pixa}{-\t@pixa}}
\def\fcir f:#1 r:#2 {\coordtopix {#2}\t@pixa
                     \writeps {\the\t@pixa\space #1 fc}%
                     \r@elupd \t@pixa \t@pixa
                     \r@elupd {-\t@pixa}{-\t@pixa}}
\def\lellip rx:#1 ry:#2 {\coordtopix {#1}\t@pixa
                     \coordtopix {#2}\t@pixb
                     \writeps {\the\t@pixa\space \the\t@pixb\space el}%
                     \r@elupd \t@pixa \t@pixb
                     \r@elupd {-\t@pixa}{-\t@pixb}}
\def\fellip f:#1 rx:#2 ry:#3 {\coordtopix {#2}\t@pixa
                     \coordtopix {#3}\t@pixb
                     \writeps {\the\t@pixa\space \the\t@pixb\space #1 fe}%
                     \r@elupd \t@pixa \t@pixb
                     \r@elupd {-\t@pixa}{-\t@pixb}}
\def\larc r:#1 sd:#2 ed:#3 {\coordtopix {#1}\t@pixa
                            \writeps {\the\t@pixa\space #2 #3 ar}}

\def\ifill f:#1 {\writeps {#1 fl}}     
\def\lfill f:#1 {\writeps {#1 fp}}     

\def\htext #1{\def\testit {#1}%
              \ifx \testit\l@paren
                \let\next=\h@move
              \else
                \let\next=\h@text
              \fi
              \next {#1}}

\def\rtext td:#1 #2{\def\testit {#2}%
                    \ifx \testit\l@paren
                      \let\next=\r@move
                    \else
                      \let\next=\r@text
                    \fi
                    \next td:#1 {#2}}

\def\textref h:#1 v:#2 {\ifx #1R%
                          \edef\l@stuff {\hss}\edef\r@stuff {}%
                        \else
                          \ifx #1C%
                            \edef\l@stuff {\hss}\edef\r@stuff {\hss}%
                          \else  
                            \edef\l@stuff {}\edef\r@stuff {\hss}%
                          \fi
                        \fi
                        \ifx #2T%
                          \edef\t@stuff {}\edef\b@stuff {\vss}%
                        \else
                          \ifx #2C%
                            \edef\t@stuff {\vss}\edef\b@stuff {\vss}%
                          \else  
                            \edef\t@stuff {\vss}\edef\b@stuff {}%
                          \fi
                        \fi}

\def\avec (#1 #2){\getpos (#1 #2)\a@rgx\a@rgy
                  \s@etpospix \a@rgx \a@rgy
                  \writeps {\the\x@pix\space \the\y@pix\space (\a@type) %
                            \the\a@lenpix\space \the\a@widpix\space av}}

\def\ravec (#1 #2){\getpos (#1 #2)\a@rgx\a@rgy
                   \r@elpospix \a@rgx \a@rgy
                   \writeps {\the\x@pix\space \the\y@pix\space (\a@type) %
                             \the\a@lenpix\space \the\a@widpix\space av}}

\def\arrowheadsize l:#1 w:#2 {\coordtopix{#1}\a@lenpix
                              \coordtopix{#2}\a@widpix}
\def\arrowheadtype t:#1 {\edef\a@type{#1}}

\def\clvec (#1 #2)(#3 #4)(#5 #6)%
           {\getpos (#1 #2)\a@rgx\a@rgy
            \coordtopix \a@rgx\t@pixa
            \advance \t@pixa by \x@segoffpix
            \coordtopix \a@rgy\t@pixb
            \advance \t@pixb by \y@segoffpix
            \getpos (#3 #4)\a@rgx\a@rgy
            \coordtopix \a@rgx\t@pixc
            \advance \t@pixc by \x@segoffpix
            \coordtopix \a@rgy\t@pixd
            \advance \t@pixd by \y@segoffpix
            \getpos (#5 #6)\a@rgx\a@rgy
            \s@etpospix \a@rgx \a@rgy
            \writeps {\the\t@pixa\space \the\t@pixb\space
                      \the\t@pixc\space \the\t@pixd\space
                      \the\x@pix\space \the\y@pix\space cv}}

\def\drawbb {\bsegment
               \drawdim bp
               \linewd 0.24       
               \setunitscale {\p@sfactor}
               \writeps {\the\xminpix\space \the\yminpix\space mv}%
               \writeps {\the\xminpix\space \the\ymaxpix\space lv}%
               \writeps {\the\xmaxpix\space \the\ymaxpix\space lv}%
               \writeps {\the\xmaxpix\space \the\yminpix\space lv}%
               \writeps {\the\xminpix\space \the\yminpix\space lv}%
             \esegment}

\def\getpos (#1 #2)#3#4{\g@etargxy #1 #2 {} \\#3#4%
                        \c@heckast #3%
                        \ifa@st
                          \g@etsympix #3\t@pixa
                          \advance \t@pixa by -\x@segoffpix
                          \pixtocoord \t@pixa #3%
                        \fi
                        \c@heckast #4%
                        \ifa@st
                          \g@etsympix #4\t@pixa
                          \advance \t@pixa by -\y@segoffpix
                          \pixtocoord \t@pixa #4%
                        \fi}

\def\getsympos (#1 #2)#3#4{\g@etargxy #1 #2 {} \\#3#4%
                     \c@heckast #3%
                     \ifa@st \else
                       \t@xderror {TeXdraw: invalid symbolic coordinate}%
                     \fi
                     \c@heckast #4%
                     \ifa@st \else
                       \t@xderror {TeXdraw: invalid symbolic coordinate}%
                     \fi}

\def\listtopix (#1)#2{\def #2{}%
                      \edef\l@ist {#1 }
                      \m@oretrue
                      \loop
                        \expandafter\g@etitem \l@ist \\\a@rgx\l@ist
                        \a@pppix \a@rgx #2%
                        \ifx \l@ist\empty
                          \m@orefalse
                        \fi
                      \ifm@ore
                      \repeat}

\def\realmult #1#2#3{\dimen0=#1pt
                     \dimen2=#2\dimen0
                     \edef #3{\expandafter\c@lean\the\dimen2}}

\def\intdiv #1#2#3{\t@counta=#1
                   \t@countb=#2
                   \ifnum \t@countb<0
                      \t@counta=-\t@counta
                      \t@countb=-\t@countb
                   \fi
                   \t@countd=1                    
                   \ifnum \t@counta<0
                      \t@counta=-\t@counta
                      \t@countd=-1
                   \fi
                   \t@countc=\t@counta  \divide \t@countc by \t@countb
                   \t@counte=\t@countc  \multiply \t@counte by \t@countb
                   \advance \t@counta by -\t@counte
                   \t@counte=-1
                   \loop
                     \advance \t@counte by 1
                   \ifnum \t@counte<16                  
                       \multiply \t@countc by 2           
                       \multiply \t@counta by 2           
                       \ifnum \t@counta<\t@countb \else   
                         \advance \t@countc by 1          
                         \advance \t@counta by -\t@countb 
                       \fi
                   \repeat
                   \divide \t@countb by 2         
                   \ifnum \t@counta<\t@countb     
                     \advance \t@countc by 1
                   \fi
                   \ifnum \t@countd<0             
                     \t@countc=-\t@countc
                   \fi
                   \dimen0=\t@countc sp           
                   \edef #3{\expandafter\c@lean\the\dimen0}}

\def\coordtopix #1#2{\dimen0=#1\d@dim
                     \dimen2=\d@sc\dimen0
                     \t@counta=\dimen2          
                     \t@countb=\s@ppix
                     \divide \t@countb by 2
                     \ifnum \t@counta<0         
                       \advance \t@counta by -\t@countb
                     \else
                       \advance \t@counta by \t@countb
                     \fi
                     \divide \t@counta by \s@ppix
                     #2=\t@counta}

\def\pixtocoord #1#2{\t@counta=#1%
                     \multiply \t@counta by \s@ppix
                     \dimen0=\d@sc\d@dim
                     \t@countb=\dimen0
                     \intdiv \t@counta \t@countb #2}

\def\pixtodim #1#2{\t@countb=#1%
                   \multiply \t@countb by \s@ppix
                   #2=\t@countb sp\relax}

\def\pixtobp #1#2{\dimen0=\p@sfactor pt
                  \t@counta=\dimen0
                  \multiply \t@counta by #1%
                  \ifnum \t@counta < 0             
                    \advance \t@counta by -32768
                  \else
                    \advance \t@counta by 32768
                  \fi
                  \divide \t@counta by 65536
                  \edef #2{\the\t@counta}}
                  
\newcount\t@counta    \newcount\t@countb   
\newcount\t@countc    \newcount\t@countd
\newcount\t@counte
\newcount\t@pixa      \newcount\t@pixb     
\newcount\t@pixc      \newcount\t@pixd

\newdimen\t@xpos      \newdimen\t@ypos

\newcount\xminpix      \newcount\xmaxpix
\newcount\yminpix      \newcount\ymaxpix

\newcount\a@lenpix     \newcount\a@widpix

\newcount\x@pix        \newcount\y@pix
\newcount\x@segoffpix  \newcount\y@segoffpix
\newcount\x@savepix    \newcount\y@savepix

\newcount\s@ppix       

\newcount\d@bs

\newcount\t@xdnum
\global\t@xdnum=0

\newbox\t@xdbox

\newwrite\drawfile

\newif\ifm@pending
\newif\ifp@ath
\newif\ifa@st
\newif\ifm@ore
\newif \ift@extonly
\newif\ifp@osinit

\newtoks\everytexdraw

\def\l@paren{(}
\def\a@st{*}

\catcode`\%=12
  \def\p@b {
\catcode`\%=14
\catcode`\{=12  \catcode`\}=12  \catcode`\u=1 \catcode`\v=2
  \def\l@br u{v  \def\r@br u}v
\catcode `\{=1  \catcode`\}=2   \catcode`\u=11 \catcode`\v=11

{\catcode`\p=12 \catcode`\t=12
 \gdef\c@lean #1pt{#1}}

\def\sppix#1/#2 {\dimen0=1#2 \s@ppix=\dimen0
                 \t@counta=#1%
                 \divide \t@counta by 2
                 \advance \s@ppix by \t@counta
                 \divide \s@ppix by #1
                 \t@counta=\s@ppix
                 \multiply \t@counta by 65536       
                 \advance \t@counta by 32891        
                 \divide \t@counta by 65782         
                 \dimen0=\t@counta sp
                 \edef\p@sfactor {\expandafter\c@lean\the\dimen0}}

\def\g@etargxy #1 #2 #3 #4\\#5#6{\def #5{#1}%
                           \ifx #5\empty
                             \g@etargxy #2 #3 #4 \\#5#6
                           \else
                             \def #6{#2}%
                             \def\next {#3}%
                             \ifx \next\empty \else
                               \t@xderror {TeXdraw: invalid coordinate}%
                             \fi
                           \fi}

\def\c@heckast #1{\expandafter
                  \c@heckastll #1\\}
\def\c@heckastll #1#2\\{\def\testit {#1}%
                        \ifx \testit\a@st
                          \a@sttrue
                        \else
                          \a@stfalse
                        \fi}

\def\g@etsympix #1#2{\expandafter
               \ifx \csname #1\endcsname \relax
                 \t@xderror {TeXdraw: undefined symbolic coordinate}%
               \fi
               #2=\csname #1\endcsname}

\def\s@etcsn #1#2{\expandafter
                  \xdef\csname#1\endcsname {#2}}

\def\g@etitem #1 #2\\#3#4{\edef #4{#2}\edef #3{#1}}
\def\a@pppix #1#2{\edef\next {#1}%
                  \ifx \next\empty \else
                    \coordtopix {#1}\t@pixa
                    \ifx #2\empty
                      \edef #2{\the\t@pixa}%
                    \else
                      \edef #2{#2 \the\t@pixa}%
                    \fi
                  \fi}

\def\s@etpospix #1#2{\coordtopix {#1}\x@pix
                     \advance \x@pix by \x@segoffpix
                     \coordtopix {#2}\y@pix
                     \advance \y@pix by \y@segoffpix
                     \u@pdateminmax \x@pix \y@pix}

\def\r@elpospix #1#2{\coordtopix {#1}\t@pixa
                     \advance \x@pix by \t@pixa
                     \coordtopix {#2}\t@pixa
                     \advance \y@pix by \t@pixa
                     \u@pdateminmax \x@pix \y@pix}

\def\r@elupd #1#2{\t@counta=\x@pix
                  \advance\t@counta by #1%
                  \t@countb=\y@pix
                  \advance\t@countb by #2%
                  \u@pdateminmax \t@counta \t@countb}

\def\u@pdateminmax #1#2{\ifnum #1>\xmaxpix
                          \global\xmaxpix=#1%
                        \fi
                        \ifnum #1<\xminpix
                          \global\xminpix=#1%
                        \fi
                        \ifnum #2>\ymaxpix
                          \global\ymaxpix=#2%
                        \fi
                        \ifnum #2<\yminpix
                          \global\yminpix=#2%
                        \fi}

\def\s@avemove #1#2{\x@savepix=#1\y@savepix=#2%
                    \m@pendingtrue
                    \ifp@osinit \else
                      \global\p@osinittrue
                      \global\xminpix=\x@savepix \global\yminpix=\y@savepix
                      \global\xmaxpix=\x@savepix \global\ymaxpix=\y@savepix
                    \fi}

\def\f@lushmove {\global\p@osinittrue
                 \ifm@pending
                   \writetx {\the\x@savepix\space \the\y@savepix\space mv}%
                   \m@pendingfalse
                   \p@athfalse
                 \fi}

\def\f@lushbs {\loop
                 \ifnum \d@bs>0
                   \writetx {bs}%
                   \global\advance \d@bs by -1
               \repeat}
               
\def\h@move #1#2 #3)#4{\move (#2 #3)%
                       \h@text {#4}}
\def\h@text #1{\pixtodim \x@pix \t@xpos
               \pixtodim \y@pix \t@ypos
               \vbox to 0pt{\normalbaselines
                            \t@stuff
                            \kern -\t@ypos
                            \hbox to 0pt{\l@stuff
                                         \kern \t@xpos
                                         \hbox {#1}%
                                         \kern -\t@xpos
                                         \r@stuff}%
                            \kern \t@ypos
                            \b@stuff\relax}}

\def\r@move td:#1 #2#3 #4)#5{\move (#3 #4)%
                             \r@text td:#1 {#5}}
\def\r@text td:#1 #2{\vbox to 0pt{\pixtodim \x@pix \t@xpos
                                  \pixtodim \y@pix \t@ypos
                                  \kern -\t@ypos
                                  \hbox to 0pt{\kern \t@xpos
                                               \rottxt {#1}{\z@sb {#2}}%
                                               \hss}%
                                  \vss}}
\def\z@sb #1{\vbox to 0pt{\normalbaselines
                          \t@stuff
                          \hbox to 0pt{\l@stuff \hbox {#1}\r@stuff}%
                          \b@stuff}}

\ifx \rotatebox\@undefined
  \def\rottxt #1#2{\bgroup
                     #2%
                   \egroup}
\else
  \let\rottxt=\rotatebox
\fi

\ifx \t@xderror\@undefined
  \let\t@xderror=\errmessage
\fi

\def\t@exdrawdef {\sppix 300/in            
                  \drawdim in              
                  \edef\u@nitsc {1}
                  \setsegscale 1           
                  \arrowheadsize l:0.16 w:0.08
                  \arrowheadtype t:T
                  \textref h:L v:B }

\ifx \includegraphics\@undefined
  \def\t@xdinclude [#1,#2][#3,#4]#5{%
    \begingroup                           
      \message {<#5>}%
      \leavevmode
      \t@counta=-#1
      \t@countb=-#2%
      \setbox0=\hbox{%
        \includegraphics{#5}}%
      \t@ypos=#4 bp%
        \advance \t@ypos by -#2 bp%
      \t@xpos=#3 bp%
        \advance \t@xpos by -#1 bp%
      \dp0=0pt \ht0=\t@ypos  \wd0=\t@xpos
      \box0%
    \endgroup}
\else
  \let\t@xdinclude=\includegraphics
\fi

\def\writeps #1{\f@lushbs
                \f@lushmove
                \p@athtrue
                \writetx {#1}}
\def\writetx #1{\ift@extonly
                  \global\t@extonlyfalse
                  \t@xdpsfn \p@sfile
                  \t@dropen \p@sfile
                \fi
                \w@rps {#1}}
\def\w@rps #1{\immediate\write\drawfile {#1}}

\def\t@xdpsfn #1{%
  \global\advance \t@xdnum by 1
  \ifnum \t@xdnum<10
    \xdef #1{\jobname.ps\the\t@xdnum}
  \else
    \xdef #1{\jobname.p\the\t@xdnum}
  \fi
}
\def\t@dropen #1{%
  \immediate\openout\drawfile=#1%
  \w@rps {\p@b PS-Adobe-3.0 EPSF-3.0}%
  \w@rps {\p@p BoundingBox: (atend)}%
  \w@rps {\p@p Title: TeXdraw drawing: #1}%
  \w@rps {\p@p Pages: 1}%
  \w@rps {\p@p Creator: \TeXdrawId}%
  \w@rps {\p@p CreationDate: \the\year/\the\month/\the\day}%
  \w@rps {50 dict begin}%
  \w@rps {/mv {stroke moveto} def}%
  \w@rps {/lv {lineto} def}%
  \w@rps {/st {currentpoint stroke moveto} def}%
  \w@rps {/sl {st setlinewidth} def}%
  \w@rps {/sd {st 0 setdash} def}%
  \w@rps {/sg {st setgray} def}%
  \w@rps {/bs {gsave} def /es {stroke grestore} def}%
  \w@rps {/fl \l@br gsave setgray fill grestore}%
  \w@rps    { currentpoint newpath moveto\r@br\space def}%
  \w@rps {/fp {gsave setgray fill grestore st} def}%
  \w@rps {/cv {curveto} def}%
  \w@rps {/cr \l@br gsave currentpoint newpath 3 -1 roll 0 360 arc}%
  \w@rps    { stroke grestore\r@br\space def}%
  \w@rps {/fc \l@br gsave setgray currentpoint newpath}%
  \w@rps    { 3 -1 roll 0 360 arc fill grestore\r@br\space def}%
  \w@rps {/ar {gsave currentpoint newpath 5 2 roll arc stroke grestore} def}%
  \w@rps {/el \l@br gsave /svm matrix currentmatrix def}%
  \w@rps    { currentpoint translate scale newpath 0 0 1 0 360 arc}%
  \w@rps    { svm setmatrix stroke grestore\r@br\space def}%
  \w@rps {/fe \l@br gsave setgray currentpoint translate scale newpath}%
  \w@rps    { 0 0 1 0 360 arc fill grestore\r@br\space def}%
  \w@rps {/av \l@br /hhwid exch 2 div def /hlen exch def}%
  \w@rps    { /ah exch def /tipy exch def /tipx exch def}%
  \w@rps    { currentpoint /taily exch def /tailx exch def}%
  \w@rps    { /dx tipx tailx sub def /dy tipy taily sub def}%
  \w@rps    { /alen dx dx mul dy dy mul add sqrt def}%
  \w@rps    { /blen alen hlen sub def}%
  \w@rps    { gsave tailx taily translate dy dx atan rotate}%
  \w@rps    { (V) ah ne {blen 0 gt {blen 0 lineto} if} {alen 0 lineto} ifelse}%
  \w@rps    { stroke blen hhwid neg moveto alen 0 lineto blen hhwid lineto}%
  \w@rps    { (T) ah eq {closepath} if}%
  \w@rps    { (W) ah eq {gsave 1 setgray fill grestore closepath} if}%
  \w@rps    { (F) ah eq {fill} {stroke} ifelse}%
  \w@rps    { grestore tipx tipy moveto\r@br\space def}%
  \w@rps {\p@sfactor\space \p@sfactor\space scale}%
  \w@rps {1 setlinecap 1 setlinejoin}%
  \w@rps {3 setlinewidth [] 0 setdash}%
  \w@rps {0 0 moveto}%
}

\def\t@drclose {%
  \bgroup
    \w@rps {stroke end showpage}%
    \w@rps {\p@p Trailer:}%
    \pixtobp \xminpix \l@lxbp  \pixtobp \yminpix \l@lybp
    \pixtobp \xmaxpix \u@rxbp  \pixtobp \ymaxpix \u@rybp
    \w@rps {\p@p BoundingBox: \l@lxbp\space \l@lybp\space
                              \u@rxbp\space \u@rybp}%
    \w@rps {\p@p EOF}%
  \egroup
  \immediate\closeout\drawfile
}

\catcode`\@=\catamp

\newenvironment{texdraw}{\leavevmode\btexdraw}{\etexdraw}
\def\real{{\bf{R}}}
\def\uno{{\bf{1}}}

\begin{document}
\bibliographystyle{unsrt}

\title{Geometric invariant theory approach to the determination of
ground states of D-wave condensates in isotropic space}

\author{Yu.M. Gufan$^1$, Al.V. Popov$^1$, G. Sartori$^{2,4}$, V. Talamini$^{3,4}$,
G. Valente$^{2,4}$ and E.B. Vinberg$^5$\\
\small
$^{1}$Rostov State University, Rostov on Don, Russia\\
\small$^2$Dipartimento di Fisica, Universit\`a di Padova, Italy\\
\small $^{3}$Dipartimento di Ingegneria Civile, Universit\`a di
Udine, Italy\\ $^{4}$INFN, Sezione di Padova, Italy\\ \small
$^{5}$ Moscow State University, Moscow, Russia}



\maketitle

\begin{abstract}
A complete and rigorous determination of the possible ground
states for D-wave pairing Bose condensates is presented, using a
geometrical invariant theory approach to the problem. The order
parameter is argued to be a vector, transforming according to a ten
dimensional real representation of the group $G=${\bf
O}$_3\otimes${\bf U}$_1\times \langle{\cal T}\rangle$. We determine
the equalities and inequalities defining the orbit space of this
linear group and its symmetry strata, which are in a one-to-one
correspondence with the possible distinct phases of the system. We find 15 allowed phases (besides the unbroken one),
with different symmetries, that we thoroughly determine. The
group-subgroup relations between bordering phases are pointed out.
The perturbative sixth degree corrections to the minimum of a
fourth degree polynomial $G$-invariant free energy, calculated by
Mermin, are also determined.

\


\end{abstract}


\section{Introduction}
Superfluidity and superconductivity are justified on the basis of
the macroscopic condensation of Bose quasi-particles. The
classical Bardeen, Cooper and Schrieffer theory for
superconductivity dates 1957. Soon after, a BCS-type transition
was proposed for the Fermi system $^{3}\mbox{\rm He}$ by Anderson
and Morel \cite{1}. Cooper pair formation was thought to occur in
an $L \neq 0$ state, to take into account the hard core nature of
$^{3}\mbox{\rm He}$ atom interaction. The superfluid phases were
actually observed \cite{1a}, and the nature of p-wave pairing is
now well established for $^{3}\mbox{\rm He}$. The theory of $L
\neq 0$-superfluids is relevant for ``unconventional''
superconductivity too. The high temperature superconducting (HTS) oxides are
anomalous in their non-Fermi liquid normal state properties and in
many cases
share with heavy fermion superconductors unconventional d-wave
pairing behaviour \cite{2,8}. In fact, experiments probing the
phase and the nodes of the gap show a sign reversal of the order
parameter, compatible with the d-wave scenario. Nevertheless,
other experiments on the same HTS compounds point in the direction
of a significant s-wave component. In the presence of external
magnetic fields, defects and interface phenomena there are also
evidences of a mixed pairing symmetry (see {\em e.g.}, the Introduction of
\cite{5bis} and references therein). Such a controversial
experimental situation is reflected in the fact that
the underlying microscopic mechanism inducing
superconductivity in HTS materials is still unclear and is one
of today's major challenges \cite{10a,10b}.

Such a situation has motivated the efforts at studying the
macroscopic properties of unconventional superconductors through
the Landau theory of phase transitions \cite{GufPR12} and
the identification of the order parameter symmetry has become a preliminary task in the construction of viable
models describing the attractive nature of the pairing
interaction.

Direct experimental evidence of an order parameter unconventional
structure lies on multiple phase transitions. The heavy fermion
compounds U$_{1-x}$Th$_x$Be$_{13}$ display four different
superconducting phases in the $T$-$x$ phase diagram and UPt$_3$
displays three superconducting phases in a $T$-$H$ phase diagram
\cite{2}. Furthermore, from power-law temperature behaviour of
thermodynamic and transport properties (e.g. specific heat,
magnetic penetration depth), a non-trivial node
structure for the gap function may be inferred, compatible with high
$L$-pairing. That is the case also for HTS oxides, for which, at
present, clear proofs are lacking of the existence of more than
one superconducting state.

Actually, it must be pointed out that, unlike
superfluid $^{3}\,\mbox{He}$, which is an isotropic fermion system,
Bloch electrons in a superconductor crystal lattice
exhibit, in general, a reduced finite symmetry; in addition, the existence of
imperfections, even in the cleanest samples, can partly destroy the
gap node structure at low energy. So, the very fact that the
experimental observation cannot at present completely unravel the
node structure of the gap function, reinforces the necessity of classifying
all the possible symmetry breaking schemes.

Moreover, the role of Fermi isotropic space beyond a zero-th
order approximation in phenomenological theories, is still
supported by some recent studies: In HTS oxides, low symmetrical
crystal fields (tetragonal and orthorombic), as well as
spontaneous strain, have weak influence on the temperature of the
superconducting phase transition \cite{11} or on the penetration depth
\cite{12}. So, even if the spectrum of the interactions which are
responsible for the pairing (whatever there origins may be) must be
anisotropic in the crystal, it is worth analyzing the possibility
that they act in a way not directly dependent on the crystal
structure (like in isotropic space).

The possible ground states of a high $L$ superfluid has been the
object of intense investigations during the 60's and the 70's.
>From the solution of the state equations, Anderson and Morel
\cite{1} and Mermin \cite{14} identified five different phases, through the minimization of a 4-th
degree Landau potential. Schakel and Bias \cite{16} analyzed the
problem using only group theoretical arguments. Capel and Schakel
\cite{15}, taking advantage of the results in \cite{16}, computed
the possible ground states of condensates driven by p-waves. They
also investigated the consequences of lowering the residual
symmetry by means of
strong spin-orbit interactions. Since, in this case, the Cooper
pair is in a $J=2$ state $(S=1, L=1)$, the order parameter is
represented by a traceless symmetric tensor. The same holds true
for an $(S=0, L=2)$ state order parameter, so their results may be
directly applied to D-wave pairing. According to the authors of
ref. \cite {15}, eleven different phases are allowed. Their
analysis, however, is questionable, since time reversal symmetry
cannot be neglected in the theory of condensate states and the use
of a 4-th degree polynomial Landau free energy strongly limits the
number of phases that would be allowed by the symmetry of the
system.

Our aim is to give a definitive answer to the classification of
possible symmetry breaking patterns in D-wave pairing Bose
condensate, in the framework of the Landau theory of phase
transitions. In this paper, we shall mainly be concerned with the
mathematical aspects of the problem, that will be solved making
use of the geometrical invariant theory approach, proposed in
ref.~\cite{SA}. The strategy is to exploit a
set of basic invariant polynomials of the symmetry group of the
system, as fundamental variables in the description of the phase
space of the system and in the minimization procedure of the free
energy \cite{23}. In the realization of this program, we shall meet
some substantial computational difficulties, that will be overcome
by means of innovative procedures.

\section{The geometrical invariant theory approach to spontaneous symmetry breaking}

Let us briefly recall some basic elements of the geometrical
invariant theory approach, or orbit space approach (see \cite{NC} and references
therein) to
the determination of possible patterns of spontaneous symmetry
breaking.

Let $x\in {\bf R}^n$ be a vector order parameter, transforming
linearly and orthogonally under the compact real symmetry group
$G$, and let $\Phi(\alpha; x)$ be the $G$-invariant free energy,
expressed in terms, also, of state variables $\alpha$. The points
$x_0(\alpha)$, where the function $\phi_\alpha(x) = \Phi(\alpha;
x)$ takes on its absolute minimum, determine the stable phase of
the system, whose residual symmetry is defined by the isotropy
subgroup of $G$ at $x_0$, $G_{x_0}$. Owing to its $G$-invariance,
the free energy is a constant along each $G$-orbit, so, each of its
stationary points is degenerate along the $G$-orbit through it. Since the
isotropy subgroups of $G$ at points of the same orbit are
conjugate in $G$, only the conjugate class, $[G_{x_0}]$, of
$G_{x_0}$ in $G$, i.e. the {\em orbit-type} (or {\em symmetry}) of
the orbit through $x_0$, is physically relevant.

The set of all $G$-orbits, endowed with the quotient topology and
differentiable structure, forms the {\em orbit space}, ${\bf
R}^n/G$, of $G$ and the subset of all the $G$-orbits with the same
orbit-type forms a {\em stratum} of ${\bf R}^n/G$. Phase
transitions take place when, by varying the values of the
$\alpha$'s, the point $x_0(\alpha)$ is shifted to an orbit lying
on a different stratum.

If $\Phi(\alpha;x)$ is a sufficiently general function of the
$\alpha$'s, by varying these parameters, the point $x_0(\alpha)$ can be
shifted to any stratum of ${\bf R}^n/G$. So, {\em the strata are
in a one-to-one correspondence with the symmetry phases allowed by
the $G$-invariance of the free energy}. On the contrary, extra
restrictions on the form of the free energy function, not coming
from G-symmetry requirements (e.g., the assumption that the free
energy is a polynomial of low degree), can limit the number of
allowed phases for the system in its ground state.

Being constant along each $G$-orbit, the free energy may be
conveniently thought of as a function defined in the orbit space
of $G$. This fact can be formalized using some basic results of
invariant theory. In fact, the G-invariant polynomial functions
separate the $G$-orbits, meaning that, for any two distinct
$G$-orbits, there is at least a polynomial $G$-invariant function
assuming different values on them. Moreover, every $G$-invariant polynomial
can be built as a real polynomial function of a {\em finite} set,
$\{p_1(x), \dots ,p_q(x)\}$, of basic polynomial invariants ({\em
integrity basis of the ring of $G$-invariant polynomials}).
Thus, the elements of an integrity basis can be conveniently
used as coordinates of the orbit space points. They
need not, for general compact groups, be algebraically
independent, but can, and will, be chosen to be homogeneous
polynomials in $x$. The number of algebraically independent
elements in a {\em minimal} set of basic polynomial invariants is $n -
\nu$, where $\nu$ is the dimension of a generic (principal) orbit
of $G$. Information on the number and degrees of a minimal set of
basic invariants, and the degrees of the algebraic relations ({\em
syzygies}) among them, can be inferred from the M\"olien function of
$G$.

Let us call $q_0$ the number of independent elements of the set
$\{p\}$. The range of the {\em orbit map}, $x \mapsto p(x) =
(p_1(x), \dots ,p_q(x))\in {\bf R}^q$, yields a realization of the
orbit space of the linear group $G$, as a connected semi-algebraic
surface, i.e. a subset of ${\bf R}^q$, determined by algebraic
equations and inequalities. The orbit space of $G$ will, therefore,
be identified with a closed and connected region, $\widehat S$, of a
$q_0$-dimensional algebraic surface, delimited by lower dimensional
semi-algebraic surfaces.

If an integrity basis has been determined, the equations and
inequalities defining the orbit space of a compact group can
be obtained from a simple recipe. It has been shown, in fact,
that the orbit space of a reductive linear group
can be identified with the semialgebraic variety
formed by the points $p\in {\bf R}^q$, satisfying the following
conditions i) and ii) \cite{SA,PS}:

\begin{itemize}
\item[i)] $p$ lies on the surface, $Z$, defined by the syzygies;
\item[ii)] the $q\times q$ matrix $\widehat P(p)$,
defined by the relations

\begin{equation}
\widehat P_{ab}(p(x)) = \sum_{j=1}^n\partial_j p_a(x)\,\partial_j
p_b(x),\quad \forall x\in {\bf R}^n
\end{equation}
is positive semi-definite and has rank $q_0$ at $p$.
\end{itemize}

The minimum of $\Phi(\alpha;x)$ can be computed as a {\em
constrained} minimum of the function, $\widehat \Phi(\alpha;p)$,
$p\in\widehat S$, defined by

\begin{equation}
\widehat \Phi(\alpha;p(x))= \Phi(\alpha;x),\;\forall x\in \real ^n.\label{1}
\end{equation}

It has also been shown \cite{SA} that the state equation

$$\partial \Phi(\alpha;x)/\partial x_j = 0, \qquad  j=1,\dots
,n,$$
determining the extremal points of $\Phi(\alpha;x)$, is equivalent
to the following equation in orbit space:

\begin{equation}
\sum_{b=1}^q\widehat P_{ab}(p)\,\partial_b \widehat{\Phi}(\alpha;
p)=0, \qquad a=1,\dots ,q,\quad p\in\widehat S.\label{min}
\end{equation}

Like all semi-algebraic sets, the orbit space of $G$ presents a
natural {\em stratification}. It can, in fact,  be considered as the
disjoint union of a {\em finite number} of semi-algebraic subsets
of decreasing dimensions ({\em geometrical strata}), each
geometrical stratum being in the border of a higher dimensional
one, but for the highest dimensional stratum, which is unique
({\em principal stratum}). The {\em geometrical strata} are the connected components of the {\em
symmetry strata}. The symmetries of two bordering strata are
related by a group--subgroup relation and the orbit-type of the
lower dimensional stratum is larger.

In order to determine the minimum of $\widehat \Phi(\alpha;p)$,
$p\in \widehat S$, as a constrained minimum, one needs the
relations defining the geometrical strata of $\real^{n}/G$. These can be obtained from positivity
and rank conditions on the matrix $\widehat P(p)$ and from the
syzygies, as already stated.

The geometrical and symmetry strata are determined in the
following way. Let $\widehat W^{(d)}$ denote the (generally non
connected) $d$ dimensional algebraic subset of $Z$, defined by the relation

\begin{equation}
\widehat W^{(d)}=\{p\in Z\mid {\rm rank}(\widehat P(p))=d,\ \widehat P(p) \ge 0\}, \label{W}
\end{equation}
then, the geometrical $d$
dimensional strata are the connected components of $\widehat
W^{(d)}$ and each $d$ dimensional symmetry stratum, $\widehat
S^{(d,\alpha)}$, $\alpha=1,\dots $, is the union of all the
geometrical $d$ dimensional strata, $\widehat W^{(d,\alpha,r)}$,
$r=1,\dots k_\alpha$, with the same orbit-type, $[G^{(d,r)}]$.
A representative in $[G^{(d,r)}]$ can be obtained
as the isotropy subgroup at an arbitrarily chosen point $x^{(d,r)}$ of the image in $\real^{n}$,
$W^{(d,r)}$, of the geometric stratum $\widehat W^{(d,r)}$. The
point  $x^{(d,r)}$ can be obtained as a solution of the equation\footnote{For
$p^{(0)}\in\widehat S$, all the solutions, $x$, of the equation $\tilde p(x)=\tilde p^{(0)}$
form a unique $G$-orbit. The equation has no solutions for $p^{(0)}\not\in \widehat S$.}
$p(x)=p^{(d,r)}$, where $p^{(d,r)}$ is an arbitrarily chosen point
in $\widehat W^{(d,r)}$.

In the following, we shall classify and characterize all the
allowed symmetry phases and possible phase transitions between
contiguous phases, for
$D$-wave phase condensates in isotropic space. In particular, in
Section III we shall identify the linear symmetry group, $G$, of
these systems and a minimal set of basic polynomial invariants of
$G$. In Section IV, we shall determine the geometrical features of
the orbit space of $G$, {\em i.e.}, its stratification (including
connection properties and bordering relations of the strata) and
the orbit-types of its strata. In section V, using an innovative method, we shall provide relatively simple expressions
for the equalities and inequalities determining each stratum. These
are essential for establishing the connection properties of
the strata and make much easier the determination of the minima of
a specific free energy function. A simple example of the
determination of the minima of a general 4-th degree $G$-invariant
polynomial and their stability against 6-th order perturbations,
will be given in Section V.

\section{Symmetry of the allowed D-wave condensate states in isotropic space}

The formation of D-wave condensate ground states breaks the symmetry of
the isotropic 3-dimensional space, which corresponds to the group
${\rm\bf O}_3\otimes{\rm\bf U}_1\times \left\langle{\cal T}\right\rangle$, where
${\rm\bf O}_3$ is the complete rotation group, ${\rm\bf U}_1$ is
the gauge group and $\left\langle{\cal T}\right\rangle$ is the
group\footnote{Here and in the following we denote by
$\left\langle g_1,\dots ,g_n\right\rangle $ the group generated by
the set $\{ g_1,\dots ,g_n\}$.} generated by the time reversal operator ${\cal
T}$.

The symmetry of the allowed D-wave condensate ground states is
defined by the relative values of the complex coefficients in the
decomposition of the gap-function, $\Delta$, in terms of spherical
harmonics with $L = 2$:

$$\Delta(\theta,\phi) = \sum_{m=-2}^2 D_m\,Y_2^m(\theta,\phi).$$

The set of functions $\{Y_2^m,Y_2^{m\,*}\}$ yields a basis of a
ten-dimensional (10\,D) space, hosting a real representation of the
symmetry group {\bf O}$_3\otimes${\bf U}$_1\times \left\langle {\cal T}\right\rangle$.
A general element, $\gamma$, of the group will be denoted
by a triple

$$\gamma =(\rho,\,e^{i\phi},\,\epsilon),$$
where, $\rho \in {\rm\bf O}_3$, $0\le\phi <2\pi$ and $\epsilon =- 1$, or $+1$
according as a time reflection is involved in the transformation, or not. In the following, we shall also use the
shortened notations $\rho$ for $(\rho,1,1)$,
$U_1(\phi)$ for $({\uno}_3,U_1(\phi),1)$ and ${\cal T}$ for $({\uno}_3,1,-1)$.

The action of $G$ can be transferred to a real irreducible action
on the 10\,D vector formed by the coefficients $\{D_{2},\dots ,D_{-2},
D_{-2}^*,\dots ,D_2^*\}$. The representation of $G$ thus obtained can
be realized in the 10\,D real vector space generated by a couple of two
independent, real, second rank, symmetric, traceless tensors,
$X^{(1)}_{ij}$ and $X^{(2)}_{ij}$, $i,j=1,2,3$, which can be
considered as the real and imaginary parts of a complex $3\times
3$ matrix $\psi$, whose elements will be written in terms of five complex coordinates, $z_j$:

\begin{equation}
z_j = x_j + i\, x_{5+j},\qquad j=1,\dots ,5; \qquad x_i\in{\bf R},
\end{equation}

\begin{equation}
\psi=\frac 1{\sqrt{2}}\left(
\begin{array}{ccc}
z_2 + \frac{\displaystyle z_5}{\sqrt 3} & z_1                                      & z_4  \\
z_1                                     & -z_2 + \frac{\displaystyle z_5}{\sqrt{3}}& z_3  \\
z_4                                     &  z_3                                     &  -\frac{\displaystyle 2\,z_5}{\sqrt{3}}
\end{array} \right).
\end{equation}
The coordinates $D_\alpha$ are connected to the $z_j$ by the following relations:

\begin{equation}
D_2    = {\displaystyle -\frac{i\,z_1 + z_2}{\sqrt 2}},\quad D_1 = {\displaystyle \frac{i\,z_3 + z_4}{\sqrt
2}},\quad D_0 = z_5,\quad D_{-1} = {\displaystyle \frac{i\,z_3 - z_4}{\sqrt
2}},\quad D_{-2} = {\displaystyle \frac{i\,z_1 - z_2}{\sqrt 2}}\,.
\end{equation}

The matrix $\psi$ transforms in the following way under a general transformation, $\gamma=(\rho,\phi,\epsilon)\in G$:

\begin{equation}
 \gamma\cdot\psi = e^{i\phi}\,\rho\,\psi' \,\rho^{\rm T},
 \label{trasf}
 \end{equation}
where $\psi'= \psi$ or $\psi^*$, according as $\epsilon =+1$ or $-1$ and the apex T denotes transposition.
As a consequence, the group $G$ acts as a group of
linear, real, {\em orthogonal} transformations on the vector order parameter
$x=(x_1,\dots ,x_{10})\in {\bf R}^{10}$.

The kernel of the representation of $G$ just defined is the group generated by the space reflection, so, it will
not be restrictive to assume that the symmetry group is

$$G = {\rm\bf SO}_3\otimes{\rm\bf U}_1\times \left\langle{\cal T}\right\rangle$$
and, when referring to $G$, in the following, we shall always mean this linear group acting in the vector space
$\real^{10}$.

The group $G$ has a trivial principal isotropy subgroup
(the isotropy subgroup of generic points of ${\bf R}^{10}$), thus
the principal $G$-orbits have the same dimensions, four, as $G$ and its {\em orbit space}, ${\bf R}^{10}/G$, has
dimensions $q_0 = 10 - 4 = 6$.

The M\"olien function  of $G$, $M_G(\eta)$, can be calculated in the
form of an invariant Haar integral over $G$ (see, for instance,
\cite{33,3334}):

$$M_G(\eta ) = \int_G\frac {d\mu(g)}{{\rm det}\left(\uno -
\eta\,g\right)},\qquad |\eta| < 1,$$
where $\mu(g)$ is a normalized invariant measure on the group $G$, the integration is over the whole group $G$ and $g\in G$.
In terms of the Euler angles $\phi_1,\theta,\phi_2$ and of the gauge angle $\alpha$, the integral can be written in the form
of a sum over the two connected components of $G$

\begin{equation}
\begin{array}{l}
M_G(\eta ) = {\displaystyle\frac 1{16\pi^2} \int_0^{2\pi}d\phi_1\int_0^{2\pi}d\phi_2\int_0^{\pi}\sin\theta\, d\theta}\,
\times \\
\\
\ \ \left\{{\displaystyle  \int_0^{2\pi}\,\frac{d\alpha}{2\,\pi}\, \prod_{k=-2}^2\left[\left(1 - \eta\,e^{i(k\,\chi  +
\alpha)}
\right) \left(1 - \eta\,e^{i(k\,\chi  - \alpha)}\right)\right]^{-1} + \prod_{k=-2}^2\left[\left(1 - \eta\,e^{i\,k\,\chi}
\right)\left(1 + \eta\,e^{i\,k\,\chi}\right)\right]^{-1} }\right\},\label{Molien}
\end{array}
\end{equation}
where $\{e^{i\,k\,\chi}\}_{-2\le k\le 2}$ are the distinct eigenvalues of the 10\,D rotation matrix and $\cos\chi$
is a function of the Euler angles\footnote{It has to be noted that, in principle, the denominators in (\ref{Molien}),
are polynomials in $\sin\chi$ and $\cos\chi$; but, by they definition, they are also even functions of $\chi$, so, they
can be expressed as polynomials only in $\cos\chi$.}:

\begin{equation}
\cos\chi = \frac 12\left[\cos\theta + \cos(\phi_1 + \phi_2) + \cos\theta\,\cos(\phi_1 + \phi_2) - 1\right].\label{Molien'}
\end{equation}
An explicit calculation of the integrals (see Appendix A) leads to

\begin{equation}
M_G(\eta ) = \frac{1 + \eta ^8 +\eta ^{10} + \eta ^{12} +  \eta ^{20}}
{(1 - \eta ^2)(1 - \eta ^4)^2(1 - \eta ^6)^2(1 - \eta ^8)},\label{10}
\end{equation}
a relation that can also be written in the following equivalent suggestive form:

\begin{equation}
M_G(\eta ) = \frac{1 - \eta^{16} - \eta^{18} - \eta^{20} - \eta^{22} - \eta^{24} + \eta^{26} +
  \eta^{28} + \eta^{30} + \eta^{32} + \eta^{34} - \eta^{50}}{(1 - \eta ^2)(1 - \eta ^4)^2(1 - \eta ^6)^2(1 - \eta
^8)^2(1-\eta^{10})(1-\eta^{12})}.\label{10'}
\end{equation}

Equations (\ref{10},\ref{10'}) yield the following indications, whose
validity has been checked through direct calculations\footnote{To
justify the statements listed in the following items
1)--3), starting from (\ref{10},\ref{10'}), one has to
check that any $G$-invariant homogeneous polynomial of degree 20 in
$x\in\real^{10}$ can be written as a polynomial in $p_1(x),\dots,
p_9(x)$.}:

\begin{enumerate}
\item A minimal integrity basis (IB) for the linear group $G$ contains nine
elements, $\{p_1,\dots , p_9\}$ with degrees $(d_1,\dots ,d_9)=(2,\, 4,\, 4,\, 6,\, 6,\,
8,\, 8,\, 10,\, 12)$.

\item The invariants $p_i$ are connected by five independent syzygies of
degrees 16, 18, 20, 22 and 24. The group is, therefore, {\em non coregular}.

\item It is possible to find a
{\em non minimal} homogeneous IB, $\{\xi_1,\dots,\xi_6,\theta_1,\dots
,\theta_4\}$, with degrees $(2,\, 4,\, 4,\, 6,\, 6,\, 8,\, 8,\, 10,\, 12,\, 20)$,
such that any invariant polynomial, $F$, can be
written in the form \cite{Stanley}

\begin{equation}
F = Q_0(\xi) + \sum_{i=1}^4\, Q_i(\xi)\, \theta_i,\label{Q}
\end{equation}
where the $Q$'s, are polynomials in the {\em algebraically
independent} polynomial invariants $(\xi_1,\dots,\xi_6)$.
\end{enumerate}

A possible choice for the IB's $\{p\}$ and $\{\xi,\theta\}$ is the following:

\begin{equation}
\begin{array}{ll}
\begin{array}{rcl}
p_1 &=& {\rm Tr}(\psi\psi^*)=\sum_{i=1}^{10}x_i^2,\\
p_2&=& {\rm Tr}\left[(\psi\psi^*)^2\right],\\
p_3 &=& |{\rm Tr}(\psi^2)|^2, \\
p_4 &=&|{\rm Tr}(\psi^3)|^2,\\
p_5 &=& |{\rm Tr}(\psi^2\psi^*)|^2,
\end{array} & \qquad\qquad
\begin{array}{rcl}
p_6 &=& \Re\left[{\rm Tr}(\psi^2){\rm Tr}(\psi^2\psi^*){\rm Tr}(\psi^{*3})\right],\\
p_7 &=& \Re\left[{\rm Tr}(\psi^{*2})\left({\rm Tr}(\psi^2\psi^*)\right)^2\right],\\
p_8 & =& \Re\left[{\rm Tr}(\psi^2\psi^*){\rm Tr}(\psi^3)\left({\rm Tr}(\psi^{*2})\right)^2\right],\\
p_9 &=& \Re\left[\left({\rm Tr}(\psi^2)\right)^3\,\left({\rm Tr} (\psi^{*3})\right)^2\right.]\,.
\end{array}
\end{array}\label{IB}
\end{equation}
and

\begin{equation}
\begin{array}{l}
\xi_i = p_i,\quad i=1,\dots ,5,  \qquad\qquad \xi_6 = p_6 + p_7, \\
\theta_1 = p_6 - p_7,\qquad \theta_2 = p_8,\qquad \theta_3 = p_9,
\qquad  \theta_4 = (p_6 - p_7) p_9.
\end{array}\label{newIB}
\end{equation}
Thus, the most general non-equilibrium polynomial Landau
potential, $\widehat\Phi(\alpha;p)$, can be written in the form

\begin{equation}
\widehat\Phi = Q_0 + Q_1\, p_7 + Q_2\, p_8 + Q_3\, p_9 + Q_4\, p_7\,p_9.
\end{equation}

Using the above definitions, the explicit form of the syzygies and of
the $\widehat P$--matrix elements can be obtained through the
following simple procedure.

After defining the {\em weight} of an element, $p_i$, of the IB to be equal to the degree, $d_i$, of the homogeneous
polynomial $p_i(x)$, and the degree of $\prod_i\,p_i^{n_i}$ as $\sum_i\,n_i\,d_i$, one
writes down the most general homogeneous polynomial, $\widehat\Theta^{(d)}(\xi,\theta)$, of the relevant weight, $d$, in the
weighted variables $\xi_i$, $i=1,\dots ,6$ and $\theta_i$,
$i=1,\dots,4$. With the definitions $\theta_0=1$, $d_0=0$, the polynomial $\widehat\Theta^{(d)}(\xi,\theta)$ can be written
in the compact form

$$\widehat\Theta^{(d)}(\xi,\theta) = \sum_{j=0}^3\,\,\delta_{N_j,d}\, A^{(j,d)}_{n_1,\dots,
n_6}\,\theta_j\,\prod_{k=1}^6\,\xi^{n_k},$$
where, $N_j=\sum_{i=1}^6 n_i\,d_i-d_j$, $\delta_{N_j,d}$ is a Kronecker symbol and the $A$'s are real coefficient to be
determined through the following conditions i) and ii):

\begin{itemize}

\item[i)] in the case of the syzygies $\widehat s_d(\xi,\theta)$, $d=16,\,18,\,20,\,22,\,24$:\quad
The expression $\widehat s_d(\xi(x),\theta(x))$ has to vanish identically in $x$;

\item[ii)] in the case of the matrix elements $\widehat P_{a,b}(\xi,\theta)$, $a,b = 1,\dots
,9$,  $d=d_a + d_b - 2$: \quad The expression $\widehat P_{ab}(\xi(x),\theta(x)) = \sum_{i=1}^{10}\,\partial_i p_a(x)\,
\partial_ip_b(x)$ has to be an identity in $x$.

\end{itemize}

The explicit expressions of the syzygies are reported in Appendix B and
the elements of the matrix $\widehat P(p)$ in Appendix C. In the
following we shall always use the minimal IB $\{p\}$.

On the algebraic surface, $Z$, defined by the syzygies in the 9\,D
space of the $p$'s, the $9\times 9$ symmetric matrix $\widehat
P(p)$ has rank 6, as it has to be, since the
orbit space is six dimensional.

\section{The orbit space of the linear group $G$}

As recalled in the Introduction, the orbit space of $G$ can be
identified with the range of the orbit map $x\mapsto p(x)$, which
is the same as the set of values of $p\in Z$, that render the
matrix $\widehat P(p)$ positive semi-definite.

The analytic conditions determining the geometrical strata, $\widehat W^{(d,\alpha,r)}$, are not difficult to specify, in
principle, but, in the case of the higher dimensional strata, their explicit expressions (as sets of equalities
and inequalities in the $p$'s) become difficult to obtain, too large to be written down extensively and
uneasy to handle, if they are derived only from rank and positivity conditions of the matrix $\widehat P(p)$. An
elegant way to bypass this difficulty is the following.

One starts with the relatively easy determination of the {\em geometrical} strata of dimensions $\le 2$ ({\em i.e.},
the connected components of the sets $\widehat W^{(d)}$, $d=1,2$, defined in (\ref{W})), starting
from the positivity and
rank properties of the matrix $\widehat P(p)$ and from the  syzygies. The subsequent step
consists in the identification of the orbit-type of each geometrical stratum. This allows
to identify the connected components, $\widehat W^{(d,\alpha,r)}$,
$r=1,\dots $,
with the same orbit-type, $[G^{(d,\alpha)}]$, and
belonging, therefore, to the same {\em symmetry}
stratum\footnote{There is a unique zero dimensional stratum, corresponding to
the point $p=p(0)=0$, whose symmetry is $[G]$. It will be
ignored in the following.}, $\widehat S^{(d,\alpha)}$.  All the
groups $G^{(2,\alpha)}$ turn out to be finite groups of
low order ($\le\, 8$).

The determination of the higher dimensional strata bordered by a
given 2\,D stratum, $\widehat S^{(2,\alpha)}$, can be obtained
starting from the identification of their orbit-types, which are
necessarily contained in the orbit-type of $\widehat
S^{(2,\alpha)}$. The task is easily accomplished by selecting, out of
the set of maximal subgroups of each $G^{(2,\alpha)}$, those which
are isotropy subgroups of $G$. To this end, the following
criterion can be used. Let $H$ be one of the maximal subgroups of
a $G^{(d,\alpha)}$ and denote by $V_H$ the linear subspace formed
by all the points $x\in\real^{10}$, which are stable under $H$.
Then, $H$ is an isotropy subgroup of $G$, if and only if the following condition is
satisfied:

$$\raisebox{-1.3ex}{$\stackrel{\displaystyle\rm Max}{\scriptstyle
x\in V_H}$}\, {\rm rank}(P(x))\,<\,d.$$
If this condition is satisfied, when $x$ spans
$V_H$, the point $p(x)\in\real^9$ spans the
topological closure of a stratum $\widehat S^{(d',\alpha')}$ and

$${\rm dim}(\widehat S^{(d',\alpha')})\,=
\,\raisebox{-1.3ex}{$\stackrel{\displaystyle\rm Max}{\scriptstyle
x\in V_H}$}\,\, {\rm rank}(P(x)).$$

To get the full set of higher dimensional strata, the last part of the procedure just described has to be repeated
for each newly determined stratum.

The derivation of the equalities and inequalities determining each
of the higher dimensional strata requires a more sophisticated
method \cite{SV}, which we shall explain and exploit in the following section.

Before resuming the results obtained with the realization of the program just presented,
let us note that, owing to the linearity of $G$, its isotropy
subgroups coincide at the points $x\in\real^{10}$ and $\kappa\,x$, $0\ne \kappa\in\real$.
Thus, taking also into account the homogeneity of the polynomials
$p_i(x)$, the points $(p_1, \dots, p_9)$ and
$(1,p_2/p_1^{d_2/2},\dots ,p_9/p_1^{d_9/2})$ lie on the same
stratum and the equalities and inequalities defining the strata
can be written as homogeneous relations in the weighted variables
$p_i$. It will be useful, therefore, to introduce the definitions

\begin{equation}
\tilde p_i = p_i/p_1^{d_i/2},\qquad \tilde p=(\tilde p_2,\dots ,\tilde p_9).
\end{equation}

The relations determining the 1\,D and 2\,D strata can be easily obtained
by selecting those solutions, $\bar p$, of the condition ${\rm rank}
(\widehat P(p))= 1$ and, respectively, ${\rm rank}(\widehat P(p))= 2$, at
which the matrix $\widehat P(\bar p)$ is positive semi-definite. In this
way, one obtains five connected components, $\widehat W^{(1,A)}$, $A=1,
\dots ,5$, of $\widehat W^{(1)}$ and seven connected components, $\widehat
W^{(2,A)}$, $A=1_+,1_-,2,3_+,3_-,4,5$, of $\widehat W^{(2)}$. The parametric
equations defining this semi-algebraic sets are listed in Tables~\ref{I} and \ref{II}.

In order to identify the distinct 1\,D and 2\,D symmetry strata, for each
connected component $\widehat W^{(d,A)}$, $d=1,2$, we have picked up a
point $p^{(d,A)}$ and, for each $p^{(d,A)}$, a ''simple" solution, $x^{(k,A)}$,
of the equation $\tilde p(x)=\tilde p^{(d,A)}$.
Then, at each point $x^{(d,A)}$ we
have determined, and compared for conjugation, the isotropy subgroup of $G$, $G^{(d,A)}$. These subgroups turn
out to be non conjugate in $G$, but for $G^{(2,1_+)}$, which is conjugate to $G^{(2,1_-)}$, and $G^{(2,3_+)}$, which is
conjugate to $G^{(2,3_-)}$. Thus, there are five distinct 1\,D and five 2\,D symmetry strata,
$\widehat S^{(d,\alpha)}$, with orbit-types $[G^{(d,\alpha)}]$, $d=1,2$, $\alpha=1,\dots ,5$.

After choosing an element, $H$, in each class $[G^{(d,\alpha)}]$, it
is easy to determine the subspace $V_H\subset\real^{10}$, formed
by the vectors which are $H$-invariant.

The results we have obtained are listed in Tables \ref{III} and \ref{IV} and the bordering relations among the strata
are illustrated in Figure~\ref{F1}. The definitions of the
rotation matrices appearing in the tables and in the following text are
recalled in Appendix D. We have used standard notations for the
corresponding geometrical transformations (see, for instance, \cite{Cornwell}).

\section{Relations defining the strata with dimensions $\ge 3$}

In order to derive the relations defining the higher dimensional
strata we shall exploit the following results \cite{SV}, that we shall briefly recall, without proof. Let $H$
be an isotropy subgroup of $G$, $S_{[H]}$ and $\widehat S_{[H]}$
the associated strata in $\real^{10}$ and, respectively,
$\real^{10}/G$ and let Stab$(H,G)$ be the stabilizer of $H$ in
$G$:

$${\rm Stab}(H,G) = \{g\in G\mid g\,H\,g^{-1} = H\}.$$
We have denoted by $V_H$ the linear subspace formed by the $H$-invariant vectors of $\real^{10}$:

\begin{equation}
V_H = \{x\in\real^{10}\mid h\,x=x,\ \forall h\in H\},
\end{equation}
and we shall denote by $V_H^{(H)}$ the subset of $V_H$ formed by the points at which the isotropy
subgroup of $G$ is $H$:

\begin{equation}
V_H^{(H)} = \{x\in\real^{10}\mid G_x = H\}.
\end{equation}
The set $V_H^{(H)}$ is the intersection of $V_H$ with the stratum of orbit-type $[H]$.

The group Stab$(H,G)$ acts linearly on $V_H$; let us call $\tilde
H$ the linear group defined by this action. The $\tilde H$-orbit through
$x\in V_H$ is the intersection of the $G$-orbit through $x$ with
$V_H$. There is, therefore, a one-to-one correspondence between
the $G$-orbits lying in the closure of $\widehat S_{[H]}$ and the $\tilde
H$ orbits. This correspondence can be restricted to a one-to-one
correspondence between the interior points of the closure of
$\widehat S_{[H]}$ and the points of the principal stratum of
$V_H/\tilde H$. If $\widehat S_{[H]}$ is connected, the one-to-one
correspondence is between the principal stratum of $V_H/\tilde H$
and $\widehat S_{[H]}$, otherwise the inverse image of $\widehat
S_{[H]}$ in the correspondence, reduces to the points of the
principal stratum of $V_H/\tilde H$, corresponding to $\tilde H$-orbits through points of $V_H^{(H)}$.
Thus an IB, $\{\lambda\},$
of $\tilde H$ can be used to parameterize the points of
$\widehat S_{[H]}$:

$$p=p(\lambda)).$$
The range of the parameters $\lambda$ will
be determined by the positivity conditions of the $\widehat
P$-matrix associated to the IB $\{\lambda\}$, $\widehat P^{(H)}(\lambda)$, and convenient
additional conditions obtained from positivity and rank conditions of $\widehat P(p)$, if $\widehat S_{[H]}$ is not
connected. To this end it will be worth noting that, for $p\in\widehat S_{[H]}$, the matrix $\widehat P(p)$ can be
written in the following form:

\begin{equation}
\widehat P(p(\lambda)) = J^{\rm T}(\lambda)\,\widehat P^{(H)}(\lambda)\,J(\lambda),\label{J}
\end{equation}
where, $J(\lambda)$ is the Jacobian matrix

\begin{equation}
J_{ai}(\lambda) = \frac {\partial p_a(\lambda)}{\partial \lambda_i}.
\end{equation}
Owing to a well known theorem in matrix theory, from
(\ref{J}) one obtains the following upper limit:

\begin{equation}
{\rm rank}\left(\widehat P(p(\lambda))\right)\,\le\,{\rm rank}(J(\lambda)).\label{limit}
\end{equation}

Below, we shall use this approach to derive rational parametric equations for all the
strata of $\widehat S$ with dimensions three and four. The results obtained in this way, starting from each
of the strata of dimensions $\ge 2$, are resumed in Table~\ref{IV} and in the lower part of Table~\ref{III}.

\subsubsection{Stratum $\widehat S^{(3,1)}$}

The stabilizer in $G$ of the group $H=G^{(3,1)}=\left\langle
C_{2z}, C_{2x}{\cal T}\right\rangle $, is the group
Stab$(H)=\left\langle C_{2x},C_{2a},U_1(\pi),{\cal T}\right\rangle
$.

An easy calculation shows that the space $V_H$ is defined by the
equations $x_i=0$, $i=1,\,3,\,4,\,7,\,8,\,9,\,10$. The elements of
$\tilde H$ act on the $x_i$, $i=2,5,6$ by changing
the signs of one or two or all of these coordinates, so, an IB of
$\tilde H$ can be chosen to be

\begin{equation}
\lambda_1 = x_5^2,\qquad \lambda_2 = x_2^2,\qquad \lambda_3=x_6^2
\end{equation}
and the expression of the $p$'s in terms of the $\lambda$'s turns
out to be the following:

\begin{equation}
\begin{array}{rcl}
p_1 &=& \lambda_1 + \lambda_2 + \lambda_3,\\ p_2 &=&
(3\,\lambda_1^2 + 6\,\lambda_1\,\lambda_2 + 3\,\lambda_2^2 +
      2\,\lambda_1\,\lambda_3 + 18\,\lambda_2\,\lambda_3 + 3\,\lambda_3^2)/6,\\
p_3 &=& (\lambda_1 + \lambda_2 - \lambda_3)^2,\\
p_4 &=& \lambda_1\,(\lambda_1 - 3\,\lambda_2 + 3\,\lambda_3)^2/6,\\
p_5 &=& \lambda_1\,(\lambda_1 - 3\,\lambda_2 - \lambda_3)^2/6,\\
p_6 &=& \lambda_1\,(\lambda_1 - 3\,\lambda_2 - \lambda_3)\,
      (\lambda_1 + \lambda_2 - \lambda_3)\,(\lambda_1 - 3\,\lambda_2 + 3\,\lambda_3)/6,\\
p_7 &=&\lambda_1\,(\lambda_1 - 3\,\lambda_2 -
\lambda_3)^2\,(\lambda_1 + \lambda_2 - \lambda_3)/6,\\
p_8 &=& \lambda_1\,(\lambda_1 - 3\,\lambda_2 - \lambda_3)\,(\lambda_1 +
\lambda_2 - \lambda_3)^2\,(\lambda_1 - 3\,\lambda_2 + 3\,\lambda_3)/6,\\
p_9 &=& \lambda_1\,(\lambda_1 + \lambda_2 -
\lambda_3)^3\,(\lambda_1 - 3\,\lambda_2 + 3\,\lambda_3)^2/6,
\label{S31}
\end{array}
\end{equation}
with

\begin{equation}
\lambda_1,\,\lambda_2,\,\lambda_3\, >\, 0.\label{S31'}
\end{equation}
For values of the $\lambda$'s satisfying (\ref{S31'}), the values of the $p$'s defined in (\ref{S31})
render $\widehat P(p)$ everywhere positive semi-definite and of rank
three. We can conclude, therefore, that they yield parametric equations for
the 3\,D stratum $\widehat S^{(3,1)}$. Being the continuous image of a connected set, the
stratum is {\em connected}.

By eliminating the parameters $\lambda$'s from (\ref{S31},\ref{S31'}), one obtains the
following relations:

\begin{equation}
\def\arraystretch{1.3}
\begin{array}{rcl}
p_4 &=&9\,\left( 2\,{p_1}^2 - 2\,p_2 - p_3
\right) \,{\left( 2\,{p_1}^2 - 2\,p_2 + 2\,p_1\,{\sqrt{p_3}} + p_3 \right) }^2\,\tau,\\
p_5 &=&\left( 2\,{p_1}^2 - 2\,p_2 - p_3 \right) \, {\left( 2\,{p_1}^2 - 6\,p_2 - 2\,p_1\,{\sqrt{p_3}} - p_3\right) }^2
\,\tau,\\
p_6 &=&-3\,{\sqrt{p_3}}\,\left( 2\,{p_1}^2 - 2\,p_2 - p_3\right) \,\left( 2\,{p_1}^2 - 6\,p_2 - 2\,p_1\,{\sqrt{p_3}} - p_3
          \right) \,\left( 2\,{p_1}^2 - 2\,p_2 + 2\,p_1\,{\sqrt{p_3}} + p_3 \right)\,\tau,\\
p_7 &=&-{\sqrt{p_3}}\, \left( 2\,{p_1}^2 - 2\,p_2 - p_3\right)\,{\left( 2\,{p_1}^2 - 6\,p_2 - 2\,p_1\,{\sqrt{p_3}} - p_3
        \right) }^2\,\tau,\\
p_8 &=&3\,p_3\,\left( 2\,{p_1}^2 - 2\,p_2 - p_3\right) \,\left( 2\,{p_1}^2 - 6\,p_2 - 2\,p_1\,{\sqrt{p_3}} - p_3
          \right) \,\left( 2\,{p_1}^2 - 2\,p_2 + 2\,p_1\,{\sqrt{p_3}} + p_3 \right)\,\tau ,\\
p_9 &=&- 9\,\sqrt{p_3^3}\,\left( 2\,{p_1}^2 - 2\,p_2 - p_3\right) \,{\left( 2\,{p_1}^2 - 2\,p_2 +
        2\,p_1\,{\sqrt{p_3}} + p_3 \right) }^2\,\tau,
\end{array}
\end{equation}

\begin{equation}
\left\{ \begin{array}{l} 0\,< p_3\, < p_1^2,\\ 2\,p_1^2 + p_3\, <
6\,p_2 \, < \, 3\left( 2\,p_1^2 - p_3\right),
\end{array}\right.
\end{equation}
where

\begin{equation}
\tau^{-1}=64\,{\left( p_1 + {\sqrt{p_3}} \right) }^3,
\end{equation}
and the square root of $p_3$ has to be intended in an algebraic sense.

\subsubsection{Stratum $\widehat S^{(3,2)}$}

The stabilizer in $G$ of the group $H = G^{(3,2)}=\left\langle
C_{2x}, C_{2z}\right\rangle $, is a group of $\infty$ order:
Stab$(H)=\left\langle C_{4x},C_{4z},{\cal T}\right\rangle \times
{\rm\bf U}_1$.

The space $V_H$ is defined by the equations $x_i=0$,
$i=1,\,3,\,4,\,6,\,8,\,9$. The transformation properties of the coordinates $x_i$, $i=2,\,5,\,7,\,10$, under the
transformations of $\tilde H$ can be conveniently described in the following way. Let us define:

\begin{equation}
\eta_1 = x_2 + i\,x_5 + i(x_7  + i\, x_{10}), \qquad
\eta_2 = x_2 - i\,x_5 + i(x_7 - i\,x_{10}),
\end{equation}
then

\begin{itemize}
\item[i)] Under a gauge transformation, U$_1(\phi)$:\quad
$(\eta_1,\eta_2)\,\rightarrow\,(e^{i\,\phi}\eta_1,e^{i\,\phi}\eta_2)$;

\item[ii)] under time reversal ${\cal T}$:\quad $(\eta_1,\eta_2)\,\rightarrow\,(\eta_2^*,\eta_1^*)$;
\item[iii)] under a transformation $C_{4x}$:\quad $(\eta_1,\eta_2)\,\rightarrow\,(e^{i\,\pi/3}\,\eta_2,e^{-i\,\pi/3}\,\eta_1)$;
\item[iv)] under a transformation $C_{4z}$:\quad $(\eta_1,\eta_2)\,\rightarrow\,(-\eta_2,-\eta_1)$.
\end{itemize}
The group $\tilde H$ is coregular and admits the following minimal IB:

\begin{equation}
\begin{array}{rcl}
\lambda_1 &=& \left(|\eta_1|^2 + |\eta_2|^2\right)/2\,=\, {x_2}^2 + {x_5}^2 + {x_7}^2 + {x_{10}}^2,\\
\lambda_2 &=& (|\eta_1|^2 - |\eta_2|^2)^2/16 \,=\, {\left( x_5\,x_7 - x_2\,x_{10} \right) }^2,\\
\lambda_3 &=& \Re\left[(\eta_1\,\eta_2^*)^3\right] \,=\, \left( {x_2}^2 - {x_5}^2 + {x_7}^2 - {x_{10}}^2 \right)
              \,\left( {x_2}^4 - 14\,{x_2}^2\,{x_5}^2 + {x_5}^4 + 2\,{x_2}^2\,{x_7}^2 - \right.\\&&
              2\,{x_5}^2\,{x_7}^2 + {x_7}^4 - \left. 24\,x_2\,x_5\,x_7\,x_{10} - 2\,{x_2}^2\,{x_{10}}^2 +
              2\,{x_5}^2\,{x_{10}}^2 - 14\,{x_7}^2\,{x_{10}}^2 + {x_{10}}^4 \right).
\end{array}
\end{equation}
It has to be noted that the invariant $\left(\Im\left[(\eta_1\,\eta_2^*)^3\right]\right)^2$ can be
expressed as a polynomial in $\lambda_1,\lambda_2,\lambda_3$.

The corresponding $\widehat P$-matrix turns out to be

\begin{equation}
\left(\begin{array}{ccc}
 4\,\lambda_1  & 8\,\lambda_2        & 12\,\lambda_3 \\
 8\,\lambda_2  & 4\,\lambda_1\,\lambda_2 & 0         \\
 12\,\lambda_3 & 0               & 36\,\lambda_1\left(\lambda_1^2 - 4\,\lambda_2\right)^2 \\
\end{array}\right)
\end{equation}
and the $p$'s can be expressed in terms of the $\lambda$'s in the following form:

\begin{equation}
\begin{array}{rcl}
p_1 &=& \lambda_1,\\
p_2 &=& \left(3\,{\lambda_1}^2 - 4\,\lambda_2\right)/6,\\
p_3 &=& {\lambda_1}^2 - 4\,\lambda_2,\\
p_4 &=& \left({\lambda_1}^3 + 12\,\lambda_1\,\lambda_2 - \lambda_3\right)/12,\\
p_5 &=& \left({\lambda_1}^3 - 4\,\lambda_1\,\lambda_2 - \lambda_3\right)/12,\\
p_6 &=& \left({\lambda_1}^4 - 16\,{\lambda_2}^2 - \lambda_1\,\lambda_3\right)/12,\\
p_7 &=& \left({\lambda_1}^4 - 8\,{\lambda_1}^2\,\lambda_2 + 16\,{\lambda_2}^2 - \lambda_1\,\lambda_3\right)/12,\\
p_8 &=& \left({\lambda_1}^5 - 8\,{\lambda_1}^3\,\lambda_2 + 16\,\lambda_1\,{\lambda_2}^2 - {\lambda_1}^2\,\lambda_3 -
        4\,\lambda_2\,\lambda_3\right)/12,\\
p_9 &=&\left({\lambda_1}^6 - 12\,{\lambda_1}^4\,\lambda_2 + 48\,{\lambda_1}^2\,{\lambda_2}^2 - 64\,{\lambda_2}^3 -
         {\lambda_1}^3\,\lambda_3 - 12\,\lambda_1\,\lambda_2\,\lambda_3 \right)/12,
\label{S32}
\end{array}
\end{equation}
where the parameters $\lambda$ have to satisfy the conditions:

\begin{equation}
\left\{ \begin{array}{l}
\lambda_1 >  0,\\
0\,<\,4\, \lambda_2\,<\,\lambda_1^2,\\
\lambda_3^2\,< \,\left(\lambda_1^2- 4\,\lambda_2\right)^3.
\end{array}\right.\label{S32'}
\end{equation}
For values of the $\lambda$'s satisfying (\ref{S32'}), the values of the $p$'s defined in (\ref{S32})
render $\widehat P(p)$ everywhere positive semi-definite and of rank
three. We can conclude, therefore, that they yield parametric equations for
the 3\,D stratum $\widehat S^{(3,1)}$. Being the continuous image of a connected set, the
stratum is {\em connected}.

By eliminating the parameters $\lambda$ from (\ref{S32},\ref{S32'}) one obtains the following relations:

\begin{equation}
\begin{array}{rcl}
p_2 &=& \left(2\,{p_1}^2 + p_3\right)/6,\\
p_5 &=& \left(-{p_1}^3 + p_1\,p_3 + 3\,p_4\right)/3,\\
p_6 &=& \left(-4\,{p_1}^4 + 5\,{p_1}^2\,p_3 - {p_3}^2 + 12\,p_1\,p_4\right)/12,\\
p_7 &=& \left(-4\,{p_1}^4 + 3\,{p_1}^2\,p_3 + {p_3}^2 + 12\,p_1\,p_4\right)/12,\\
p_8 &=& \left(-4\,{p_1}^5 + 5\,{p_1}^3\,p_3 - p_1\,{p_3}^2 + 12\,{p_1}^2\,p_4 - 6\,p_3\,p_4\right)/6,\\
p_9 &=& \left(-16\,{p_1}^6 + 24\,{p_1}^4\,p_3 - 9\,{p_1}^2\,{p_3}^2 + {p_3}^3 + 48\,{p_1}^3\,p_4 - 36\,p_1\,p_3\,p_4
           \right)/12
\end{array}
\end{equation}
and

\begin{equation}
\left\{ \begin{array}{l} p_1,\,p_3 > 0,\\
p_3^3 <\, \left[12\,p_4 - (4\,\,p_1^2 - 3\,p_3)p_1\right]^2.
\end{array}\right.
\end{equation}

\subsubsection{Stratum $\widehat S^{(4,1)}$}

The stabilizer in $G$ of the group $H =G^{(4,1)}=\left\langle
C_{2x},{\cal T} \right\rangle $, is the group Stab$(H,G)={\rm\bf
O}_2^x\times\left\langle U_1(\pi),{\cal T}\right\rangle  $.

The space $V_H$ is defined by the equations $x_i=0$,
$i=1,\,4,\,7,\,8,\,10$. The transformation properties of the
coordinates, $x_i$, $i=2,\,3,\,5,\,6,\,9$ of $V_H$ are easily derived
from (\ref{trasf}) and can be put in the following advantageous
form. Let us define

\begin{equation}
\eta_1 =\frac{\sqrt{3}\,x_2 + x_5}2,\qquad
\eta_2 = x_3 + i\,\frac{x_2 - \sqrt{3}\,x_5}2 ,\qquad
\eta_3 = x_6 - i \,x_9,
\end{equation}
then,

\begin{itemize}
\item[i)] under a proper rotation, O$_2^x(\phi)$:\quad $(\eta_1,\eta_2,\eta_3) \rightarrow (\eta_1,e^{-2\, i\,\phi}\, \eta_2,
e^{\,i\,\phi}\, \eta_3)$;
\item[ii)] under a gauge transformation, U$_1(\pi)$:\quad $(\eta_1,\eta_2,\eta_3) \rightarrow (-\eta_1,-\eta_2,-\eta_3)$;
\item[iii)] under a transformation $C_{2z}$:\quad $(\eta_1,\eta_2,\eta_3) \rightarrow (\eta_1, -\eta_2^*, \eta_3^*)$;
\item[iv)] under time reversal  ${\cal T}$:\quad $(\eta_1,\eta_2,\eta_3) \rightarrow (\eta_1, \eta_2, -\eta_3)$.
\end{itemize}

The group $\tilde H$ is not coregular and admits the following minimal IB:

\begin{equation}
\begin{array}{rcl}
\lambda_1 &=& 3\,|\eta_1|^2 =\frac 34 {\left(\sqrt{3}\,x_2 + x_5 \right)}^2,\\
\lambda_2 &=& |\eta_2|^2 = \frac 14 \left(x_2^2 + 4\,x_3^2 - 2\,\sqrt{3}\,x_2\,x_5 + 3\,x_5^2\right),\\
\lambda_3 &=& |\eta_3|^2 = {x_6}^2 + {x_9}^2,\\
\lambda_4 &=& \sqrt 3\,\eta_1\,\Im\left(\eta_2\,\eta_3^2\right) = \frac{\sqrt 3}4 \left({\sqrt{3}}\,x_2 + x_5 \right) \,
            \left(x_2\,{x_6}^2 - {\sqrt{3}}\,x_5\,{x_6}^2 - 4\,x_3\,x_6\,x_9 -\right.\\
          &\ &  \left. x_2\,{x_9}^2 + {\sqrt{3}}\,x_5\,{x_9}^2  \right),\\
\lambda_5 &=& \left[\Im\left(\eta_2\,\eta_3^2\right)\right]^2 = \frac
            14\, {\left(x_2\,{x_6}^2 - {\sqrt{3}}\,x_5\,{x_6}^2 - 4\,x_3\,x_6\,x_9 - x_2\,{x_9}^2 + {\sqrt{3}}\,x_5\,{x_9}^2
           \right)}^2.
\end{array}
\end{equation}
with the syzygy

\begin{equation}
\lambda_4^2 - \lambda_1\lambda_5 = 0. \label{syS41}
\end{equation}
and the $\widehat P$-matrix

\begin{equation}
\left(\begin{array}{ccccc}
12\,\lambda_1\, & \,0\,            & \,0\,            & \,6\,\lambda_4\, & \,0                                                                                                           \\
0\,             & \,4\,\lambda_2\, & \,0\,            & \,2\,\lambda_4\, & \,4\,\lambda_5                                                                                                \\
0\,             & \,0\,            & \,4\,\lambda_3\, & \,4\,\lambda_4\, & \,8\,\lambda_5                                                                                                \\
6\,\lambda_4\,  & \,2\,\lambda_4\, & \,4\,\lambda_4\, & \,4\,\lambda_1\,\lambda_2\,\lambda_3 + \lambda_1\,{\lambda_3}^2 + 2\,\lambda_3\left( 4\,\lambda_2 + \lambda_3 \right)\lambda_4   \\
0\,             & \,4\,\lambda_5\, & \,8\,\lambda_5\, & \,2\,\lambda_3\left(4\,\lambda_2 + \lambda_3 \right) \,\lambda_4 &4\,\lambda_3\left( 4\,\lambda_2 + \lambda_3 \right)\,\lambda_5 \\
\end{array}\right)
\end{equation}
The conditions $\widehat P(\lambda)\ge 0$ and rank$\left(\widehat P(\lambda)\right) =
4$, added to (\ref{syS41}), yield the following restrictions
on the acceptable range for the $\lambda$'s:

\begin{equation}
\left\{\begin{array}{l} \lambda_1,\lambda_2,\lambda_3\, \ge\, 0,\\
0 \le \lambda_5 \,<\, \lambda_2\lambda_3^2,\\ \lambda_1 +
\lambda_5 \,>\,0.
\end{array}\right.
\label{S41'}
\end{equation}

The $p$'s can be expressed in terms of the $\lambda$'s in the following form:

\begin{equation}
\begin{array}{rcl}
p_1 &=& \left(\lambda_1 + 3\,\lambda_2 + 3\,\lambda_3\right)/3,\\
p_2 &=& \left({\lambda_1}^2 + 6\,\lambda_1\,\lambda_2 + 9\,{\lambda_2}^2 + 14\,\lambda_1\,\lambda_3 +
        18\,\lambda_2\,\lambda_3 + 9\,{\lambda_3}^2 +  24\,\lambda_4\right)/18\\
p_3 &=& {\left( \lambda_1 + 3\,\lambda_2 - 3\,\lambda_3 \right)}^2/9,\\
p_4 &=& \left(4\,{\lambda_1}^3 - 72\,{\lambda_1}^2\,\lambda_2 + 324\,\lambda_1\,{\lambda_2}^2 - 36\,{\lambda_1}^2\,\lambda_3 +
    324\,\lambda_1\,\lambda_2\,\lambda_3 + \right.\\&&
    \left. 81\,\lambda_1\,{\lambda_3}^2 + 108\,\lambda_1\,\lambda_4 - 972\,\lambda_2\,\lambda_4 - 486\,\lambda_3\,\lambda_4 +
    729\,\lambda_5\right)/648,\\
p_5 &=& \left(4\,{\lambda_1}^3 - 72\,{\lambda_1}^2\,\lambda_2 + 324\,\lambda_1\,{\lambda_2}^2 + 12\,{\lambda_1}^2\,\lambda_3 -
    108\,\lambda_1\,\lambda_2\,\lambda_3 + \right.\\&&
    \left. 9\,\lambda_1\,{\lambda_3}^2 - 36\,\lambda_1\,\lambda_4 + 324\,\lambda_2\,\lambda_4 - 54\,\lambda_3\,\lambda_4 +
    81\,\lambda_5\right)/648,\\
p_6 &=& \left( \lambda_1 + 3\,\lambda_2 - 3\,\lambda_3 \right)\,\left( 4\,{\lambda_1}^3 - 72\,{\lambda_1}^2\,\lambda_2 +
      324\,\lambda_1\,{\lambda_2}^2 - 12\,{\lambda_1}^2\,\lambda_3 + \right. \\&&
      \left. 108\,\lambda_1\,\lambda_2\,\lambda_3 - 27\,\lambda_1\,{\lambda_3}^2 + 36\,\lambda_1\,\lambda_4 - 324\,\lambda_2\,\lambda_4 +
      162\,\lambda_3\,\lambda_4 - 243\,\lambda_5 \right) /1944,\\
p_7 &=& \left( \lambda_1 + 3\,\lambda_2 - 3\,\lambda_3 \right) \,\left( 4\,{\lambda_1}^3 - 72\,{\lambda_1}^2\,\lambda_2 +
      324\,\lambda_1\,{\lambda_2}^2 + 12\,{\lambda_1}^2\,\lambda_3 - \right. \\&&
      \left. 108\,\lambda_1\,\lambda_2\,\lambda_3 + 9\,\lambda_1\,{\lambda_3}^2 - 36\,\lambda_1\,\lambda_4 + 324\,\lambda_2\,\lambda_4 -
      54\,\lambda_3\,\lambda_4 + 81\,\lambda_5 \right) /1944,\\
p_8 &=& {\left( \lambda_1 + 3\,\lambda_2 - 3\,\lambda_3 \right)}^2\,\left( 4\,{\lambda_1}^3 - 72\,{\lambda_1}^2\,\lambda_2 +
      324\,\lambda_1\,{\lambda_2}^2 - 12\,{\lambda_1}^2\,\lambda_3 + \right. \\&&
      \left. 108\,\lambda_1\,\lambda_2\,\lambda_3 - 27\,\lambda_1\,{\lambda_3}^2 + 36\,\lambda_1\,\lambda_4 - 324\,\lambda_2\,\lambda_4 +
      162\,\lambda_3\,\lambda_4 - 243\,\lambda_5 \right) /5832,\\
p_9 &=& {\left( \lambda_1 + 3\,\lambda_2 - 3\,\lambda_3 \right)}^3\, \left( 4\,{\lambda_1}^3 - 72\,{\lambda_1}^2\,\lambda_2 +
      324\,\lambda_1\,{\lambda_2}^2 - 36\,{\lambda_1}^2\,\lambda_3  + \right.\\&&
      \left. 324\,\lambda_1\,\lambda_2\,\lambda_3 +  81\,\lambda_1\,{\lambda_3}^2 + 108\,\lambda_1\,\lambda_4 - 972\,\lambda_2\,\lambda_4 -
      486\,\lambda_3\,\lambda_4 + 729\,\lambda_5\right) /17496.
\end{array}\label{S41}
\end{equation}
For values of the $\lambda$'s satisfying (\ref{syS41}) and
(\ref{S41'}), the values of the $p$'s defined in (\ref{S41})
render $\widehat P(p)$ everywhere positive semi-definite and of rank
four. We can conclude, therefore, that they yield parametric equations for
the 4\,D stratum $\widehat S^{(4,1)}$. Being the continuous image of a connected set, the
stratum is {\em connected}.

For $\lambda_1\ne 0$ [$\lambda_5\ne 0$], the syzygy can be solved with respect to
$\lambda_5$ [$\lambda_1$] and the expression one obtains can be substituted into
(\ref{S41}), so, in this region, $(\lambda_1,\dots ,\lambda_4)$ [$(\lambda_2,\dots
,\lambda_5)$] play the role of local coordinates for the manifold underlying the
stratum. Since, owing to (\ref{S41'}), $\lambda_1$ and
$\lambda_5$ cannot vanish simultaneouly, the whole stratum is
covered by two rational charts of local coordinates.

The elimination of the parameters $\lambda$ from (\ref{S41},\ref{S41'}) leads
to cumbersome relations, which it is not worth writing down.

\subsubsection{Stratum $\widehat S^{(4,2)}$}

The stabilizer in $G$ of the isotropy subgroup $H = G^{(4,2)}=
\left\langle  C_{2z}\right\rangle $, is the group Stab$(H)={\rm\bf
O}_2^z\times{\rm\bf U}_1$.

The space $V_H$ is defined by the equations $x_i=0$,
$i=3,\,4,\,8,\,9$. The transformation properties of the
coordinates, $x_i$, $i=1,\,2,\,5,\,6,\,7,\,10$ of $V_H$ can be put in the
following advantageous form. Let us define

\begin{equation}
\eta_1 = x_5 + i x_{10},\qquad \eta_2 = \frac{x_7 + i\, x_6 + i\,(x_2 + i\, x_1)}{\sqrt{2}},\qquad
\eta_3 = \frac{x_7 + i\, x_6 - i\,(x_2 + i\, x_1)}{\sqrt{2}},
\end{equation}
then,
\begin{itemize}
\item[i)] under a proper rotation, O$_2^z(\phi)$:\quad $(\eta_1, \eta_2,\eta_3) \rightarrow (\eta_1,e^{-2\, i\,
\phi}, e^{-2\,i\,\phi}\, \eta_3)$;

\item[ii)] under a gauge transformation, U$_1(\phi)$:\quad $(\eta_1, \eta_2,\eta_3) \rightarrow
(e^{i\,\phi}\, \eta_1, e^{-i\,\phi}\, \eta_2, e^{i\,\phi} \,\eta_3)$;

\item[iii)] under a transformation $C_{2x}$:\quad $(\eta_1, \eta_2,\eta_3)\rightarrow (\eta_1, \eta_3^*,\eta_2^*)$;

\item[iv)] under time reversal  ${\cal T}$:\quad $(\eta_1, \eta_2,\eta_3) \rightarrow (\eta_1^*, -\eta_3, -\eta_2)$.
\end{itemize}

A minimal IB of $\tilde H$, which is coregular, can be chosen to be

\begin{equation}
\begin{array}{rcl}
\lambda_1 &=& \sum_{j=1}^3\,|\eta_j|^2 = {x_1}^2 + {x_2}^2 + {x_5}^2 + {x_6}^2 + {x_7}^2 + {x_{10}}^2,\\
\lambda_2 &=& |\eta_1|^2 = x_5^2 + x_{10}^2,\\
\lambda_3 &=& (|\eta_2|^2 - |\eta_3|^2)^2/4 = {\left( x_2\,x_6 - x_1\,x_7 \right) }^2,\\
\lambda_4 &=& 2\,\Re\left(\eta_1^2\,\eta_2\,\eta_3^*\right) \,=\,
      -x_1^2\, x_5^2 - x_2^2\, x_5^2 + x_5^2\, x_6^2 + x_5^2\, x_7^2 -\\
 &\ & 4\, x_1\, x_5\, x_6\, x_{10} - 4\, x_2\, x_5\, x_7\, x_{10} + x_1^2\, x_{10}^2 + x_2^2\, x_{10}^2 -
      x_6^2\, x_{10}^2 - x_7^2\, x_{10}^2.
\end{array}
\end{equation}

The $\widehat P$ matrix relative to the IB $\{\lambda\}$ has the following form:

\begin{equation}
\widehat P(\lambda)= 4\,\left(\begin{array}{cccc}
\lambda_1    & \lambda_2 & 2\,\lambda_3                       & 2\,\lambda_4 \\
\lambda_2    & \lambda_2 &  0                                  & \lambda_4    \\
2\,\lambda_3 &     0     & (\lambda_1 - \lambda_2) \,\lambda_3 &  0           \\
2\,\lambda_4 & \lambda_4 &  0                                  & \lambda_2\,(\lambda_ 1^2-\lambda_1\,\lambda_2-4\,\lambda_3)
\end{array}\right)
\end{equation}
The conditions $\widehat P(\lambda) > 0$ and rank$\left(\widehat P(p)\right) =
4$ yield the following restrictions on the acceptable values of the $\lambda$'s:

\begin{equation}
\left\{\begin{array}{l} 0\,<\,\lambda_2\, <\, \lambda_1,\\
 0\, <\, 4\,\lambda_3\, <\, \left(\lambda_1 - \lambda_2\right)^2,\\
\lambda_4^2\, <\, \lambda_2^2\left[\left(\lambda_1 - \lambda_2\right)^2 - 4\,\lambda_3\right].
\end{array}\right.
\label{S42'}
\end{equation}

The expressions of the $p$'s in terms of the $\lambda$'s turn out
to be the following:

\begin{equation}
\begin{array}{rcl}
p_1 &=& \lambda_1,\\ p_2 &=& \left(3\,{\lambda_1}^2 - 2\,\lambda_1\,\lambda_2 + 2\,{\lambda_2}^2 + 12\,\lambda_3 -
    2\,\lambda_4\right)/6,\\
    p_3 &=& {\lambda_1}^2 - 2\,\lambda_1\,\lambda_2 + 2\,{\lambda_2}^2 - 4\,\lambda_3 - 2\,\lambda_4,\\
p_4 &=& \lambda_2\,\left( 9\,{\lambda_1}^2 - 18\,\lambda_1\,\lambda_2 + 10\,{\lambda_2}^2 - 36\,\lambda_3 + 6\,\lambda_4
    \right)/6,\\
p_5 &=& \left(5\,{\lambda_1}^2\,\lambda_2 - 14\,\lambda_1\,{\lambda_2}^2 + 10\,{\lambda_2}^3 - 4\,\lambda_2\,\lambda_3 -
           4\,\lambda_1\,\lambda_4 + 6\,\lambda_2\,\lambda_4\right)/6,\\
p_6 &=& \left(6\,{\lambda_1}^3\,\lambda_2 - 17\,{\lambda_1}^2\,{\lambda_2}^2 + 14\,\lambda_1\,{\lambda_2}^3 -
           2\,{\lambda_2}^4 - 24\,\lambda_1\,\lambda_2\,\lambda_3 + 20\,{\lambda_2}^2\,\lambda_3 -\right. \\
    &\ &   3\,{\lambda_1}^2\,\lambda_4 + 2\,\lambda_1\,\lambda_2\,\lambda_4 +
            4\,{\lambda_2}^2\,\lambda_4 + \left. 12\,\lambda_3\,\lambda_4 -  2\,{\lambda_4}^2\right)/6,\\
p_7 &=&  \left( 4\,{\lambda_1}^3\,\lambda_2 - 11\,{\lambda_1}^2\,{\lambda_2}^2 + 6\,\lambda_1\,{\lambda_2}^3 +
         2\,{\lambda_2}^4 - 16\,\lambda_1\,\lambda_2\,\lambda_3 + 28\,{\lambda_2}^2\,\lambda_3 - \right. \\
    &\ & 5\,{\lambda_1}^2\,\lambda_4 + 10\,\lambda_1\,\lambda_2\,\lambda_4 - 4\,{\lambda_2}^2\,\lambda_4 + 4\,\lambda_3\,
         \lambda_4 + \left.  2\,{\lambda_4}^2\right)/6,\\
p_8 &=& \left(3\,{\lambda_1}^4\,\lambda_2 + 2\,{\lambda_1}^3\,{\lambda_2}^2 - 36\,{\lambda_1}^2\,{\lambda_2}^3 +
          52\,\lambda_1\,{\lambda_2}^4 - 20\,{\lambda_2}^5 - 24\,{\lambda_1}^2\,\lambda_2\,\lambda_3 -\right. \\
    &\ &  8\,\lambda_1\,{\lambda_2}^2\,\lambda_3 + 80\,{\lambda_2}^3\,\lambda_3 + 48\,\lambda_2\,{\lambda_3}^2 -
         6\,{\lambda_1}^3\,\lambda_4 + 16\,{\lambda_1}^2\,\lambda_2\,\lambda_4 - 16\,\lambda_1\,{\lambda_2}^2\,\lambda_4 + \\
    &\ & \left. 8\,{\lambda_2}^3\,\lambda_4 + 24\,\lambda_1\,\lambda_3\,\lambda_4 - 16\,\lambda_2\,\lambda_3\,\lambda_4 -
         4\,\lambda_1\,{\lambda_4}^2 + 12\,\lambda_2\,{\lambda_4}^2\right)/6,\\
p_9 &=& \left(33\,{\lambda_1}^4\,{\lambda_2}^2 - 132\,{\lambda_1}^3\,{\lambda_2}^3 + 168\,{\lambda_1}^2\,{\lambda_2}^4 -
          72\,\lambda_1\,{\lambda_2}^5 + 4\,{\lambda_2}^6 - 264\,{\lambda_1}^2\,{\lambda_2}^2\,\lambda_3 + \right.\\
    &\ &  528\,\lambda_1\,{\lambda_2}^3\,\lambda_3 - 144\,{\lambda_2}^4\,\lambda_3 + 528\,{\lambda_2}^2\,{\lambda_3}^2 -
        9\,{\lambda_1}^4\,\lambda_4 + 36\,\lambda_1^3\,\lambda_2\,\lambda_4 - 60\,\lambda_1^2\,\lambda_2^2\,\lambda_4 + \\
    &\ &  48\,\lambda_1\,{\lambda_2}^3\,\lambda_4 - 12\,{\lambda_2}^4\,\lambda_4 + 72\,{\lambda_1}^2\,\lambda_3\,
          \lambda_4 - 144\,\lambda_1\,\lambda_2\,\lambda_3\,\lambda_4 + 96\,{\lambda_2}^2\,\lambda_3\,\lambda_4 -\\
    &\ & 144\,{\lambda_3}^2\,\lambda_4 - 12\,{\lambda_1}^2\,{\lambda_4}^2 + 24\,\lambda_1\,\lambda_2\,{\lambda_4}^2 +
         12\,{\lambda_2}^2\,{\lambda_4}^2 + 48\,\lambda_3\,{\lambda_4}^2 - \left. 4\,{\lambda_4}^3\right)/6.
    \label{S42}
\end{array}
\end{equation}

In the range of values for the $\lambda$'s defined in
(\ref{S42'}), the values of the $p$'s defined in (\ref{S42})
render $\widehat P(p)$ everywhere positive semi-definite and of rank
four. We can conclude, therefore, that they yield parametric equations for
the 4\,D stratum $\widehat S^{(4,2)}$. Being the continuous image of a connected set, the
stratum is {\em connected}.

The elimination of the parameters $\lambda$ from (\ref{S42},\ref{S42'}) leads
to cumbersome relations, which it is not worthwhile writing down.

\subsubsection{Principal stratum $\widehat S^{(6,1)}$}

Implicit equations for the principal stratum are yielded in
$\real^9$ by the syzygies relating the invariants $p_i$,
$i=1,\dots ,9$, which can be easily obtained from the syzygies (\ref{B1}-\ref{B5}), using (\ref{newIB}.
At the points of $\widehat S_{\rm p}=\widehat S^{(6,1)}$ the matrix
$\widehat P(p)$ has to be positive semi-definite and its rank has
to be 6.

The method used for the singular strata does not help to obtain parametric equations for
$\widehat S_{\rm p}$ and to attain the goal one has to resort to less elegant and more pedantic
procedures. A possibility is to use the invariants of an IB of the subgroup {\bf SO}$_3$ of $G$
as parameters in terms of which to express the $p$'s, but in this case, great care has to be paid to the determination
of the range of the parameters that render
one-to-one the correspondence with the points $p\in\widehat S_{\rm p}$.

The M\"olien function for the subgroup {\bf SO}$_3$ of $G$ can be obtained as a byproduct of the calculation of the
M\"olien function for $G$ (see Appendix A):

\begin{equation}
M_{{\rm\bf SO}_3}(\eta) = \frac {1 + \eta^4 + \eta^8}{\left(1-\eta^2 - \eta^2\right)^3\,\left(1-\eta^3\right)^4},
\label{Molien3-7}
\end{equation}
which, after multiplying numerator and denominator by $\left(1 - \eta^4\right)$, becomes

\begin{equation}
M_{{\rm\bf SO}_3}(\eta) = \frac {1 - \eta^{12}}{\left(1 - \eta^2\right)^3\,\left(1 - \eta^3\right)^4\,
\left(1 - \eta^4\right)}.\label{Molien3-8}
\end{equation}
Equations (\ref{Molien3-7},\ref{Molien3-8}) indicate that a minimal IB for the subgroup {\bf SO}$_3$ of $G$ is formed by
eight invariants, with degrees $(2,\, 2,\, 2,\, 3,\, 3,\, 3,\, 3,\, 4)$ related by only one syzygy of degre 12.
A possible choice is the following:

\begin{equation}
\begin{array}{ll}
\begin{array}{rcl}
\lambda_1 &=& {\rm Tr}\left(\psi\,\psi^*\right),\\
\lambda_2 &=& \Re\left[{\rm Tr}\left(\psi^2\right)\right],\\
\lambda_3 &=& \Im\left[{\rm Tr}\left(\psi^2\right)\right],\\
\lambda_4 &=& \Re\left[{\rm Tr}\left(\psi^3\right)\right]),
\end{array}
& \qquad\qquad\begin{array}{rcl}
\lambda_5 &=& \Im\left[{\rm Tr}\left(\psi^3\right)\right],\\
\lambda_6 &=& \Re\left[{\rm Tr}\left(\psi^2\,\psi^*\right)\right],\\
\lambda_7 &=& \Im\left[{\rm Tr}\left(\psi^2\,\psi^*\right)\right],\\
\lambda_8 &=& {\rm Tr}\left(\psi\,\psi^*\,\psi\,\psi^*\right).
\end{array}\end{array}\label{IBSO3}
\end{equation}
Only the first seven elements of the basis are algebraically independent. The explicit expression of the syzygy relating
the $\lambda$'s is reported in Appendix B (see~(\ref{sySO3})).

The $\widehat P$-matrix relative to the IB $\{\lambda_1,\dots ,\lambda_8\}$ is is the following and
for values of the $\lambda$'s satisfying the syzygy (\ref{sySO3}), it has rank seven:

\begin{equation}
\begin{array}{ll}
\begin{array}{rcl}
\widehat P_{1i} &=& 2\,d^{(\lambda)}_i\,\lambda_i                                   , \\
\widehat P_{22} &=& 4\lambda_1                                                      , \\
\widehat P_{23} &=& 0                                                               , \\
\widehat P_{24} &=& 6\lambda_6                                                      , \\
\widehat P_{25} &=& 6\lambda_7                                                      , \\
\widehat P_{26} &=& 2\left( \lambda_4 + 2\lambda_6 \right)                          , \\
\widehat P_{27} &=& 2\left( \lambda_5 - 2\lambda_7 \right)                          , \\
\widehat P_{28} &=& 4\lambda_1\lambda_2                                             , \\
\widehat P_{33} &=& 4\lambda_1                                                      , \\
\widehat P_{34} &=& -6\lambda_7                                                     , \\
\widehat P_{35} &=& 6\lambda_6                                                      , \\
\widehat P_{36} &=& 2\left( \lambda_5 + 2\lambda_7 \right)                          , \\
\widehat P_{37} &=& -2\left( \lambda_4 - 2\lambda_6 \right)                         , \\
\widehat P_{38} &=& 4\lambda_1\lambda_3                                             , \\
\widehat P_{44} &=& \frac 34\left(6\lambda_1^2-\lambda_2^2-\lambda_3^2-6\lambda_8\right), \\
\end{array}
&
\begin{array}{rcl}
d^{(\lambda)}           &=&  (2,\, 2,\, 2,\, 3,\, 3,\, 3,\, 3,\, 4),\quad i=1,\dots ,8,                                                                                            \\
\widehat P_{45} &=&  0                                                                                                                                                ,   \\
\widehat P_{46} &=&  \frac 12\left(2\lambda_1\lambda_2 + {\lambda_2}^2 - {\lambda_3}^2\right)                                                          , \\
\widehat P_{47} &=& -\left(\lambda_1 - \lambda_2 \right) \lambda_3                                                                                                     ,  \\
\widehat P_{48} &=& 2\left( 2\lambda_1\lambda_4 + \lambda_2\lambda_6 - \lambda_3\lambda_7 \right)                                                                      ,  \\
\widehat P_{55} &=& \frac 34 \left(6{\lambda_1}^2 - {\lambda_2}^2 - {\lambda_3}^2 - 6\lambda_8 \right)                                                 ,  \\
\widehat P_{56} &=& \left(\lambda_1 + \lambda_2 \right)\lambda_3                                                                                                       ,  \\
\widehat P_{57} &=& \frac 12\left(2\lambda_1\lambda_2 - {\lambda_2}^2 + {\lambda_3}^2\right)                                                          , \\
\widehat P_{58} &=& 2\left( 2\lambda_1\lambda_5 + \lambda_3\lambda_6 + \lambda_2\lambda_7  \right)                                                                     ,  \\
\widehat P_{66} &=& \frac 1{12}\left(2{\lambda_1}^2 + 8\lambda_1\lambda_2 + 5{\lambda_2}^2 + 5{\lambda_3}^2 + 6\lambda_8\right)                       , \\
\widehat P_{67} &=& \frac 23 \lambda_1\lambda_3                                                                                                        , \\
\widehat P_{68} &=& -\frac 23 \left( 3\lambda_2\lambda_4 + 3\lambda_3\lambda_5 - 14\lambda_1\lambda_6 + 2\lambda_2\lambda_6 + 2\lambda_3\lambda_7\right), \\
\widehat P_{77} &=&  \frac 1{12}\left(2{\lambda_1}^2 - 8\lambda_1\lambda_2 + 5{\lambda_2}^2 + 5{\lambda_3}^2 + 6\lambda_8\right)                          ,\\
\widehat P_{78} &=& \frac 23\left(3\lambda_3\lambda_4 - 3\lambda_2\lambda_5 - 2\lambda_3\lambda_6 + 14\lambda_1\lambda_7 + 2\lambda_2\lambda_7\right)   ,\\
\widehat P_{88} &=& -\frac 83\left(3{\lambda_1}^3 - 2{\lambda_4}^2 - 2{\lambda_5}^2 +  2{\lambda_6}^2 + 2{\lambda_7}^2 - 9\lambda_1\lambda_8\right)     . \\
\end{array}
\end{array}\label{matricePSO3}
\end{equation}

The connection between the $p$'s and the $\lambda$'s can be immediately obtained from their very definitions:

\begin{equation}
\begin{array}{rcl}
p_1 &=& \lambda_1,\\
p_2 &=& \lambda_8,\\
p_3 &=& \lambda_2^2 +  \lambda_3^2,\\
p_4 &=& \lambda_4^2 + \lambda_5^2,\\
p_5 &=& \lambda_6^2 + \lambda_7^2,\\
p_6 &=& \lambda_2\,\lambda_4\,\lambda_6 + \lambda_3\,\lambda_5\,\lambda_6 - \lambda_3\,\lambda_4\,\lambda_7 +
  \lambda_2\,\lambda_5\,\lambda_7,\\
p_7 &=& \lambda_2\,{\lambda_6}^2 + 2\,\lambda_3\,\lambda_6\,\lambda_7 - \lambda_2\,{\lambda_7}^2,\\
p_8 &=& {\lambda_2}^2\,\lambda_4\,\lambda_6 - {\lambda_3}^2\,\lambda_4\,\lambda_6 +
  2\,\lambda_2\,\lambda_3\,\lambda_5\,\lambda_6 + 2\,\lambda_2\,\lambda_3\,\lambda_4\,\lambda_7 -
  {\lambda_2}^2\,\lambda_5\,\lambda_7 + {\lambda_3}^2\,\lambda_5\,\lambda_7,\\
p_9 &=& {\lambda_2}^3\,{\lambda_4}^2 - 3\,\lambda_2\,{\lambda_3}^2\,{\lambda_4}^2 +
  6\,{\lambda_2}^2\,\lambda_3\,\lambda_4\,\lambda_5 - 2\,{\lambda_3}^3\,\lambda_4\,\lambda_5 -
  {\lambda_2}^3\,{\lambda_5}^2 + 3\,\lambda_2\,{\lambda_3}^2\,{\lambda_5}^2.\end{array}\label{Plambda}
\end{equation}
As already noted, there is not one-to-one correspondence between the points $\lambda=(\lambda_1,\dots ,\lambda_8)$,
lying on the surface determined by the syzygy (\ref{sySO3}) and rendering positive semi-definite and
of rank seven the $\widehat P$-matrix
associated to the integrity basis $\{\lambda\}$, and the points $p$ of the principal stratum of $\real^{10}/G$. In fact, the action of the group $G$ in $\real^{10}$ induces a linear
action of $G$ on the variables $\lambda_i$. With the definitions

\begin{equation}
\eta_1 = \lambda_2 + i\,\lambda_3,\qquad \eta_2 = \lambda_4 +
i\,\lambda_5,\qquad \eta_3 = \lambda_6 + i\,\lambda_7,
\end{equation}
the variables $\lambda$ and $\eta$ have the following transformation properties:

\begin{itemize}
\item[i)] $\lambda_1$ and $\lambda_8$ are invariant under the whole group;

\item[ii)] under gauge transformations, U$_1(\phi)$:\quad $(\eta_1,\eta_2,\eta_3) \rightarrow
(e^{2\,i\,\phi}\,\eta_1,e^{3\,i\,\phi}\,\eta_2, e^{i\,\phi}\,\eta_3)$;

\item[iii)] under time reversal  ${\cal T}$:\quad $(\eta_1,\eta_2,\eta_3)\rightarrow (\eta_1^*,\eta_2^*,\eta_3^*)$.
\end{itemize}

The one-to-one correspondence problem can be solved by finding a criterion enabling to select a unique point in each orbit
of the action of $G$ in the $\lambda$ space. To this end, let us meke to preliminary remarks;

\begin{itemize}
\item[i)] A direct check shows that for $\eta_1=0$ and general values of $\eta_2$ and $\eta_3$
the rank of $\widehat P(p(\lambda))$ is six.

\item[ii)] Using (\ref{limit}) one easily realizes that rank$\left(\widehat P(p(\lambda))\right)\,<\,$rank$\left(J(\lambda)
\right)\,<\,6$, for $\lambda_3=\lambda_5=\lambda_7=0$.
\end{itemize}

Now, let us choose any orbit, $\bar\Omega$, in the principal stratum of $\real^{10}/G$, {\em i.e.}, a $\bar p\in\widehat S_{\rm p}$, and
let $\bar\lambda$ be such that $p(\bar\lambda)=\bar p$. At
least one of the parameters $\bar\eta_i$, $i=1,2,3$, has to be different from zero. Let us say that $\bar \eta_{\bar i}$,
is the first one, in lexicographic order\footnote{In a lexicographic ordering, $(v_1,\dots ,v_k)\,>\,(v'_1,\dots ,v'_k)$
means that the first non vanishing component $v_i-v'_i$, $i=1,\dots ,k$, is positive.}.
Then, on $\bar\Omega$ there will be a point $\bar\lambda'$, such that
$\bar\eta'_{\bar i}$ is real and positive; this point can be attained from $\bar\lambda$ by means of a convenient gauge
transformation. Since at least a $\bar\eta_j$, $j=1,2,3$, has to be $\ne 0$, this means $\bar i<3$. At this point we still
have at our
disposal time reversal transformations to fix the sign of the
imaginary part of one of the $\bar\eta_i$, $i> \bar i$ (recall that at least one of the $\Im(\bar\eta_j)$ has to be
$\ne 0$). So, if $\bar j$ is such that $\Im(\bar\eta_{\bar j})$ is the first (in lexicographic order)  non vanishing
imaginary part of the $\bar\eta$'s, we can choose $\Im(\bar\eta_{\bar j})\,>\,0$.

After introducing a lexicographic order
in the real vector spaces generated by the vectors formed, respectively, with the
real and imaginary parts of the $\eta$'s, the criterion that we have devised can be resumed in the following simple
additional conditions on these vectors:

\begin{equation}
\lambda_3 = 0,\qquad (\lambda_2,\lambda_4,\lambda_6)\,>\,0,\qquad (\lambda_3,\lambda_5,\lambda_7)\,>\,0,\qquad {\rm and}
\ \lambda_5=0 \ {\rm for}\ \lambda_2=0.
\end{equation}

Using these conditions, the expressions (\ref{Plambda}) of the $p$'s can be simplified as follows:

\begin{equation}
\begin{array}{ll}
\begin{array}{rcl}
p_1 &=& \lambda_1,\\ p_2 &=& \lambda_8,\\ p_3 &=& \lambda_2^2,\\
p_4 &=& \lambda_4^2 + \lambda_5^2,\\ p_5 &=& \lambda_6^2 +
\lambda_7^2,
\end{array}
& \qquad\qquad\begin{array}{rcl} p_6 &=&
\lambda_2\left(\lambda_4\,\lambda_6 +
\lambda_5\,\lambda_7\right),\\ p_7 &=& \lambda_2\left(\lambda_6^2
- \lambda_7^2\right),\\ p_8 &=&
\lambda_2^2\left(\lambda_4\,\lambda_6 -
\lambda_5\,\lambda_7\right),\\ p_9 &=& \lambda_3\left(\lambda_4^2
- \lambda_5^2\right).
\end{array}\end{array}\label{p'}
\end{equation}
where the $\lambda$'s are required to satisfy the syzygy (\ref{sySO3}), the conditions rendering the matrix
$\widehat P(p)$ positive semi-definite and of rank six and $(\lambda_5, \lambda_7) \,>\, 0$.

\section{An example: minima of a general fourth degree Landau polynomial}

As a simple example, we have also calculated the minimum of a general fourth degree polynomial free energy

\begin{equation}
\widehat\Phi^{(4)}(p)=\frac{\alpha_0}2 \,p_1^2 + \sum_{j=1}^3 \alpha_j p_j,\qquad \alpha_i\in\real; \ i=0,\dots, 3,
\label{11}
\end{equation}
in the additional assumptions that it is bounded below and has a local maximum at the origin ($\alpha_1<0$). In
(\ref{11}) the $\alpha$'s are phenomenological parameters.

Recalling the definition $\tilde p_i=p_i/p_1^{d_i/2}$,
the polynomial $\widehat\Phi^{(4)}(p)$ can be put in the following form:

\begin{equation} \label{Delta}
\widehat\Phi^{(4)}(p) = \frac {p_1^2}2\,\Delta + \alpha_1\, p_1,
\end{equation}
where we have defined

\begin{equation}
\Delta = \alpha_0 + 2\,\alpha_2\,\tilde p_2 + 2\,\alpha_3\,\tilde p_3.
\end{equation}
Since, owing to its definition (\ref{IB}), $p_1$ ranges over the whole non negative real numbers, in the orbit space, the polynomial
$\widehat\Phi^{(4)}(p)$ is bounded below (in the assumption $\alpha_1<0$) if and
only if the minimum of $\Delta$, $\delta$, evaluated in the section $p_1=1$ of the orbit space, is positive.
Being the minimum of the r.h.s. of (\ref{Delta}), thought of as a function only of
$p_1 \ge 0$,  equal to $-\alpha_1/(2\Delta)$, the absolute minimum of $\widehat\Phi^{(4)}(p)\big|_{p\in
S}$ is $-\alpha_1^2/(2\delta)$.
In this way, the determination of the minimum of $\widehat \Phi^{(4)}(p)$ is reduced to the calculation of $\delta$.

\begin{center}
\begin{figure}
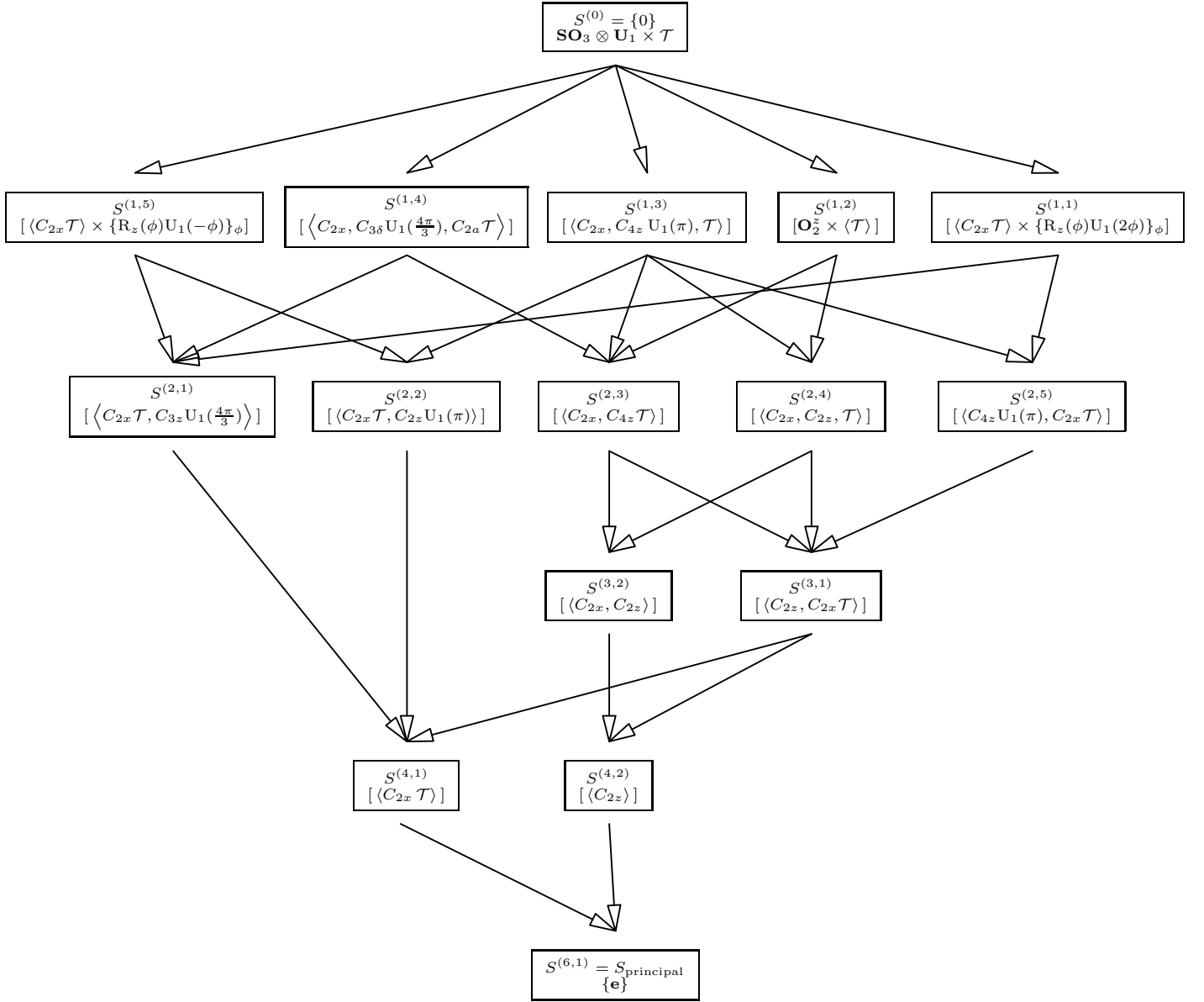

\caption{Localization of the absolute minimum of a fourth degree $G$-invariant polynomial,
$\widehat\Phi^{(4)}(\alpha,p)=\alpha_0 p_1^2/2 + \sum_{j=1}^3 \alpha_j p_j$, $\alpha_1<0$, as a function of its
coefficients. For values of $(\alpha_0,\alpha_2,\alpha_3)$ in ${\cal R}_1$ or in ${\cal R}_2$ or in ${\cal R}_3$, the
absolute minimum lies, respectively, in the strata $S^{(1,4)}$ or $\{S^{(1,2)},S^{(1,3)},S^{(2,4)}\}$ (degenerate minimum)
or $S^{(1,1)}$. For particular values of the $\alpha$'s, see Table~\ref{IV}.}\label{F2}

\begin{texdraw}

\drawdim mm

\setunitscale 0.6

\move(100 40) \avec(100 180) \move(34 100) \avec(180 100)
\textref h:C v:C \htext(95 171){$\small\alpha_2$}
\textref h:C v:C \htext(171 95){$\small\alpha_3$}

\move(100 100) \linewd 0.8 \lvec(90 160)
\move(100 100) \linewd 0.8 \lvec(165 100)
\move(100 100) \linewd 0.8 \lvec(70 40)

\textref h:C v:C \htext(60 105){${\cal R}_2$:\ ${\small\alpha_0 > -2\,\alpha_2/3}$}
\textref h:C v:C \htext(140 130){${\cal R}_1:\ {\small\alpha_0 > -2\,\alpha_2}$}
\textref h:C v:C \htext(136 70){${\cal R}_3:\ {\small\alpha_0 > -\alpha_2-2\,\alpha_3}$}

\textref h:C v:C \rtext td:280 (90 140){$\scriptstyle\alpha_2=-6\,\alpha_3$}
\textref h:C v:C \rtext td:65 (74 55){$\scriptstyle\alpha_2=2\,\alpha_3$}
\textref h:C v:C \htext(135 103){$\scriptstyle\alpha_2=0$}

\setgray 1 \lvec(100 195)

\end{texdraw}
\end{figure}
\end{center}

The absolute minimum of $\delta$ in each singular stratum can be easily computed using the equations of the strata.
For the principal stratum the determination of the minimum of $\widehat\Phi^{(4)}(p)\big|_{p_1=1}$ as a constrained
minimum is difficult and it is easier to solve equation (\ref{1}) and to check subsequently if the solutions lie in the
principal stratum.
A comparison of the values of the minima in the different strata, in order to determine the absolute minimum, leads to the
results resumed in Table~\ref{V} and illustrated in Figure~\ref{F2}, where the denomination of the strata introduced in
 the text has been used. Owing to the
low degree of the polynomial defining $\widehat\Phi^{(4)}(p)$ in (\ref{11}), and the consequent low number of free
parameters $\alpha$, the absolute minimum exhibits strong degeneracy, particularly for special values of the $\alpha$'s.
If these special values are excluded, spontaneous breaking of the symmetry can generate only five distinct phases out of fifteen
permitted in the $G$-symmetry; some of them are unstable. For no non trivial values of $(\alpha_2,\alpha_3)$ does
the absolute minimum lie on the stratum $S^{(2,2)}$.

For general values of the $\alpha$'s, our results are in agreement with Mermin's ones \cite{14}. Let us add a few words
about the perturbative stability of the three degenerate phases in the region ${\cal R}_2$ (see Table~\ref{V}).
For $(\alpha_0,\alpha_2,\alpha_3)\in{\cal R}_2$, the addition to the free energy, $\Phi^{(4)}$, of a ''small" perturbation,
consisting in an invariant polynomial of degree six:

\begin{equation}
\Theta^{(6)}=\alpha_4\,p_4 + \alpha_5\,p_5,
\end{equation}
splits the three degenerate minima
determined by the 4-th degreee term\footnote{The inclusion of terms of 6-th degree, depending only on
$p_1,p_2$ and $p_3$ would be useless for splitting the degenerate minima, so these terms will be neglected, with no loss
of generality.}. This is easy to check, at least in the additional assumption that the perturbation leaves the absolute
minimum in one of the strata corresponding to the degenerate phases. In fact,
at the first perturbative order, one obtains from Tables~\ref{I} and \ref{II} the following shifts,
$\Theta^{(6)}_{(d,r)}$, in the values of the 6-th order free energy at the points where $\Phi^{(4)}(p)$ takes on its
degenerate absolute minimum under consideration:

\begin{equation}
\Theta^{(6)}_{(1,2)} = \frac{\alpha_1^3}{\delta^3}\,\frac{\alpha_4 + \alpha_5}6,\qquad
\Theta^{(6)}_{(1,3)} = 0,\qquad \Theta^{(6)}_{(2,4)} =  \frac{\alpha_1^3}{\delta^3}\,
\left(\alpha_4 + \alpha_5\right)\,\xi,
\end{equation}
where $0\,<\, \xi \,<\, 1/6$.

Since $-\alpha_1/\delta > 0$, the absolute minimum will be perturbatively stable on $\widehat S^{(1,2)}$ or,
respectively, $\widehat S^{(1,3)}$, according as $\alpha_4+\alpha_5$ is negative or positive.

As stressed in the Introduction, the difficulties mentioned above can be overcome if one puts less restrictive
upper limits to the degree of the polynomial describing the free energy. It is trivial, for instance, to realize
that the following class of bounded below polynomial functions \cite{NC} have a vanishing maximum at the origin of $\real^{10}$
and display an absolute minimum at the arbitrarily chosen point $\bar p\in \widehat S$:

\begin{equation}
\sum_{i=1}^9\,\alpha_i\left[\left(p_i - \bar p_i\right)^{2n_i} - \bar p_i^{2n_i}\right],
\end{equation}
where the $\alpha$'s are positive constants and the $n$'s are positive integers.

The physical implications of our results and the derivation of a more realistic form of the free energy will be
discussed in forthcoming papers. We shall limit ourselves, here, to note that, the D-wave content in terms, {\em e.g.},
of d- or extended s-waves (mentioned in the introduction) can be directly read off the fourth column of Table~\ref{III},
for each of the possible symmetry phases of the system under consideration.

\section{Conclusions}
Through a detailed and rigorous analysis, we have determined and characterized
the possible ground states of D-wave condensates in isotropic space, which are rilevant both
in high $T_c$ superconductivity and in $^3$He phase transitions. The
problem has been formulated and solved
in the framework of geometric invariant theory, that, as noted long ago \cite{SA}, is the natural
mathematical setting in which to deal with minimization of invariant potentials, which are plagued
by degeneracy problems. This enabled us to complete and/or correct results previously obtained
by other authors, using different methods.

Until now, the kind of approach we have used had only been applied to
relatively simple models, with coregular symmetry groups, in condensed matter \cite{SV1} and in
elementary particle physics \cite{ST}. In this paper, we have fully characterized the geometry of the
orbit space of the linear group {\bf SO}$_3\otimes${\bf U}$_1\times\langle{\cal T}\rangle$, acting in
the space $\real^{10}$, a highly non trivial group, with a non coregular minimal integrity basis,
formed by nine elements with degrees up to 12. For this group we have determined the explicit
form of the elements of an integrity basis and of the syzygies relating these elements, the
equations and inequalities determining the orbit space and its stratification and the corresponding
isotropy subgroups chains. The minima of a general fourth degree polynomial invariant free energy
have been recalculated and the sixth degree polynomial perturbative corrections to its absolute
minimum have been determined. Some of our conclusions have been compared with previous results
obtained by other authors.

To overcome some computational difficulties in the full characterization of the stratification
of the orbit space, difficulties originating from the high dimensionality of the
problem, we had to devise a
sophisticated procedure, allowing to obtain {\em rational} parametric equations also for the
higher dimensional singular strata. The procedure is absolutely general and is reminiscent, but
in a different context and in higher dimensions, of classical techniques for parameterizing plane algebraic curves
\cite{Walker}.

\acknowledgments This paper is partially supported by RFBR, INFN and MURST

\appendix

\section{Calculation of M\"olien functions}

In this Appendix, we shall explicitly calculate the two integrals, $I_1$ and $I_2$, appearing in the expression of
$M_G(\eta)$ (see (\ref{Molien})):

\begin{equation}
\begin{array}{rcl}
I_1 &=& {\displaystyle\frac 1{32\pi^3} \int_0^{2\pi}d\phi_1\int_0^{2\pi}d\phi_2\int_0^{\pi}\sin\theta d\theta
\int_0^{2\pi}\,\frac{d\alpha}{ f_1(\cos\chi, \alpha)}},\\
&&\\
I_2 &=& {\displaystyle\frac 1{16\pi^2} \int_0^{2\pi}d\phi_1\int_0^{2\pi}d\phi_2\int_0^{\pi}\frac{\sin\theta\, d\theta}
{f_2(\cos\chi)}},
\end{array}\label{Molien1}
\end{equation}
where

\begin{equation}
\begin{array}{rcl}
f_1(\cos\chi, \alpha) &=& {\displaystyle
\prod_{k=-2}^2\left[\left(1 - \eta\,e^{i(k\,\chi  + \alpha)}\right)\left(1 - \eta\,e^{i(k\,\chi  - \alpha)}\right)\right]}\\
&&\\
f_2(\cos\chi) &=& {\displaystyle
\prod_{k=-2}^2\left[\left(1 - \eta\,e^{i\,k\,\chi} \right)\left(1 + \eta\,e^{i\,k\,\chi}\right)\right]}
\end{array}\label{Molien2}
\end{equation}
and $\cos \chi$ is defined in (\ref{Molien'}).

Using as new integration variables
$\phi_2$, $u = \cos(\phi_1 + \phi_2)$ and $v = \cos\chi$, the integral over $\phi_2$ can be immediately performed and
one remains with

\begin{equation}
\begin{array}{rcl}
I_1 &=& {\displaystyle\frac 1{2\pi^2} \int_0^{2\pi}\,d\alpha \int_{-1}^{1}\, \frac{dv}{f_1(v,\alpha)}
\int_v^1\,\frac{du}{(1 + u)\sqrt{1-u^2}}}\\
&&\\
&=& {\displaystyle\frac 1{2\pi^2} \int_0^{2\pi}\,d\alpha \int_{-1}^{1}\,\frac{dv}{f_1(v,\alpha)}
\,\frac{\sqrt{1-v^2}}{1 + v}} \\
&&\\
&=& {\displaystyle\frac 1{2\pi^2} \int_0^{2\pi}\,d\alpha \int_{0}^{2\pi}\,d\chi\,\frac{\sin^2(\chi/2)}
{f_1(\cos\chi,\alpha)}}
\end{array}\label{Molien3-1}
\end{equation}
and, analogously,

\begin{equation}
I_2 =  \frac 1{\pi} \int_{0}^{2\pi}d\chi\,\frac{\sin^2(\chi/2)}{f_2(\cos\chi)}.
\label{Molien3-2}
\end{equation}

After defining

\begin{equation}
\zeta_1 = e^{i\,\chi},\qquad \zeta_2 = e^{i\,\alpha},
\end{equation}
the remaining integrals can be conveniently thought of as integrals on unit circles of complex planes
of the variables $\zeta_1$ and $\zeta_2$

\begin{equation}
\begin{array}{rcl}
I_1 &=& -{\displaystyle \frac 1{8\,\pi^2\,\eta^4} \oint \frac{d\zeta_2}
{\left(\zeta_2 - \eta\right)\left(\zeta_2\,\eta - 1\right)}
\oint \,{d\zeta_1}\,F_1\left(\zeta_1,\zeta_2,\eta\right)  }\\
&&\\
I_2 &=& {\displaystyle \frac 1{4\,\pi \,i\,\eta^4\,\left(\eta^2 - 1\right)} \oint \,d\zeta_1\,
F_2\left(\zeta_1,\eta\right),}\label{Molien3-3}
\end{array}
\end{equation}
where

\begin{equation}
\begin{array}{l}
F_1\left(\zeta_1,\zeta_2,\eta\right) = \\
\ \ {\displaystyle \frac{\zeta_1^4\,\left(\zeta_1 -1\right)^2}
               {\left(\zeta_1^2 -\frac{\eta}{\zeta_2} \right) \,
                \left(\zeta_1^2 - \zeta_2\,\eta \right) \,
                \left(\zeta_1 -\frac{\eta}{\zeta_2} \right) \,
                \left(\zeta_1 - \zeta_2\,\eta \right) \,
                \left(\zeta_1 - \frac{\zeta_2}{\eta} \right) \,
                 \left(\zeta_1 - \frac 1{\zeta_2\,\eta} \right) \,
                 \left(\zeta_1^2 -\frac{\zeta_2}{\eta} \right) \,
                 \left(\zeta_1^2 -\frac 1{\zeta_2\,\eta} \right) }  }\\
\\
F_2\left(\zeta_1,\eta\right) = {\displaystyle \frac{\zeta_1^4\,\left(\zeta_1 - 1\right)^2}{\left(\zeta_1^2 - \eta^2\right)
\left(\zeta_1^4 - \eta^2\right)\left(\zeta_1^2 - \frac 1{\eta^2}\right)\left(\zeta_1^4 - \frac 1{\eta^2}\right)}.}
\end{array}\label{Molien3-4}
\end{equation}

Using the Cauchy theorem and recalling that $|\eta|<1$, the calculation reduces to a computation of residues at the
internal poles to the unit circles. So, one ends up with the following expressions for $I_1$ and $I_2$:

\begin{equation}
\begin{array}{rcl}
I_1 &=& {\displaystyle \frac{1 + 3\,\eta ^8 + 2\,\eta^{10} + 3\,\eta ^{12} + \eta ^{20}}{2\,\left(1 - \eta^2\right)^3\,
\left(1 - \eta^4\right)^2\,\left(1 - \eta^8\right)\,\left(1 - \eta + \eta^2\right)^2\,
\left(1 + \eta + \eta^2\right)^2},}\\
&&\\
I_2 &=& {\displaystyle \frac{1 - \eta^2 +\eta ^4}{2\,\left(1 - \eta ^2\right)^3
\,\left(1 - \eta ^4\right)\,\left(1 - \eta + \eta^2\right)\,\left(1 + \eta + \eta^2\right)}}.
\end{array}\label{Molien3-5}
\end{equation}
The sum of $I_1$ and $I_2$ gives to the expression of $M_G(\eta)$ reported in (\ref{10}).

As a byproduct of the calculation, one finds also the following expression for the M\"olien function of the subgroup
{\bf SO}$_3$ of $G$:

\begin{equation}
\begin{array}{rcl}
M_{{\rm\bf SO}_3}(\eta) &=& {\displaystyle\frac 1{8\pi^2} \int_0^{2\pi\,}d\phi_1\,\int_0^{2\pi}\,d\phi_2\,
\int_0^{\pi}\,d\theta\,\sin\theta\,f_1(\cos\chi, 0)   }\\
&&\\
&=& -{\displaystyle \frac 1{4\,\pi\,i\,\eta^4\,\left(1 - \eta^2\right)}\,\oint \,{d\zeta_1}\,F_1\left(\zeta_1,1,\eta\right)
  } \\
&&\\
&=& {\displaystyle \frac {1 + \eta^4 + \eta^8}{\left(1 - \eta^2\right)^3\,\left(1 - \eta^3\right)^4}.}
\end{array}\label{Molien3-6}
\end{equation}

\section{Syzygies}
\subsection{Explicit form of the syzygies for the group $G$}

The ten elements of the non minimal IB $\{\xi,\theta\}$ defined in (\ref{IB},\ref{newIB}) are related by the following
syzygies:

\begin{equation}
\begin{array}{rcl}
{\theta _1}^2 &=&
   \frac{1}{192}\left[36\,{\xi _1}^6\,\xi _3 - 144\,{\xi _1}^4\,\xi _2\,\xi _3 + 180\,{\xi _1}^2\,{\xi _2}^2\,\xi _3 -
       72\,{\xi _2}^3\,\xi _3 - 36\,{\xi _1}^4\,{\xi _3}^2 + 108\,{\xi _1}^2\,\xi _2\,{\xi _3}^2 -\right.\\
  &\ & 72\,{\xi _2}^2\,{\xi _3}^2 + 9\,{\xi _1}^2\,{\xi _3}^3 - 18\,\xi _2\,{\xi _3}^3 - 80\,{\xi _1}^3\,\xi _3\,\xi _4 +
       144\,\xi _1\,\xi _2\,\xi _3\,\xi _4 + 24\,\xi _1\,{\xi _3}^2\,\xi _4 + 48\,\xi _3\,{\xi _4}^2 +\\
  &\ & 144\,{\xi _1}^3\,\xi _3\,\xi _5 - 144\,\xi _1\,\xi _2\,\xi _3\,\xi _5 + 72\,\xi _1\,{\xi _3}^2\,\xi _5 -
       288\,\xi _3\,\xi _4\,\xi _5 - 144\,\xi _3\,{\xi _5}^2 -96\,{\xi _1}^2\,\xi _3\,\xi _6 -\\
  &\ & 24\,{\xi _3}^2\,\xi _6 + 192\,{\xi _6}^2 + \left. 24 \,\theta _1\,\left( 8\,{\xi _1}^2 - 12\,\xi _2 - \xi _3 \right)
       \,\xi _3 +
       96\,\theta _2\,\left( \xi _1\,\xi _3 - 4\,\xi _5 \right) - 16\,\theta _3\,\xi _3 \right];
\end{array}\label{B1}
\end{equation}

\begin{equation}
\begin{array}{rcl}
\theta _1\,\theta _2 &=& -\frac{1}{2}\left(\xi _3\,\xi _4\,\xi _6 + \theta _1\,\xi _3\,\xi _4 - 2\,\theta _2\,\xi _6 +
       2\,\theta _3\,\xi _5 \right);
\end{array}\label{B2}
\end{equation}

\begin{equation}
\begin{array}{rcl}
{\theta _2}^2 &=&
   \frac{1}{768}\left[ -36\,{\xi _1}^6\,{\xi _3}^2 + 144\,{\xi _1}^4\,\xi _2\,{\xi _3}^2 -
   180\,{\xi _1}^2\,{\xi _2}^2\,{\xi _3}^2 + 72\,{\xi _2}^3\,{\xi _3}^2 + 36\,{\xi _1}^4\,{\xi _3}^3 -\right.\\
&\ & 108\,{\xi _1}^2\,\xi _2\,{\xi _3}^3 + 72\,{\xi _2}^2\,{\xi _3}^3 - 9\,{\xi _1}^2\,{\xi _3}^4 + 18\,\xi _2\,{\xi _3}^4 +
     80\,{\xi _1}^3\,{\xi _3}^2\,\xi _4 - 144\,\xi _1\,\xi _2\,{\xi _3}^2\,\xi _4 -\\
&\ & 24\,\xi _1\,{\xi _3}^3\,\xi _4 - 48\,{\xi _3}^2\,{\xi _4}^2 - 44\,{\xi _1}^3\,{\xi _3}^2\,\xi _5 +
     144\,\xi _1\,\xi _2\,{\xi _3}^2\,\xi _5 - 72\,\xi _1\,{\xi _3}^3\,\xi _5 + \\
&\ & 1056\,{\xi _3}^2\,\xi _4\,\xi _5 +144\,{\xi _3}^2\,{\xi _5}^2 + 96\,{\xi _1}^2\,{\xi _3}^2\,\xi _6 + 24\,{\xi _3}^3\,\xi _6 - 384\,\xi _3\,{\xi _6}^2 +\\
&\ & 24\,\theta _1\,\xi _3\,\left( -8\,{\xi _1}^2\,\xi _3 + 12\,\xi _2\,\xi _3 + {\xi _3}^2 - 16\,\xi _6 \right) -
     96\,\theta _2\,\xi _3\, \left( \xi _1\,\xi _3 - 4\,\xi _5 \right) +\\
&\ &\left. 16\,\theta _3\,\left( {\xi _3}^2 + 24\,\xi _6\right) - 384\,\theta _4\right] ;
\end{array}\label{B3}
\end{equation}

\begin{equation}
\begin{array}{rcl}
\theta _2\,\theta _3 &=&
   \frac{1 }{128}\left[- 36\,{\xi _1}^7\,{\xi _3}^2 + 144\,{\xi _1}^5\,\xi _2\,{\xi _3}^2 -
     180\,{\xi _1}^3\,{\xi _2}^2\,{\xi _3}^2 + 72\,\xi _1\,{\xi _2}^3\,{\xi _3}^2 + 36\,{\xi _1}^5\,{\xi _3}^3 - \right.\\
&\ & 108\,{\xi _1}^3\,\xi _2\,{\xi _3}^3 + 72\,\xi _1\,{\xi _2}^2\,{\xi _3}^3 -
     9\,{\xi _1}^3\,{\xi _3}^4 + 18\,\xi _1\,\xi _2\,{\xi _3}^4 + 144\,{\xi _1}^6\,\xi _3\,\xi _4 -
     576\,{\xi _1}^4\,\xi _2\,\xi _3\,\xi _4 + \\
&\ & 720\,{\xi _1}^2\,{\xi _2}^2\,\xi _3\,\xi _4 - 288\,{\xi _2}^3\,\xi _3\,\xi _4 - 64\,{\xi _1}^4\,{\xi _3}^2\,\xi _4 +
     288\,{\xi _1}^2\,\xi _2\,{\xi _3}^2\,\xi _4 - 288\,{\xi _2}^2\,{\xi _3}^2\,\xi _4 + \\
&\ & 12\,{\xi _1}^2\,{\xi _3}^3\,\xi _4 - 72\,\xi _2\,{\xi _3}^3\,\xi _4 - 320\,{\xi _1}^3\,\xi _3\,{\xi _4}^2 +
     576\,\xi _1\,\xi _2\,\xi _3\,{\xi _4}^2 + 48\,\xi _1\,{\xi _3}^2\,{\xi _4}^2 + 192\,\xi _3\,{\xi _4}^3 +\\
&\ & 144\,{\xi _1}^6\,\xi _3\,\xi _5 - 576\,{\xi _1}^4\,\xi _2\,\xi _3\,\xi _5 + 720\,{\xi _1}^2\,{\xi _2}^2\,\xi _3\,\xi _5 -
     288\,{\xi _2}^3\,\xi _3\,\xi _5 - 288\,{\xi _1}^4\,{\xi _3}^2\,\xi _5 + \\
&\ & 576\,{\xi _1}^2\,\xi _2\,{\xi _3}^2\,\xi _5 - 288\,{\xi _2}^2\,{\xi _3}^2\,\xi _5 - 36\,{\xi _1}^2\,{\xi _3}^3\,\xi _5 -
     72\,\xi _2\,{\xi _3}^3\,\xi _5 + 256\,{\xi _1}^3\,\xi _3\,\xi _4\,\xi _5 + \\
&\ & 1440\,\xi _1\,{\xi _3}^2\,\xi _4\,\xi _5 - 960\,\xi _3\,{\xi _4}^2\,\xi _5 + 576\,{\xi _1}^3\,\xi _3\,{\xi _5}^2 -
     576\,\xi _1\,\xi _2\,\xi _3\,{\xi _5}^2 + 432\,\xi _1\,{\xi _3}^2\,{\xi _5}^2 -\\
&\ & 4800\,\xi _3\,\xi _4\,{\xi _5}^2 - 576\,\xi _3\,{\xi _5}^3 + 96\,{\xi _1}^3\,{\xi _3}^2\,\xi _6 +
     24\,\xi _1\,{\xi _3}^3\,\xi _6 + 384\,{\xi _1}^2\,\xi _3\,\xi _4\,\xi _6 - 1152\,\xi _2\,\xi _3\,\xi _4\,\xi _6 - \\
&\ & 192\,{\xi _3}^2\,\xi _4\,\xi _6 - 384\,{\xi _1}^2\,\xi _3\,\xi _5\,\xi _6 - 96\,{\xi _3}^2\,\xi _5\,\xi _6 -
     384\,\xi _1\,\xi _3\,{\xi _6}^2 + 1536\,\xi _4\,{\xi _6}^2 + 1536\,\xi _5\,{\xi _6}^2 -\\
&\ & 24\,\theta _1\,\left( 8\,{\xi _1}^3\,{\xi _3}^2 - 12\,\xi _1\,\xi _2\,{\xi _3}^2 - \xi _1\,{\xi _3}^3 -
     32\,{\xi _1}^2\,\xi _3\,\xi _5 + 48\,\xi _2\,\xi _3\,\xi _5 + 4\,{\xi _3}^2\,\xi _5 + 16\,\xi _1\,\xi _3\,\xi _6 +  \right.\\
&\ &\left.\qquad  64\,\xi _4\,\xi _6 - 64\,\xi _5\,\xi _6 \right) +\\
&\ & 8\,\theta _2\,\left( 36\,{\xi _1}^6 - 144\,{\xi _1}^4\,\xi _2 + 180\,{\xi _1}^2\,{\xi _2}^2 - 72\,{\xi _2}^3 -
     36\,{\xi _1}^4\,\xi _3 + 108\,{\xi _1}^2\,\xi _2\,\xi _3 - 72\,{\xi _2}^2\,\xi _3 - \right.\\
&\ & \qquad 3\,{\xi _1}^2\,{\xi _3}^2 - 18\,\xi _2\,{\xi _3}^2 - 80\,{\xi _1}^3\,\xi _4 + 144\,\xi _1\,\xi _2\,\xi _4 + 72\,\xi _1\,\xi _3\,\xi _4 +
     48\,{\xi _4}^2 + 144\,{\xi _1}^3\,\xi _5 - \\
&\ & \qquad 144\,\xi _1\,\xi _2\,\xi _5 + 168\,\xi _1\,\xi _3\,\xi _5 - \left. 480\,\xi _4\,\xi _5 - 336\,{\xi _5}^2 -
     288\,{\xi _1}^2\,\xi _6 + 288\,\xi _2\,\xi _6 \right) + \\
&\ & \left. 16\,\theta _3\,\left( \xi _1\,{\xi _3}^2 - 4\,\xi _3\,\xi _4 + 96\,{\xi _1}^2\,\xi _5 - 144\,\xi _2\,\xi _5 -
    16\,\xi _3\,\xi _5 + 24\,\xi _1\,\xi _6 \right) - 384\,\theta _4\,\xi _1 \right];
\end{array}\label{B4}
\end{equation}

\begin{equation}
\begin{array}{rcl}
{\theta _3}^2 &=&
   \frac{1}{256}\left[-144\,{\xi _1}^8\,{\xi _3}^2 + 144\,{\xi _1}^6\,\xi _2\,{\xi _3}^2 + 1008\,{\xi _1}^4\,{\xi _2}^2\,{\xi _3}^2 -
       1872\,{\xi _1}^2\,{\xi _2}^3\,{\xi _3}^2 + 864\,{\xi _2}^4\,{\xi _3}^2 + \right.\\
&\ &   108\,{\xi _1}^6\,{\xi _3}^3 + 144\,{\xi _1}^4\,\xi _2\,{\xi _3}^3 -
       1188\,{\xi _1}^2\,{\xi _2}^2\,{\xi _3}^3 + 936\,{\xi _2}^3\,{\xi _3}^3 - 144\,{\xi _1}^2\,\xi _2\,{\xi _3}^4 +
       288\,{\xi _2}^2\,{\xi _3}^4 - \\
&\ &   9\,{\xi _1}^2\,{\xi _3}^5 + 18\,\xi _2\,{\xi _3}^5 + 1728\,{\xi _1}^7\,\xi _3\,\xi _4 -
       6912\,{\xi _1}^5\,\xi _2\,\xi _3\,\xi _4 + 8640\,{\xi _1}^3\,{\xi _2}^2\,\xi _3\,\xi _4 - \\
&\ &   3456\,\xi _1\,{\xi _2}^3\,\xi _3\,\xi _4 - 1408\,{\xi _1}^5\,{\xi _3}^2\,\xi _4 +
       5568\,{\xi _1}^3\,\xi _2\,{\xi _3}^2\,\xi _4 - 5184\,\xi _1\,{\xi _2}^2\,{\xi _3}^2\,\xi _4 +
       416\,{\xi _1}^3\,{\xi _3}^3\,\xi _4 - \\
&\ &   1296\,\xi _1\,\xi _2\,{\xi _3}^3\,\xi _4 - 24\,\xi _1\,{\xi _3}^4\,\xi _4 - 3840\,{\xi _1}^4\,\xi _3\,{\xi _4}^2 +
       6912\,{\xi _1}^2\,\xi _2\,\xi _3\,{\xi _4}^2 + 960\,{\xi _1}^2\,{\xi _3}^2\,{\xi _4}^2 -\\
&\ &   576\,\xi _2\,{\xi _3}^2\,{\xi _4}^2 - 48\,{\xi _3}^3\,{\xi _4}^2 +  2304\,\xi _1\,\xi _3\,{\xi _4}^3 +
       1728\,{\xi _1}^7\,\xi _3\,\xi _5 - 6912\,{\xi _1}^5\,\xi _2\,\xi _3\,\xi _5 + \\
&\ &   8640\,{\xi _1}^3\,{\xi _2}^2\,\xi _3\,\xi _5 - 3456\,\xi _1\,{\xi _2}^3\,\xi _3\,\xi _5 -
       2304\,{\xi _1}^5\,{\xi _3}^2\,\xi _5 + 4032\,{\xi _1}^3\,\xi _2\,{\xi _3}^2\,\xi _5 - \\
&\ &   1728\,\xi _1\,{\xi _2}^2\,{\xi _3}^2\,\xi _5 - 1584\,\xi _1\,\xi _2\,{\xi _3}^3\,\xi _5 -
       72\,\xi _1\,{\xi _3}^4\,\xi _5 + 3072\,{\xi _1}^4\,\xi _3\,\xi _4\,\xi _5 +
       14976\,{\xi _1}^2\,{\xi _3}^2\,\xi _4\,\xi _5 +\\
&\ &   3456\,\xi _2\,{\xi _3}^2\,\xi _4\,\xi _5 + 288\,{\xi _3}^3\,\xi _4\,\xi _5 -
       11520\,\xi _1\,\xi _3\,{\xi _4}^2\,\xi _5 + 6912\,{\xi _1}^4\,\xi _3\,{\xi _5}^2 -
       6912\,{\xi _1}^2\,\xi _2\,\xi _3\,{\xi _5}^2 + \\
&\ &   4032\,{\xi _1}^2\,{\xi _3}^2\,{\xi _5}^2 + 1728\,\xi _2\,{\xi _3}^2\,{\xi _5}^2 + 144\,{\xi _3}^3\,{\xi _5}^2 -
       57600\,\xi _1\,\xi _3\,\xi _4\,{\xi _5}^2 - 6912\,\xi _1\,\xi _3\,{\xi _5}^3 +\\
&\ &   864\,{\xi _1}^6\,\xi _3\,\xi _6 - 3456\,{\xi _1}^4\,\xi _2\,\xi _3\,\xi _6 +
       4320\,{\xi _1}^2\,{\xi _2}^2\,\xi _3\,\xi _6 - 1728\,{\xi _2}^3\,\xi _3\,\xi _6 -
       480\,{\xi _1}^4\,{\xi _3}^2\,\xi _6 + \\
&\ &   3744\,{\xi _1}^2\,\xi _2\,{\xi _3}^2\,\xi _6 - 1728\,{\xi _2}^2\,{\xi _3}^2\,\xi _6 +
       408\,{\xi _1}^2\,{\xi _3}^3\,\xi _6 - 144\,\xi _2\,{\xi _3}^3\,\xi _6 + 24\,{\xi _3}^4\,\xi _6 +\\
&\ &   2688\,{\xi _1}^3\,\xi _3\,\xi _4\,\xi _6 - 10368\,\xi _1\,\xi _2\,\xi _3\,\xi _4\,\xi _6 -
       1344\,\xi _1\,{\xi _3}^2\,\xi _4\,\xi _6 - 384\,\xi _3\,{\xi _4}^2\,\xi _6 -
       1152\,{\xi _1}^3\,\xi _3\,\xi _5\,\xi _6 - \\
&\ &   3456\,\xi _1\,\xi _2\,\xi _3\,\xi _5\,\xi _6 + 576\,\xi _1\,{\xi _3}^2\,\xi _5\,\xi _6 -
       14592\,\xi _3\,\xi _4\,\xi _5\,\xi _6 - 3456\,\xi _3\,{\xi _5}^2\,\xi _6 - 5376\,{\xi _1}^2\,\xi _3\,{\xi _6}^2 -\\
&\ &   2304\,\xi _2\,\xi _3\,{\xi _6}^2 - 768\,{\xi _3}^2\,{\xi _6}^2 + 18432\,\xi _1\,\xi _4\,{\xi _6}^2 +
       18432\,\xi _1\,\xi _5\,{\xi _6}^2 + 6144\,{\xi _6}^3 +\\
&\ &   8\,\theta _1\left( 36\,{\xi _1}^6\,\xi _3 - 144\,{\xi _1}^4\,\xi _2\,\xi _3 + 180\,{\xi _1}^2\,{\xi _2}^2\,\xi _3 -
       72\,{\xi _2}^3\,\xi _3 - 132\,{\xi _1}^4\,{\xi _3}^2 - 36\,{\xi _1}^2\,\xi _2\,{\xi _3}^2 + \right.\\
&\ &   \qquad 360\,{\xi _2}^2\,{\xi _3}^2 - 3\,{\xi _1}^2\,{\xi _3}^3 + 54\,\xi _2\,{\xi _3}^3 + 3\,{\xi _3}^4 -
       80\,{\xi _1}^3\,\xi _3\,\xi _4 + 144\,\xi _1\,\xi _2\,\xi _3\,\xi _4 - 24\,\xi _1\,{\xi _3}^2\,\xi _4 + \\
&\ &   \qquad 240\,\xi _3\,{\xi _4}^2 + 1296\,{\xi _1}^3\,\xi _3\,\xi _5 - 1872\,\xi _1\,\xi _2\,\xi _3\,\xi _5 -
       72\,\xi _1\,{\xi _3}^2\,\xi _5 - 864\,\xi _3\,\xi _4\,\xi _5 - 144\,\xi _3\,{\xi _5}^2 - \\
&\ &   \qquad 96\,{\xi _1}^2\,\xi _3\,\xi _6 - 864\,\xi _2\,\xi _3\,\xi _6 - 96\,{\xi _3}^2\,\xi _6 -
       2304\,\xi _1\,\xi _4\,\xi _6 +  2304\,\xi _1\,\xi _5\,\xi _6 + \left. 768\,{\xi _6}^2 \right) +\\
&\ &   96\,\theta _2\,\left( 36\,{\xi _1}^7 - 144\,{\xi _1}^5\,\xi _2 + 180\,{\xi _1}^3\,{\xi _2}^2 - 72\,\xi _1\,{\xi _2}^3 -
       36\,{\xi _1}^5\,\xi _3 + 108\,{\xi _1}^3\,\xi _2\,\xi _3 - 72\,\xi _1\,{\xi _2}^2\,\xi _3 +  \right.\\
&\ &   \qquad 5\,{\xi _1}^3\,{\xi _3}^2 - 30\,\xi _1\,\xi _2\,{\xi _3}^2 - \xi _1\,{\xi _3}^3 - 80\,{\xi _1}^4\,\xi _4 +
       144\,{\xi _1}^2\,\xi _2\,\xi _4 + 72\,{\xi _1}^2\,\xi _3\,\xi _4 + 48\,\xi _1\,{\xi _4}^2 + \\
&\ &   \qquad 144\,{\xi _1}^4\,\xi _5 - 144\,{\xi _1}^2\,\xi _2\,\xi _5 + 136\,{\xi _1}^2\,\xi _3\,\xi _5 +
       48\,\xi _2\,\xi _3\,\xi _5 + 4\,{\xi _3}^2\,\xi _5 - 480\,\xi _1\,\xi _4\,\xi _5 -\\
&\ &   \qquad  336\,\xi _1\,{\xi _5}^2 - 288\,{\xi _1}^3\,\xi _6 + 288\,\xi _1\,\xi _2\,\xi _6 +
       \left. 16\,\xi _1\,\xi _3\,\xi _6 + 64\,\xi _4\,\xi _6 - 64\,\xi _5\,\xi _6 \right) +\\
&\ &   16\,\theta _3\,\left( 36\,{\xi _1}^6 - 144\,{\xi _1}^4\,\xi _2 + 180\,{\xi _1}^2\,{\xi _2}^2 - 72\,{\xi _2}^3 -
       36\,{\xi _1}^4\,\xi _3 + 108\,{\xi _1}^2\,\xi _2\,\xi _3 - 72\,{\xi _2}^2\,\xi _3 + \right.\\
&\ &   \qquad 13\,{\xi _1}^2\,{\xi _3}^2 - 6\,\xi _2\,{\xi _3}^2 + {\xi _3}^3 - 80\,{\xi _1}^3\,\xi _4 +
       144\,\xi _1\,\xi _2\,\xi _4 - 24\,\xi _1\,\xi _3\,\xi _4 + 48\,{\xi _4}^2 +  \\
&\ &   \qquad 1296\,{\xi _1}^3\,\xi _5 - 1872\,\xi _1\,\xi _2\,\xi _5 - 72\,\xi _1\,\xi _3\,\xi _5 - 480\,\xi _4\,\xi _5 -
       \left.  336\,{\xi _5}^2 + 192\,{\xi _1}^2\,\xi _6 - 48\,\xi _3\,\xi _6 \right) - \\
&\ &   \left. 512\,\theta _4\,\left( 3\,{\xi _1}^2 + 9\,\xi _2 + \xi _3 \right)  \right];
\end{array}\label{B5}
\end{equation}

\begin{equation}
\begin{array}{rcl}
\theta_4&=&\theta_1\, \theta_3.
\end{array}\label{B6}
\end{equation}

\subsection{Explicit form of the syzygy for the the rotation subgroup of $G$}

The invariants $\lambda_1,\dots ,\lambda_8$ of the subgroup {\bf SO}$_3$ of $G$, defined in (\ref{IBSO3}), satisfy the
following syzygy:

\begin{equation}
\begin{array}{rcl}
\lambda_8^3 &=& \frac 1{72} \left(36\,{\lambda_1}^6 - 36\,{\lambda_1}^4\,{\lambda_2}^2 +
      9\,{\lambda_1}^2\,{\lambda_2}^4 - 36\,{\lambda_1}^4\,{\lambda_3}^2 + 18\,{\lambda_1}^2\,{\lambda_2}^2\,{\lambda_3}^2 +
      9\,{\lambda_1}^2\,{\lambda_3}^4 - \right.\\
&\ &  80\,{\lambda_1}^3\,{\lambda_4}^2 +  24\,\lambda_1\,{\lambda_2}^2\,{\lambda_4}^2 - 16\,{\lambda_2}^3\,{\lambda_4}^2 +
     24\,\lambda_1\,{\lambda_3}^2\,{\lambda_4}^2 + 48\,\lambda_2\,{\lambda_3}^2\,{\lambda_4}^2 + 48\,{\lambda_4}^4 - \\
&\ & 96\,{\lambda_2}^2\,\lambda_3\,\lambda_4\,\lambda_5 + 32\,{\lambda_3}^3\,\lambda_4\,\lambda_5 -
     80\,{\lambda_1}^3\,{\lambda_5}^2 + 24\,\lambda_1\,{\lambda_2}^2\,{\lambda_5}^2 + 16\,{\lambda_2}^3\,{\lambda_5}^2 +
     24\,\lambda_1\,{\lambda_3}^2\,{\lambda_5}^2 - \\
&\ & 48\,\lambda_2\,{\lambda_3}^2\,{\lambda_5}^2 + 96\,{\lambda_4}^2\,{\lambda_5}^2 + 48\,{\lambda_5}^4 +
     96\,{\lambda_1}^2\,\lambda_2\,\lambda_4\,\lambda_6 + 96\,\lambda_1\,{\lambda_2}^2\,\lambda_4\,\lambda_6 -
     48\,{\lambda_2}^3\,\lambda_4\,\lambda_6 - \\
&\ & 96\,\lambda_1\,{\lambda_3}^2\,\lambda_4\,\lambda_6 - 48\,\lambda_2\,{\lambda_3}^2\,\lambda_4\,\lambda_6 +
     96\,{\lambda_1}^2\,\lambda_3\,\lambda_5\,\lambda_6 + 192\,\lambda_1\,\lambda_2\,\lambda_3\,\lambda_5\,\lambda_6 -
     48\,{\lambda_2}^2\,\lambda_3\,\lambda_5\,\lambda_6 - \\
&\ & 48\,{\lambda_3}^3\,\lambda_5\,\lambda_6 + 144\,{\lambda_1}^3\,{\lambda_6}^2 -
     288\,{\lambda_1}^2\,\lambda_2\,{\lambda_6}^2 + 72\,\lambda_1\,{\lambda_2}^2\,{\lambda_6}^2 +
     72\,\lambda_1\,{\lambda_3}^2\,{\lambda_6}^2 - \\
&\ & 288\,{\lambda_4}^2\,{\lambda_6}^2 - 288\,{\lambda_5}^2\,{\lambda_6}^2 + 384\,\lambda_4\,{\lambda_6}^3 -
     144\,{\lambda_6}^4 - 96\,{\lambda_1}^2\,\lambda_3\,\lambda_4\,\lambda_7 +
     192\,\lambda_1\,\lambda_2\,\lambda_3\,\lambda_4\,\lambda_7 + \\
&\ & 48\,{\lambda_2}^2\,\lambda_3\,\lambda_4\,\lambda_7 + 48\,{\lambda_3}^3\,\lambda_4\,\lambda_7 +
     96\,{\lambda_1}^2\,\lambda_2\,\lambda_5\,\lambda_7 - 96\,\lambda_1\,{\lambda_2}^2\,\lambda_5\,\lambda_7 -
     48\,{\lambda_2}^3\,\lambda_5\,\lambda_7 + \\
&\ & 96\,\lambda_1\,{\lambda_3}^2\,\lambda_5\,\lambda_7 - 48\,\lambda_2\,{\lambda_3}^2\,\lambda_5\,\lambda_7 -
     576\,{\lambda_1}^2\,\lambda_3\,\lambda_6\,\lambda_7 + 1152\,\lambda_5\,{\lambda_6}^2\,\lambda_7 +
     144\,{\lambda_1}^3\,{\lambda_7}^2 +\\
&\ & 288\,{\lambda_1}^2\,\lambda_2\,{\lambda_7}^2 + 72\,\lambda_1\,{\lambda_2}^2\,{\lambda_7}^2 +
     72\,\lambda_1\,{\lambda_3}^2\,{\lambda_7}^2 - 288\,{\lambda_4}^2\,{\lambda_7}^2 -
     288\,{\lambda_5}^2\,{\lambda_7}^2 - \\
&\ & 1152\,\lambda_4\,\lambda_6\,{\lambda_7}^2 - 288\,{\lambda_6}^2\,{\lambda_7}^2 - 384\,\lambda_5\,{\lambda_7}^3 -
     144\,{\lambda_7}^4 - 144\,{\lambda_1}^4\,\lambda_8 + 108\,{\lambda_1}^2\,{\lambda_2}^2\,\lambda_8 - \\
&\ & 18\,{\lambda_2}^4\,\lambda_8 + 108\,{\lambda_1}^2\,{\lambda_3}^2\,\lambda_8 -
     36\,{\lambda_2}^2\,{\lambda_3}^2\,\lambda_8 - 18\,{\lambda_3}^4\,\lambda_8 + 144\,\lambda_1\,{\lambda_4}^2\,\lambda_8 +\\
&\ & 144\,\lambda_1\,{\lambda_5}^2\,\lambda_8 - 288\,\lambda_2\,\lambda_4\,\lambda_6\,\lambda_8 -
     288\,\lambda_3\,\lambda_5\,\lambda_6\,\lambda_8 - 144\,\lambda_1\,{\lambda_6}^2\,\lambda_8 +
     288\,\lambda_2\,{\lambda_6}^2\,\lambda_8 + \\
&\ & 288\,\lambda_3\,\lambda_4\,\lambda_7\,\lambda_8 - 288\,\lambda_2\,\lambda_5\,\lambda_7\,\lambda_8 +
     576\,\lambda_3\,\lambda_6\,\lambda_7\,\lambda_8 - 144\,\lambda_1\,{\lambda_7}^2\,\lambda_8 -
     288\,\lambda_2\,{\lambda_7}^2\,\lambda_8 + \\
&\ & \left. 180\,{\lambda_1}^2\,{\lambda_8}^2 - 72\,{\lambda_2}^2\,{\lambda_8}^2 - 72\,{\lambda_3}^2\,{\lambda_8}^2\right).
\label{sySO3}
\end{array}
\end{equation}


\section{Explicit form of the $\widehat P$-matrix for the group $G$}

For

\begin{equation}
(d_1,\dots ,d_9) = (2,\, 4,\, 4,\, 6,\, 6,\, 8,\, 8,\, 10,\, 12),
\end{equation}
the explicit form of the $\widehat P$-matrix related to the integrity basis $\{p\}$ of the group $G$ is the following

\begin{eqnarray*}
P_{1, i} &=& 2\,d_i\,p_i,\qquad i=1,\dots ,9,\\
P_{2, 2} &=&  -\frac{8}{3}\,\left( 3\,{p_1}^3 - 9\,p_1\,p_2 - 2\,p_4 + 2\,p_5 \right) ,\\
P_{2, 3} &=& 8\,p_1\,p_3,\\
P_{2, 4} &=&  4\,\left( 2\,p_1\,p_4 + p_6 \right),\\
P_{2, 5} &=& \frac{4}{3}\,\left(14\,p_1\,p_5 - 3\,p_6 - 2\,p_7 \right) ,\\
P_{2, 6} &=&  -\frac{2}{3}\,\left( 3\,p_3\,p_4 - 3\,p_3\,p_5 - 26\,p_1\,p_6 + 2\,p_8 \right) ,\\
P_{2, 7} &=& - \frac{4}{3}\,\left( 2\,p_3\,p_5 - 17\,p_1\,p_7 + 3\,p_8 \right) ,\\
P_{2, 8} &=&  -\frac{2}{3}\,\left( 2\,p_3\,p_6 - 3\,p_3\,p_7 - 32\,p_1\,p_8 + 3\,p_9 \right) ,\\
P_{2, 9} &=&  4\,\left( p_3\,p_8 + 5\,p_1\,p_9 \right),\\
P_{3, 3} &=& 16\,p_1\,p_3,\\
P_{3, 4} &=& 24\,p_6,\\
P_{3, 5} &=&  8\,\left( p_6 + 2\,p_7 \right),\\
P_{3, 6} &=&  4\,\left( p_3\,p_4 + 3\,p_3\,p_5 + 2\,p_1\,p_6 + 2\,p_8 \right),\\
P_{3, 7} &=&  8\,\left( 2\,p_3\,p_5 + p_1\,p_7 + p_8 \right),\\
P_{3, 8} &=&  4\,\left( 2\,p_3\,p_6 + 3\,p_3\,p_7 + 4\,p_1\,p_8 + p_9 \right),\\
P_{3, 9} &=&  24\,\left( p_3\,p_8 + p_1\,p_9 \right),\\
P_{4, 4} &=&  3\,\left( 6\,{p_1}^2 - 6\,p_2 - p_3 \right)\,p_4,\\
P_{4, 5} &=&  2\,\left( 2\,p_1\,p_6 + p_8 \right),\\
P_{4, 6} &=& \frac{1}{2}\left(4\,p_1\,p_3\,p_4 + 24\,p_4\,p_5 + 18\,{p_1}^2\,p_6 - 18\,p_2\,p_6 - 3\,p_3\,p_6 +
              2\,p_9\right),\\
P_{4, 7} &=& \frac 1{32} \left(-36\,{p_1}^6 + 144\,{p_1}^4\,p_2 - 180\,{p_1}^2\,{p_2}^2 + 72\,{p_2}^3 +
         36\,{p_1}^4\,p_3 - 108\,{p_1}^2\,p_2\,p_3 + 72\,{p_2}^2\,p_3 - \right. \\
 &\  &   9\,{p_1}^2\,{p_3}^2 + 18\,p_2\,{p_3}^2 + 80\,{p_1}^3\,p_4 - 144\,p_1\,p_2\,p_4 -
         24\,p_1\,p_3\,p_4 - 48\,{p_4}^2 - 144\,{p_1}^3\,p_5 + \\
 &\  &   144\,p_1\,p_2\,p_5 - 72\,p_1\,p_3\,p_5 + 288\,p_4\,p_5 + 144\,{p_5}^2 - 96\,{p_1}^2\,p_6 + 288\,p_2\,p_6 +
         112\,p_3\,p_6 +\\
 &\  &   \left.  288\,{p_1}^2\,p_7 - 288\,p_2\,p_7 + 32\,p_1\,p_8 + 16\,p_9\right),\\
P_{4, 8} &=&  \frac{1}{2}\left(2\,{p_3}^2\,p_4 + 48\,p_4\,p_7 + 18\,{p_1}^2\,p_8 - 18\,p_2\,p_8 - 3\,p_3\,p_8 +
              4\,p_1\,p_9\right),\\
P_{4, 9} &=&  3\,\left( 12\,p_4\,p_8 + 6\,{p_1}^2\,p_9 - 6\,p_2\,p_9 -p_3\,p_9\right),\\
P_{5, 5} &=&  \frac 1{3}\left(2\,{p_1}^2\,p_5 + 6\,p_2\,p_5 + 5\,p_3\,p_5 + 8\,p_1\,p_7\right),\\
P_{5,6} &=&  \frac 1{48}\left(-36\,{p_1}^6 + 144\,{p_1}^4\,p_2 - 180\,{p_1}^2\,{p_2}^2 + 72\,{p_2}^3 + 36\,{p_1}^4\,p_3 -
             108\,{p_1}^2\,p_2\,p_3 + 72\,{p_2}^2\,p_3 - \right. \\
         &\  & 9\,{p_1}^2\,{p_3}^2 + 18\,p_2\,{p_3}^2 + 80\,{p_1}^3\,p_4 - 144\,p_1\,p_2\,p_4 - 24\,p_1\,p_3\,p_4 -
               48\,{p_4}^2 - 144\,{p_1}^3\,p_5 + \\
         &\  & 144\,p_1\,p_2\,p_5 + 24\,p_1\,p_3\,p_5 + 480\,p_4\,p_5 + 144\,{p_5}^2 -
               80\,{p_1}^2\,p_6 + 336\,p_2\,p_6 + 88\,p_3\,p_6 + \\
         &\  & \left.288\,{p_1}^2\,p_7 - 288\,p_2\,p_7 + 48\,p_3\,p_7 - 32\,p_1\,p_8 + 16\,p_9\right),\\
P_{5, 7} &=&  \frac 1{96}\left(-36\,{p_1}^6 + 144\,{p_1}^4\,p_2 - 180\,{p_1}^2\,{p_2}^2 + 72\,{p_2}^3 + 36\,{p_1}^4\,p_3 -
               108\,{p_1}^2\,p_2\,p_3 +  \right.\\
         &\  & 72\,{p_2}^2\,p_3 - 9\,{p_1}^2\,{p_3}^2 + 18\,p_2\,{p_3}^2 + 80\,{p_1}^3\,p_4 - 144\,p_1\,p_2\,p_4 -
               24\,p_1\,p_3\,p_4 - \\
         &\  & 48\,{p_4}^2 - 144\,{p_1}^3\,p_5 + 144\,p_1\,p_2\,p_5 + 184\,p_1\,p_3\,p_5 + 288\,p_4\,p_5 + 912\,{p_5}^2 -
               96\,{p_1}^2\,p_6 +  \\
         &\  & \left. 288\,p_2\,p_6 + 48\,p_3\,p_6 + 352\,{p_1}^2\,p_7 - 96\,p_2\,p_7 + 160\,p_3\,p_7 - 96\,p_1\,p_8 +
               16\,p_9\right),\\
P_{5, 8} &=&  \frac 1{6}\left(6\,{p_3}^2\,p_5 + 8\,p_1\,p_3\,p_6 + 96\,p_5\,p_6 + 12\,p_1\,p_3\,p_7 + 48\,p_4\,p_7 +
              2\,{p_1}^2\,p_8 + 6\,p_2\,p_8 + 5\,p_3\,p_8\right),\\
P_{5, 9} &=&  2\,\left( -12\,p_3\,p_4\,p_5 + {p_3}^2\,p_6 + 24\,{p_6}^2 + 2\,p_1\,p_3\,p_8 +
           6\,p_4\,p_8\right),\\
P_{6, 6} &=& \frac 1{12}\left(2\,{p_1}^2\,p_3\,p_4 + 6\,p_2\,p_3\,p_4 + 5\,{p_3}^2\,p_4 + 54\,{p_1}^2\,p_3\,p_5 -
               54\,p_2\,p_3\,p_5 - 9\,{p_3}^2\,p_5 + 48\,p_1\,p_4\,p_5 + \right.\\
         &\  & 24\,p_1\,p_3\,p_6 + \left. 48\,p_4\,p_6 + 144\,p_5\,p_6 + 96\,p_4\,p_7 + 12\,p_3\,p_8 + 8\,p_1\,p_9\right),\\
P_{6, 7} &=&  \frac 1{96}\left(-36\,{p_1}^7 + 144\,{p_1}^5\,p_2 - 180\,{p_1}^3\,{p_2}^2 + 72\,p_1\,{p_2}^3 +
               36\,{p_1}^5\,p_3 - 108\,{p_1}^3\,p_2\,p_3 +\right.\\
         &\  & 72\,p_1\,{p_2}^2\,p_3 - 9\,{p_1}^3\,{p_3}^2 + 18\,p_1\,p_2\,{p_3}^2 + 80\,{p_1}^4\,p_4 -
               144\,{p_1}^2\,p_2\,p_4 - 24\,{p_1}^2\,p_3\,p_4 - 48\,p_1\,{p_4}^2 - \\
         &\  & 144\,{p_1}^4\,p_5 + 144\,{p_1}^2\,p_2\,p_5 - 72\,{p_1}^2\,p_3\,p_5 + 96\,{p_3}^2\,p_5 + 288\,p_1\,p_4\,p_5 +
               144\,p_1\,{p_5}^2 - 96\,{p_1}^3\,p_6 + \\
         &\  & 288\,p_1\,p_2\,p_6 + 176\,p_1\,p_3\,p_6 + 1152\,p_5\,p_6 + 288\,{p_1}^3\,p_7 - 288\,p_1\,p_2\,p_7 +
               192\,p_1\,p_3\,p_7 +  \\
         &\  & 576\,p_4\,p_7 + 576\,p_5\,p_7 - \left. 64\,{p_1}^2\,p_8 + 96\,p_2\,p_8 + 80\,p_3\,p_8 + 16\,p_1\,p_9\right),\\
P_{6, 8} &=&   \frac 1{12}\left(8\,p_1\,{p_3}^2\,p_4 + 48\,p_3\,p_4\,p_5 + 12\,{p_3}^2\,p_6 + 96\,{p_6}^2 +
               54\,{p_1}^2\,p_3\,p_7 - 54\,p_2\,p_3\,p_7 - 9\,{p_3}^2\,p_7 + \right.\\
         &\  & 96\,p_1\,p_4\,p_7 + 24\,p_1\,p_3\,p_8 + 72\,p_4\,p_8 + 216\,p_5\,p_8 + 2\,{p_1}^2\,p_9 + \left. 6\,p_2\,p_9 +
               5\,p_3\,p_9\right),\\
P_{6, 9} &=&  \frac 1{2}\left(2\,{p_3}^3\,p_4 + 24\,p_3\,p_4\,p_6 + 18\,{p_1}^2\,p_3\,p_8 - 18\,p_2\,p_3\,p_8 -
              3\,{p_3}^2\,p_8 + 24\,p_1\,p_4\,p_8 + 4\,p_1\,p_3\,p_9 +\right.\\
         &\  &\left. 12\,p_4\,p_9 + 60\,p_5\,p_9 \right),\\
P_{7, 7} &=&  \frac 1{3}\left(2\,{p_1}^2\,p_3\,p_5 + 6\,p_2\,p_3\,p_5 + 5\,{p_3}^2\,p_5 + 12\,p_1\,{p_5}^2 + 24\,p_5\,p_6 +
              8\,p_1\,p_3\,p_7 + 48\,p_5\,p_7\right),\\
P_{7, 8} &=&  \frac {1}{6}\left(12\,p_1\,{p_3}^2\,p_5 + 36\,p_3\,p_4\,p_5 - 36\,p_3\,{p_5}^2 + 2\,{p_1}^2\,p_3\,p_6 +
              6\,p_2\,p_3\,p_6 + 5\,{p_3}^2\,p_6 + 48\,p_1\,p_5\,p_6 + \right.\\
         &\  &\left. 24\,{p_6}^2 + 6\,{p_3}^2\,p_7 + 72\,{p_7}^2 + 8\,p_1\,p_3\,p_8 + 120\,p_5\,p_8\right),\\
P_{7, 9} &=&  -12\,p_1\,p_3\,p_4\,p_5 + 4\,p_1\,{p_3}^2\,p_6 + 12\,p_3\,p_4\,p_6 - 12\,p_3\,p_5\,p_6 + 24\,p_1\,{p_6}^2 +
               2\,{p_3}^2\,p_8 + 24\,p_7\,p_8 + \\
         &\  & 24\,p_5\,p_9,\\
P_{8, 8} &=&  \frac{1}{12}\left(2\,{p_1}^2\,{p_3}^2\,p_4 + 6\,p_2\,{p_3}^2\,p_4 + 5\,{p_3}^3\,p_4 +
              54\,{p_1}^2\,{p_3}^2\,p_5 - 54\,p_2\,{p_3}^2\,p_5 - 9\,{p_3}^3\,p_5 + \right.\\
         &\  & 192\,p_1\,p_3\,p_4\,p_5 + 24\,p_1\,{p_3}^2\,p_6 + 96\,p_3\,p_4\,p_6 - 288\,p_3\,p_5\,p_6 + 12\,{p_3}^2\,p_8 +
               576\,p_7\,p_8 + \\
         &\  & \left. 8\,p_1\,p_3\,p_9 + 192\,p_5\,p_9 \right),\\
P_{8, 9} &=&  \frac{1}{2}\left(4\,p_1\,{p_3}^3\,p_4 + 12\,{p_3}^2\,{p_4}^2 + 18\,{p_1}^2\,{p_3}^2\,p_6 -
               18\,p_2\,{p_3}^2\,p_6 - 3\,{p_3}^3\,p_6 + 48\,p_1\,p_3\,p_4\,p_6 -\right.\\
         &\  & 84\,p_3\,{p_6}^2 - 24\,p_3\,p_4\,p_8 + 84\,{p_8}^2 + 2\,{p_3}^2\,p_9 + \left. 48\,p_6\,p_9 +
               84\,p_7\,p_9 \right),\\
P_{9, 9} &=&  3\,\left( 6\,{p_1}^2\,{p_3}^3\,p_4 - 6\,p_2\,{p_3}^3\,p_4 - {p_3}^4\,p_4 + 12\,p_1\,{p_3}^2\,{p_4}^2 +
               288\,p_1\,{p_3}^2\,p_4\,p_5 - 1152\,p_3\,p_4\,{p_5}^2 - \right.\\
         &\  & 24\,{p_3}^2\,p_4\,p_6 - 288\,p_1\,p_3\,{p_6}^2 + 1152\,p_5\,{p_6}^2 + 144\,{p_1}^2\,p_3\,p_4\,p_7 -
               432\,p_2\,p_3\,p_4\,p_7 - 72\,{p_3}^2\,p_4\,p_7 +\\
         &\  & 1152\,p_4\,{p_7}^2 + 108\,{p_1}^6\,p_8 - 432\,{p_1}^4\,p_2\,p_8 + 540\,{p_1}^2\,{p_2}^2\,p_8 -
               216\,{p_2}^3\,p_8 - 108\,{p_1}^4\,p_3\,p_8 + \\
         &\  & 324\,{p_1}^2\,p_2\,p_3\,p_8 - 216\,{p_2}^2\,p_3\,p_8 + 27\,{p_1}^2\,{p_3}^2\,p_8 - 54\,p_2\,{p_3}^2\,p_8 -
               240\,{p_1}^3\,p_4\,p_8 + \\
         &\  & 432\,p_1\,p_2\,p_4\,p_8 + 72\,p_1\,p_3\,p_4\,p_8 + 144\,{p_4}^2\,p_8 + 432\,{p_1}^3\,p_5\,p_8 -
               432\,p_1\,p_2\,p_5\,p_8 + \\
         &\  & 216\,p_1\,p_3\,p_5\,p_8 - 864\,p_4\,p_5\,p_8 - 432\,{p_5}^2\,p_8 - 864\,{p_1}^2\,p_7\,p_8 +
               864\,p_2\,p_7\,p_8 + 144\,{p_1}^2\,p_5\,p_9 - \\
         &\  & 432\,p_2\,p_5\,p_9 - \left. 72\,p_3\,p_5\,p_9 + 288\,p_1\,p_7\,p_9 \right).
\end{eqnarray*}

\section{Explicit form of the generators of the isotropy subgroups of {\bf SO}$_3$}

In this Appendix are collected the definitions of the groups and of the elements and the generators of the subgroups of
{\bf SO}$_3$, involved in the body of the paper, in the tables and in the figures.

\subsubsection{Generators or group elements}

\begin{equation}
\begin{array}{l}
\begin{array}{lll}
C_{2x}=\left(\begin{array}{ccc}1 & 0 & 0 \\ 0 & -1 & 0 \\ 0 & 0 & -1\end{array}\right),\qquad
&
C_{2z}=\left(\begin{array}{ccc}-1 & 0 & 0 \\ 0 & -1 & 0 \\ 0 & 0 & 1\end{array}\right),\qquad
&
C_{3z}=\left(\begin{array}{ccc}-\frac 12 &  {\frac {\sqrt 3}2} & 0 \\ -{\frac {\sqrt 3}2} & -\frac 12 & 0 \\ 0 & 0 &1
\end{array}\right),
\end{array}  \\
\\
\begin{array}{lll}
C_{4z}=\left(\begin{array}{ccc}0 & 1 & 0 \\ -1 & 0 & 0 \\ 0 & 0 & 1\end{array}\right),\qquad
&
C_{4x}=\left(\begin{array}{ccc}1 & 0 & 0 \\ 0 & 0 & 1 \\ 0 & -1 & 0\end{array}\right),\qquad
&
C_{2a}=\left(\begin{array}{ccc}0 & 1 & 0 \\ 1 & 0 & 0 \\ 0 & 0 & -1\end{array}\right),
\end{array}  \\
\\
\begin{array}{lll}
C_{3\delta}=\left(\begin{array}{ccc}0 & 1 & 0 \\ 0 & 0 & 1 \\ 1 & 0 & 0\end{array}\right),\
&
{\rm R}_z(\phi)=\left(\begin{array}{ccc}\cos(\phi)  & \sin(\phi) & 0 \\
                                        -\sin(\phi) & \cos(\phi) & 0 \\
                                        0           & 0          & 1  \end{array}\right),\
&
{\rm R}_x(\phi)=\left(\begin{array}{ccc}  1  &     0       &      0     \\
                                          0  & \cos(\phi)  & \sin(\phi) \\
                                          0  & -\sin(\phi) & \cos(\phi)  \end{array}\right).
                                          \end{array}
\end{array}
\end{equation}

\subsubsection{Groups}

By ${\rm\bf SO}_3$ and ${\rm\bf O}_3$ we denote the proper and, respectively, the complete 3\,D rotation
group and, by ${\rm\bf O}_2^\alpha$, $\alpha=x,z$ the following ${\rm\bf O}_2$ subgroups of
${\rm\bf SO}_3$:
${\rm\bf O}_2^x = \{{\rm R}_x(\phi)\}_{0\le\phi < 2\pi}\cup \{C_{2z}\,{\rm R}_x(\phi)\}_{0\le\phi < 2\pi}$ and
${\rm\bf O}_2^z = \{{\rm R}_z(\phi)\}_{0\le\phi < 2\pi}\cup \{C_{2x}\,{\rm R}_z(\phi)\}_{0\le\phi < 2\pi}.$

\begin{table}
\caption{Relations defining 1\,D geometrical strata, $\widehat
S^{(1,A)} = \widehat W^{(1,A)}$  in orbit space. For $2\leq i\leq
9$, $\tilde p_i = p_i/(p_1^{d_i/2})$, $d_i$ denotes the degree of
the polynomial $p_i(x)$ in the order parameter $x$ and the
values, $p^{(1,A)}$, taken on by $\tilde p$ on $\widehat
W^{(1,A)}$ are listed in the columns from 2 to 9. \label{I}}

\begin{tabular}{ccccccccc}
$A$ & $\tilde p_2$ & $\tilde p_3$ & $\tilde p_4$ & $\tilde p_5$ & $\tilde p_6$ & $\tilde p_7$ & $\tilde p_8$ & $\tilde p_9$ \\ \hline
1   &  1           &    0         &     0        &   0          &   0          &   0          &   0          &      0       \\
2   & $1/2$        &    1         &   $1/6$      & $1/6$        & $1/6$        & $1/6$        & $1/6$        &    $1/6$     \\
3   & $1/2$        &    1         &     0        &   0          &   0          &   0          &   0          &    0         \\
4   & $1/3$        &    0         &   $1/3$      &   0          &   0          &   0          &   0          &    0         \\
5   & $1/2$        &    0         &     0        &   0          &   0          &   0          &   0          &    0         \\
\end{tabular}
\end{table}

\begin{table}
 \caption{Parametric equations defining the geometrical strata, $\widehat W^{(2,A)}$ ($A$ is listed in the head row)
 in  orbit space.
 For $2\leq i \leq 9$, $\tilde p_i=p_i/(p_1^{d_i/2})$ and $d_i$ denotes the degree of the
 polynomial $p_i(x)$, in the order parameter $x$, and $\epsilon=\pm 1$. \label{II}}

\begin{tabular}{cccccc}
$\tilde p\backslash A$ & $1_\epsilon$                       & 2
& $3_\epsilon$                              & 4             & 5
\\ \hline $\tilde p_2$           & $(2+\xi^2)/6$
& $1/2$                        & $(2+\xi^2)/6$
& $1/2$         & $\xi$          \\ $\tilde p_3$           & 0
& $\xi$                        & $\xi^2$
&  1            & $2-2\,\xi$     \\ $\tilde p_4$           &
$(2-\xi)^2(1+\xi)/12$              & 0
& $(2-\xi)^2(1+\xi)/12$                     & $\xi$         & 0
\\ $\tilde p_5$           & 0                                  & 0
& $\xi^2(1+\xi)/12$                         & $\xi$         & 0
\\ $\tilde p_6$           & 0                                  & 0
& $(2-\xi)(1+\xi)\,\xi^2/12$                & $\xi$         & 0
\\ $\tilde p_7$           & 0                                  & 0
& $(1+\xi)\,\xi^3/12$                       & $\xi$         & 0
\\ $\tilde p_8$           & 0                                  & 0
& $(2-\xi)(1+\xi)\,\xi^3/12$                & $\xi$         & 0
\\ $\tilde p_9$           & 0                                  & 0
& $(2-\xi)^2(1+\xi)\,\xi^3/12$              & $\xi$         & 0
\\ $\xi$ range            & $]0,\frac{1+3\,\epsilon}2[$        &
]0,1[                        & $]0,\epsilon [$ & $]0,\frac 16[$&
$]\frac 12,1[$ \\
\end{tabular}
\end{table}

\begin{table}
 \caption{Possible symmetry strata, $S^{(d,r)}$ for D-wave driven pairing in isotropic space ($d$ denotes the dimension
 of the stratum and $r$ is an enumeration index).
 From left to right, the columns refer to the phase reference numbers according to our ($(d,r)$) and ref.[12]
 (${\cal N}$) classification, the order ($|H|$) and the the residual symmetry group, $H$, the
 complex coordinates ($z_j=x_j+i\,x_{5+j}$, $j=1,\dots ,5$) of the $H$-invariant superconducting vector order parameter
 and corresponding values of the partial wave amplitudes $(D_2,\dots,D_{-2})$. In columns five and six, the $t$'s are
 real parameters, while the $v$'s are complex ones.
 The definitions of the group elements are recalled in Appendix D. \label{III}}
\begin{tabular}{cccccc}
$             (d,r)$ & ${\cal N}      $     & $ |H|$                   & $H$                                                                                                       & $\small       (z_1,\dots ,z_5            )$           & $\small \sqrt 2\left(D_2,D_1,D_0,D_{-1},D_{-2}\right)$                     \\ \hline
$\scriptstyle (1,1)$ & $\scriptstyle II  $  & $\scriptstyle \infty$    & $\scriptstyle \left\langle C_{2x}{\cal T}\right\rangle \times\{{\rm R}_z(\phi)\,{\rm U}_1(2\phi) \}_\phi$ & $\scriptstyle (it,-t,0,0,0               )$           & $\scriptstyle (2 t,0,0,0,0                                              )$ \\
$\scriptstyle (1,2)$ & $\scriptstyle VIII$  & $\scriptstyle \infty$    & $\scriptstyle {\rm\bf O}_2^z\otimes\left\langle {\cal T}\right\rangle$                                    & $\scriptstyle (0,0,0,0,t                 )$           & $\scriptstyle (0,0,t\sqrt 2,0,0                                         )$ \\
$\scriptstyle (1,3)$ & $\scriptstyle IX,X$  & $\scriptstyle 16$        & $\scriptstyle \left\langle C_{2x},{\cal T},C_{4z}{\rm U}_1(\pi)\right\rangle$                             & $\scriptstyle (0,t,0,0,0                 )$           & $\scriptstyle (-t,0,0,0,-t                                              )$ \\
$\scriptstyle (1,4)$ & $\scriptstyle XI  $  & $\scriptstyle 24$        & $\scriptstyle \left\langle C_{2x},C_{2a}{\cal T},C_{3\delta}\,{\rm U}_1(4 \pi/3)\right\rangle $           & $\scriptstyle (0,-it,0,0,t               )$           & $\scriptstyle (it,0,t\sqrt 2,0,it                                       )$ \\
$\scriptstyle (1,5)$ & $\scriptstyle I   $  & $\scriptstyle \infty$    & $\scriptstyle \left\langle C_{2x}{\cal T}\right\rangle\times\{{\rm R}_z(\phi)\,{\rm U}_1(- \phi) \}_\phi$ & $\scriptstyle (0,0,t,-it,0               )$           & $\scriptstyle (0,0,0,2\,it,0                                            )$ \\
$\scriptstyle (2,1)$ & $\scriptstyle V   $  & $\scriptstyle 6$         & $\scriptstyle \left\langle C_{2x}{\cal T}, C_{3z}{\rm U}_1(4\pi/3)\right\rangle$                          & $\scriptstyle (it_1,-t_1,t_2,-it_2,0     )$           & $\scriptstyle (2 t_1,0,0,2\,i\,t_2,0                                         )$ \\
$\scriptstyle (2,2)$ & $\scriptstyle IV  $  & $\scriptstyle 4$         & $\scriptstyle \left\langle C_{2x}{\cal T}, C_{2z}{\rm U}_1(\pi)\right\rangle$                             & $\scriptstyle (0,0,t_1,it_2,0            )$           & $\scriptstyle (0,it_1+i\,t_2,0,it_1-it_2                                )$ \\
$\scriptstyle (2,3)$ & $\scriptstyle --  $  & $\scriptstyle 8$         & $\scriptstyle \left\langle C_{2x}, C_{4z}{\cal T}\right\rangle$                                           & $\scriptstyle (0,it_1,0,0,t_2            )$           & $\scriptstyle (-it_1,0,t_2\sqrt 2,0,-it_1                               )$ \\
$\scriptstyle (2,4)$ & $\scriptstyle VII $  & $\scriptstyle 8$         & $\scriptstyle \left\langle C_{2x}, C_{2z}, {\cal T}\right\rangle$                                         & $\scriptstyle (0,t_1,0,0,t_2             )$           & $\scriptstyle (-t_1,0,t_2\sqrt 2,0,-t_1                                 )$ \\
$\scriptstyle (2,5)$ & $\scriptstyle VI  $  & $\scriptstyle 8$         & $\scriptstyle \left\langle C_{2x}{\cal T}, C_{4z}{\rm U}_1(\pi)\right\rangle$                             & $\scriptstyle (it_1,t_2,0,0,0            )$           & $\scriptstyle (t_1-t_2,0,0,0,-t_1-t_2                                   )$ \\
$\scriptstyle (3,1)$ & $\scriptstyle III $  & $\scriptstyle 4$         & $\scriptstyle \left\langle C_{2z}, C_{2x}{\cal T}\right\rangle$                                           & $\scriptstyle (it_1,t_2,0,0,t_3          )$           & $\scriptstyle (t_1-t_2,0,t_3\sqrt 2,0,-t_1-t_2                          )$ \\
$\scriptstyle (3,2)$ & $\scriptstyle --  $  & $\scriptstyle 4$         & $\scriptstyle \left\langle C_{2x}, C_{2z}\right\rangle$                                                   & $\scriptstyle (0,v_1,0,0,v_2             )$           & $\scriptstyle (-v_1,0,v_2\sqrt 2,0,-v_1                                 )$ \\
$\scriptstyle (4,1)$ & $\scriptstyle --  $  & $\scriptstyle 2$         & $\scriptstyle \left\langle C_{2x}{\cal T}\right\rangle$                                                   & $\scriptstyle (it_1,t_2,t_3,it_4,t_5     )$           & $\scriptstyle (t_1-t_2,it_3+it_4,t_5\sqrt 2,it_3-it_4,-t_1-t_2          )$ \\
$\scriptstyle (4,2)$ & $\scriptstyle --  $  & $\scriptstyle 2$         & $\scriptstyle \left\langle C_{2z}\right\rangle$                                                           & $\scriptstyle (v_1,v_2,0,0,v_3           )$           & $\scriptstyle (-iv_1-v_2,0,v_3\sqrt 2,0,iv_1-v_2                        )$ \\
$\scriptstyle (6,1)$ & $\scriptstyle --  $  & $\scriptstyle 1$         & $\scriptstyle \{{\bf 1} \}$                                                                               & $\scriptstyle (v_1,v_2,v_3,v_4,v_5       )$           & $\scriptstyle (-iv_1-v_2,iv_3+v_4,v_5\sqrt 2,iv_3-v_4,iv_1-v_2          )$ \\
\end{tabular}
\end{table}

\begin{table}
 \caption{Strata of dimensions $\ge 3$, bordering each stratum $\widehat S^{(d,r)}$ ($d$ denotes the dimension of the stratum and $r$ is
 an enumeration index).
 From left to right, the columns refer to the enumeration numbers, $(d,r)$, of the lower bordering
 stratum, $\widehat S^{(d,r)}$, its orbit-type, $[G^{(d,r)}]$, the maximal conjugacy classes of subgroups of
 $[G^{(d,r)}]$, $[H]_M$, the order, $|H_M|$, of $H_M$ and the indices, $(d',r')$, of the corresponding strata, if any.
 The definitions of the group elements are recalled in Appendix D. \label{IV}}

\begin{tabular}{cccccc}
$(d,r)$ & $[G^{(d,r)}]$                                                    & $[H]_M$                                             & $|H_M|$& $(d',r')$  \\ \hline
(2,1)   & $[\left\langle C_{2x}{\cal T},C_{3z}U_1(4\pi/3) \right\rangle]$  & $[\left\langle C_{3z}U_1(4\pi/3)\right\rangle]$     & 3      &  --        \\
        &                                                                  & $[\left\langle C_{2x}{\cal T}\right\rangle]$        & 2      & (4,1)      \\
(2,2)   & $[\left\langle  C_{2x}{\cal T},C_{2z}\,U_1(\pi)\right\rangle]$   & $[\left\langle C_{2x}{\cal T}\right\rangle]$        & 2      & (4,1)      \\
        &                                                                  & $[\left\langle C_{2z}{\rm U}_1(\pi)\right\rangle]$  & 2      &  --        \\
(2,3)   & $[\left\langle C_{2z}, C_{4x}{\cal T}\right\rangle]$             & $[\left\langle C_{2z},C_{2x}{\cal T}\right\rangle]$ & 4      & (3,1)      \\
        &                                                                  & $[\left\langle C_{2x},C_{2z}\right\rangle$          & 4      & (3,2)      \\
        &                                                                  & $[\left\langle C_{4z}{\cal T} \right\rangle]$       & 4      &  --        \\
(2,4)   & $[\left\langle C_{2x}, C_{2z}, {\cal T}\right\rangle ]$          & $[\left\langle C_{2x},C_{2z}\right\rangle]$         & 4      & (3,2)      \\
        &                                                                  & $[\left\langle C_{2z},C_{2x}{\cal T}\right\rangle]$ & 4      & (3,1)      \\
(2,5)   & $[\left\langle  C_{2x}{\cal T},C_{4z}\,U_1(\pi)\right\rangle]$   & $[\left\langle C_{2z},C_{2x}{\cal T}\right\rangle]$ & 4      & (3,1)      \\
        &                                                                  & $[\left\langle C_{4z}\,U_1(\pi)\right\rangle ]$     & 4      &  --        \\
(3,1)   & $[\left\langle  C_{2z}, C_{2x}\,{\cal T}\right\rangle]$          & $[\left\langle C_{2x}{\cal T}\right\rangle]$        & 2      & (4,1)      \\
        &                                                                  & $[\left\langle C_{2z}\right\rangle]$                & 2      & (4,2)      \\
(3,2)   & $[\left\langle  C_{2x},C_{2z}\right\rangle]$                     & $[\left\langle C_{2z}\right\rangle]$                & 2      & (4,2)      \\
(4,1)   & $[\left\langle C_{2x}{\cal T}\right\rangle]$                     & $[{\uno}]$                                          & 1      & (6,1)      \\
(4,2)   & $[\left\langle C_{2z}\right\rangle]$                             & $[{\uno}]$                                          & 1      & (6,1)      \\
\end{tabular}
\end{table}

\begin{table}
\caption{Absolute minimum, $\widehat\Phi^{(4)}_{\rm min}=-\alpha_1^2/(2\delta)$, of a general, bounded below, $G$-invariant
4-th degree polynomial, $\widehat\Phi^{(4)}(\alpha,p)=\alpha_0 p_1^2/2 + \sum_{j=1}^3 \alpha_j p_j$, $\alpha_1<0$, and
hosting strata, $S^{(d,r)}$, as functions of the coefficients $\alpha$.
The denomination of the strata is the same as in Table~\ref{III}. \label{V} }
\begin{tabular}{ccc}
$\alpha$ range                                                 & $\delta$                           & $(d,r)$   \\ \hline
${\cal R}_1:\ {\rm Max}(0,-3\alpha_0/2,-6\alpha_3)<\alpha_2$                & $\alpha_0 + 2\alpha_2/3$          & (1,4)          \\
${\cal R}_2:\ -6\alpha_3>\alpha_2>{\rm Max}(-\alpha_0-2\alpha_3,2\alpha_3)$ & $\alpha_0 + \alpha_2 + 2\alpha_3$ & (1,2), (1,3), (2,4) \\
${\cal R}_3:\ -\alpha_0/2 < \alpha_2 < {\rm Min}(0,2\alpha_3)$              & $\alpha_0 + 2\alpha_2$            & (1,1)          \\
${\cal R}_{13}:\ 0=\alpha_2<{\rm Min}(\alpha_0,\alpha_3)$                      & $\alpha_0$                     & (1,1), (1,4), (1,5), (2,1), (3,1), (4,1)\\
${\cal R}_{12}:\ {\rm Max}(-3\alpha_0/2,0)<\alpha_2=-6\alpha_3$                & $\alpha_0-4\alpha_3$           & (1,2), (1,3), (1,4), (2,3), (2,4), (3,2)\\
${\cal R}_{23}:\ -\alpha_0/2<\alpha_2=2\alpha_3<0$                             & $\alpha_0+4\alpha_3$           & (1,1), (1,2), (1,3), (2,4), (2,5)\\
${\cal R}_{123}:\ \alpha_2 = \alpha_3 = 0 < \alpha_0$                           & $\alpha_0$                    & all, except (0,1)   \\
\end{tabular}
\end{table}

\begin{center}
\begin{figure}
\caption{Possible phase transitions between bordering strata, connected, in the figure, by continuous
sequences of one or more arrows.  The group elements are defined in Appendix D.}
\label{F1}

\begin{texdraw}

\drawdim mm

\move(80 180) \textref h:C v:C \htext{\ }

\move(80 150) \textref h:C v:C
\htext{\framebox{\scriptsize$\def\arraystretch{0.8}\begin{array}{c}
S^{(0)}=\{0\}\\{\rm\bf SO}_3\otimes{\rm\bf U}_1\times{\cal T}\end{array}$}}

\move(80 144) \avec(4 127)
\move(80 144) \avec(47 127)
\move(80 144) \avec(85 127)
\move(80 144) \avec(115 127)
\move(80 144) \avec(150 127)

\textref h:C v:C \htext(4 120){\framebox{\scriptsize$\def\arraystretch{0.8}\begin{array}{c}  S^{(1,5)}\\
{[}\left\langle C_{2x}{\cal T}\right\rangle \times\{{\rm R}_z(\phi){\rm U}_1(-\phi)\}_\phi {]} \end{array}$}}

\textref h:C v:C \htext(47 120){\framebox{\scriptsize$\def\arraystretch{0.8}\begin{array}{c}  S^{(1,4)}\\
{[}\left\langle C_{2x},C_{3\delta}{\rm U}_1(\frac{4\pi}{3}),C_{2a}{\cal T}\right\rangle {]} \end{array}$}}

\textref h:C v:C \htext(85 120){\framebox{\scriptsize$\def\arraystretch{0.8}\begin{array}{c} S^{(1,3)}\\ {[}\left\langle
C_{2x},C_{4z}\,{\rm U}_1(\pi),{\cal T}\right\rangle {]} \end{array}$}}

\textref h:C v:C \htext(115 120){\framebox{\scriptsize$\def\arraystretch{0.8}\begin{array}{c} S^{(1,2)}\\
{[}{\rm\bf O}_2^z\times\left\langle {\cal T}\right\rangle {]} \end{array}$}}

\textref h:C v:C \htext(150 120){\framebox{\scriptsize$\def\arraystretch{0.8}\begin{array}{c} S^{(1,1)}\\
 {[}\left\langle C_{2x}{\cal T}\right\rangle  \times  \{{\rm R}_z(\phi){\rm
 U}_1(2\phi)\}_\phi
{]} \end{array}$}}

\move(4 114) \avec(10 97)\move(4 114)  \avec(47 97)
\move(47 114) \avec(10 97)\move(47 114)  \avec(79 97)
\move(85 114) \avec(47 97) \move(85 114)  \avec(79 97) \move(85 114) \avec(111 97)\move(85 114)  \avec(146 97)
\move(115 114) \avec(79 97)\move(115 114)  \avec(111 97)
\move(150 114) \avec(10 97)\move(150 114)  \avec(146 97)

\textref h:C v:C \htext(10 90){\framebox{\scriptsize$\def\arraystretch{0.8}\begin{array}{c} S^{(2,1)}\\ {[}\left\langle
C_{2x}{\cal T},C_{3z}{\rm U}_1(\frac{4\pi}{3})\right\rangle {]} \end{array}$}}

\textref h:C v:C \htext(47 90){\framebox{\scriptsize$\def\arraystretch{0.8}\begin{array}{c} S^{(2,2)}\\ {[}\left\langle
C_{2x}{\cal T},C_{2z}{\rm U}_1(\pi)\right\rangle {]}  \end{array}$}}

\textref h:C v:C \htext(79 90){\framebox{\scriptsize$\def\arraystretch{0.8}\begin{array}{c}S^{(2,3)}\\ {[}\left\langle C_{2x},C_{4z}{\cal T}
\right\rangle {]}\end{array}$}}

\textref h:C v:C \htext(111 90){\framebox{\scriptsize$\def\arraystretch{0.8}\begin{array}{c} S^{(2,4)}\\ {[}\left\langle
C_{2x},C_{2z},{\cal T}\right\rangle {]}\end{array}$}}

\textref h:C v:C \htext(146 90){\framebox{\scriptsize$\def\arraystretch{0.8}\begin{array}{c} S^{(2,5)}\\
{[}\left\langle C_{4z}{\rm U}_1(\pi),C_{2x}{\cal T}\right\rangle {]} \end{array}$}}

\move(10 83) \avec(47 37)
\move(47 83) \avec(47 37)
\move(79 83) \avec(79 67) \move(79 83)  \avec(111 67)
\move(111 83) \avec(79 67) \move(111 83)  \avec(111 67)
\move(146 83) \avec(111 67)

\textref h:C v:C \htext(79 60){\scriptsize\framebox{$\def\arraystretch{0.8}\begin{array}{c} S^{(3,2)}\\ {[}\left\langle
C_{2x},C_{2z}\right\rangle {]}
\end{array}$}}

\textref h:C v:C \htext(111 60){\scriptsize\framebox{$\def\arraystretch{0.8}\begin{array}{c} S^{(3,1)}\\ {[}\left\langle
C_{2z},C_{2x}{\cal T}\right\rangle {]}\end{array}$}}

\move(79 54) \avec(79 37)
\move(111 54) \avec(47 37) \move(111 54) \avec(79 37)

\textref h:C v:C \htext(47 30){\scriptsize\framebox{$\def\arraystretch{0.8}\begin{array}{c} S^{(4,1)}\\ {[}\left\langle
 C_{2x}\,{\cal T}\right\rangle {]}\end{array}$}}

\textref h:C v:C \htext(79 30){\scriptsize\framebox{$\def\arraystretch{0.8}\begin{array}{c}  S^{(4,2)}\\ {[}\left\langle
C_{2z} \right\rangle {]} \end{array}$}}

\move(46 24) \avec(80 7)
\move(79 24) \avec(80 7)

\textref h:C v:C \htext(80 0){\scriptsize\framebox{$\def\arraystretch{0.8}\begin{array}{c}  S^{(6,1)}=S_{\rm principal} \\
 \{{\bf e}\}\end{array}$}}

\end{texdraw}
\end{figure}
\end{center}

\newpage

\begin{thebibliography}{pippoplutominnie}
\bibitem{1} P. W. Anderson, P. Morel  Phys. Rev. {\bf  123}, 1911
(1961).
\bibitem{1a} D. D. Osheroff, R.C. Richardson, D.M. Lee, Phys. Rev.
Lett. {\bf 28}, 885 (1972).
\bibitem{2} M. Sigrist, K. Ueda, Rev. Mod. Phys. {\bf 63}, 239
(1991).
\bibitem{8}  D. J. van Harlingen,  Rev. Mod. Phys.  {\bf 67}, 515
(1995).
\bibitem{5bis} H. Ghosh, J. Phys. Condens. Matter {\bf 11}, L371
(1999).
\bibitem{10a} J. Orenstein, Nature {\bf 401}, 333 (1999).
\bibitem{10b} J.F. Annet, Physica {\bf C 317-318}, 1 (1999).
\bibitem{GufPR12} Yu. M. Gufan, G. M. Vereshkov, P. Tol\'{e}dano,
B. Mettout, R. Bouzerar and V. Lorman, Phys. Rev. B{\bf 51}, 9219
and  9228 (1995).
\bibitem{11} K. Nakamura, A. Yu. Gufan, Yu. M. Gufan and E. G.
Rudashevskii, Crystallography Reports {\bf 44}, 469 and 603
(1999).
\bibitem{12} Yu. M. Gufan, I. G. Levchenko and E. G. Rudashevskii,
Physics of the Solid State {\bf 41}, 1422 (1999) and references
therein.
\bibitem{14} N. D. Mermin, Phys. Rev. A {\bf 9}, 868 (1974).
\bibitem{16} A. M. J. Schakel, F. A. Bias, J. Phys.: Condens. Matter
{\bf 1}, 1743(1989).
\bibitem{15} H. W. Capel, A. M. J. Schakel, Physica A {\bf 160}, 409
(1989).
\bibitem{SA} M. Abud and G. Sartori, Phys. Lett. {\bf B 104}, 147
(1981) and  Ann. Phys. {\bf 150}, 307 (1983).
\bibitem{23} Yu. M. Gufan, Sov. Phys. Solid State {\bf 13}, 175
(1971).
\bibitem{NC} G. Sartori, Supplemento del Nuovo Cimento
{\bf 14}, 1 (1991).
\bibitem{PS} C. Procesi and G. W. Schwarz, Inv. Math. {\bf 81},
539 (1985).
\bibitem{33} E. B. Vinberg and V. L. Popov, {\em Invariant
Theory},
in Encyclopaedia of Mathematical Sciences, Vol. {\bf 55}
(Springer-Verlag, Berlin, 1994), pp. 123-284.
\bibitem{3334} V. L. Popov, {\em Groups, Generators, Syzygies and
Orbits in Invariant Theory}, Trans. Math. Monog. Vol. {\bf 100},
(A.M.S., Providence, 1992).
\bibitem{Stanley} R. Stanley, Bull. Am. Math. Soc. New. Ser. {\bf
1}, 475 (1979).
\bibitem{SV} G. Sartori and G. Valente, in preparation.
\bibitem{Cornwell} J. F. Cornwell, {\em Group Theory in Physics}, (Academic, London,  1984),Vol.
I.
\bibitem{SV1} G. Sartori and G. Valente, J. Phys. A
{\bf 29}, 193 (1996).
\bibitem{ST} G. Sartori and V. Talamini, J. Math. Phys. {\bf 39},
2367 (1998).
\bibitem{Walker} R. J. Walker, {\em Algebraic Curves}, (Dover,
 New York, 1950) (Theorem. 5.1 p. 67).
\end{thebibliography}
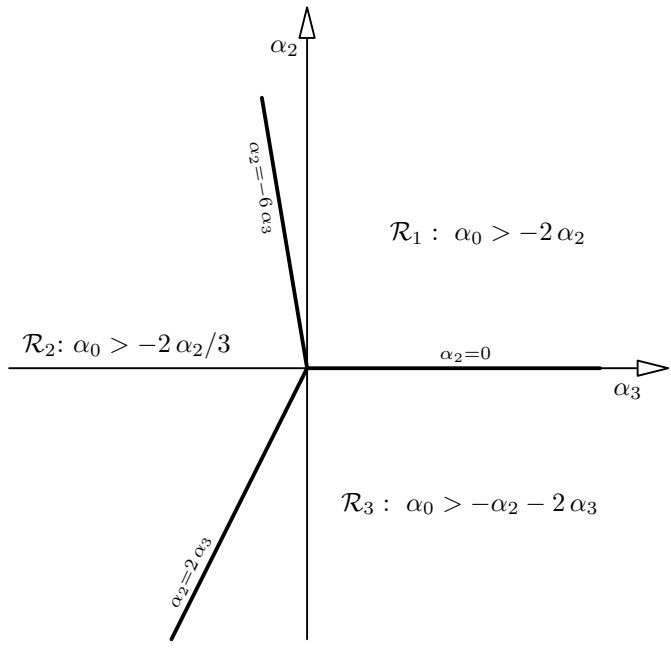
\end{document}